\documentclass[12pt,final,floatfix,eqsecnum,onecolumn,amsfonts,amsmath,amssymb,nofootinbib]{revtex4}

\usepackage[sf,bf,small]{titlesec}
\usepackage[font=footnotesize,labelfont=bf,justification=raggedright]{caption}
\usepackage{mathrsfs}
\usepackage{graphicx}
\usepackage{bm}
\usepackage{bbm}
\usepackage{xr}

\providecommand*{\I}{\mathrm{i}}                           
\providecommand*{\bra}[1]{\langle#1|}                      
\providecommand*{\ket}[1]{|#1\rangle}                      
\providecommand*{\rbra}[1]{(#1|}                           
\providecommand*{\rket}[1]{|#1)}                           
\providecommand*{\lrbra}[1]{\rbra{\widetilde{#1}}}    
\providecommand*{\lrbracket}[2]{\lrbra{#1}#2)}       
\providecommand*{\Pro}{\underline{\mathbbm{P}}}
\providecommand*{\klr}[1]{\left(#1\right)}					 
\providecommand*{\kle}[1]{\left[#1\right]}					 
\providecommand*{\klg}[1]{\left\{#1\right\}}				 

\providecommand*{\mrmd}{d}									 

\DeclareMathOperator{\Tr}{Tr}													

\newcommand*{\mcal}[1]{\mathcal{#1}}                   

\newcommand*{\umat}[1]{\underline{\mathscr{#1}}}      
\newcommand*{\umc}[1]{\underline{\mathcal{#1}}}      
\renewcommand{\vec}[1]{\bm{#1}}                       
\newcommand*{\uvec}[1]{\underline{\bm{#1}}}           
\newcommand*{\vmc}[1]{\vec{\mathcal{#1}}}							
\newcommand{\LambShift}{L}											
\newcommand{\FineStructure}{\Delta}										
\newcommand{\HyperFineSplitting}{\mcal{A}}						
\newcommand{\sez}{\vec{e}_0}													
\newcommand{\sep}{\vec{e}_+}													
\newcommand{\sem}{\vec{e}_-}													

\newcommand{\mycell}[2]{\parbox{#1}{\vskip4pt #2 \vskip4pt}}	

\newcommand{\PV}{\mathrm{PV}}													
\newcommand{\PC}{\mathrm{PC}}													
\newcommand{\um}{\underline{\mathbbm 1}}							

\renewcommand{\thesection}{\arabic{section}}
\renewcommand{\thesubsection}{\!.\arabic{subsection}}

\titleformat{name=\section}[block]{\large\bfseries\sffamily}{\thesection.}{1ex}{}
\titleformat{name=\subsection}{\bfseries\sffamily}{\arabic{section}.\arabic{subsection}.}{1ex}{}

\newcommand{\extref}[1]{I.\ref{#1}} 

\begin{document}

\title{\Large Metastable states, the adiabatic theorem\\ and parity violating geometric phases II}

\author{Timo~Bergmann}
\email{T.Bergmann@ThPhys.Uni-Heidelberg.DE}

\author{Thomas~Gasenzer}
\email{T.Gasenzer@ThPhys.Uni-Heidelberg.DE}

\author{Otto~Nachtmann}
\email{O.Nachtmann@ThPhys.Uni-Heidelberg.DE}

\affiliation{Institut f{\"u}r Theoretische Physik, Universit{\"a}t Heidelberg,\\ 
Philosophenweg 16, 69120 Heidelberg, Germany}

\date{\today}

\begin{abstract}
We discuss and calculate parity conserving (PC) and parity violating (PV)  geometric phases for the metastable $2S$ states of hydrogen and deuterium. The atoms are supposed to be subjected to slowly varying electric and magnetic fields which act as external parameters for the atoms. Geometric flux density fields are introduced which allow for an easy overview how to choose the paths in parameter space in order to obtain only PC or only PV geometric phases. The PV phases are calculated in the Standard Model of particle physics. Even if numerically they come out small they have interest of principle as a new manifestation of parity violation in atomic physics.
\\[-4ex]

\hfill
{\small HD--THEP--07--09}
\end{abstract}

\maketitle

\section{Introduction}\label{s:Introduction}

In this article we discuss the atomic states of hydrogen and deuterium with principal quantum number $n=2$ in slowly varying electric and magnetic fields. We are interested in parity (P) violating effects for the metastable $2S$ states.

In the accompanying paper \cite{BeGaNa07_I} we have already discussed the motivation for our study. There we have made a detailed investigation of the adiabatic limit for a system containing metastable and short lived states. In the following the sections of \cite{BeGaNa07_I} shall be quoted as section I.1 etc. and equation numbers as (I.1.1) etc.

Our present paper II is organised as follows. In section \ref{s:Basics} we review briefly the parity violating Hamiltonian relevant for our atomic systems. We define our notation for the $n=2$ states and give the mass matrix for these states in external electric and magnetic fields. In this paper we always consider states at rest. The case of atoms travelling in the atomic beam interferometer \cite{ABSE95} will be treated elsewhere.
In section \ref{s:AdiabaticLimit} we study the adiabatic limit for the metastable $2S$ states using the results of section \extref{I.s:MetastableStates}. We identify the P-conserving (PC) and P-violating (PV) contributions to the geometric phases.
In section \ref{s:Results} we calculate these phases for the case that the geometric phase factor matrix is diagonal. In this case we have only abelian geometric phases. We give a graphical representation of these geometric phases in terms of surface integrals over geometric flux densities in the space of electric and magnetic fields. This allows us to see in an easy way how the paths in parameter space have to be chosen in order to get either pure PC or pure PV phases. Section \ref{s:Conclusions} contains our conclusions and an outlook. In appendix \ref{s:Values} we collect tables giving the relevant numerical quantities of our systems. Appendix \ref{s:Adiabaticity} contains a discussion of the adiabaticity condition for our concrete cases and appendix \ref{s:GeometricFluxes} gives the detailed calculations for the geometric flux densities.

We use units $\hbar=c=1$ if other units are not explicitly indicated.

\section{Hamiltonian and state vectors}\label{s:Basics}

\subsection{The P-violating Hamiltonian}\label{s:HPV}

In the framework of the Standard Model (SM) \cite{Wei67,Sal68,Gla70} the effective P-violating Hamiltonian relevant for atomic physics is due to the exchange of a $Z$ boson between the atomic electrons and the quarks in the nucleus. In terms of the electron and quark field operators we have
\begin{align}\label{e1:eff.HPV}
H_\PV       &= H^{(1)}_\PV + H^{(2)}_\PV\ ,\\ \label{e1:eff.HPV.1}
H^{(1)}_\PV &= -\frac{G}{\sqrt2}\int\mrmd^3x\ 2g_A^e\bar e(\vec x)\gamma^\mu\gamma_5 e(\vec x)
\klr{\sum_q g_V^q\bar q(\vec x)\gamma_\mu q(\vec x)}\ ,\\ \label{e1:eff.HPV.2}
H^{(2)}_\PV &= -\frac{G}{\sqrt2}\int\mrmd^3x\ 2g_V^e\bar e(\vec x)\gamma^\mu e(\vec x)
\klr{\sum_q g_A^q\bar q(\vec x)\gamma_\mu \gamma_5 q(\vec x)}\ ,
\end{align}
where $q = u,d,s$, neglecting possible contributions from the heavy quarks $c,b,t$. The $g_{A,V}^{e}$ and $g_{A,V}^{q}$ are the neutral current coupling constants for the electron and the quarks, respectively, and $G$ is Fermi's constant. For the notation and the definitions see \cite{Nac90} and \cite{BoBrNa95}. In the framework of the SM the coupling constants of the weak neutral current are
\begin{align}\label{e1:gVA.e}
g_V^e &= -\tfrac12 + 2\sin^2\vartheta_W\ ,& g_A^e &= -\tfrac12\ ,\\ \label{e1:gVA.u}
g_V^u &= \phantom-\tfrac12 - \tfrac43\sin^2\vartheta_W\ ,& g_A^u &= \phantom-\tfrac12\ ,\\ \label{e1:gVA.ds}
g_V^{d,s} &= -\tfrac12 + \tfrac23\sin^2\vartheta_W\ ,& g_A^{d,s} &= -\tfrac12
\end{align}
where $\vartheta_W$ is the weak mixing angle.

For our study of atomic parity violation (APV) in light atoms, it is sufficient to consider a point-like, infinitely heavy, nucleus along with a nonrelativistic approximation of $H_\PV$ (for details see \cite{BoBrNa95}). In the nonrelativistic reduction, the effective P-violating Hamiltonians (\ref{e1:eff.HPV.1}), (\ref{e1:eff.HPV.2}) for an atom with proton number $Z$ and neutron number $N$ read
\begin{align}\label{e1:HPV1}
H^{(1)}_\PV &= \frac{G}{4\sqrt2}\frac1{m_e}Q_W^{(1)}(Z,N)\klg{\delta^3(\vec x)(\vec\sigma\cdot\vec p) + (\vec\sigma\cdot\vec p)\delta^3(\vec x)}\ ,\\ \label{e1:HPV2}
H^{(2)}_\PV &= \frac{G}{4\sqrt2}\frac1{m_e}Q_W^{(2)}(Z,N)\klg{\delta^3(\vec x)(\vec I\cdot\vec\sigma)(\vec\sigma\cdot\vec p) + (\vec\sigma\cdot\vec p)(\vec I\cdot\vec\sigma)\delta^3(\vec x)}\ .
\end{align}
Here $\vec\sigma$ and $\vec p$ are the Pauli spin matrix vector and the momentum operator for the electron, respectively, and $\vec I$ is the nuclear spin operator. The $Q_W^{(1,2)}(Z,N)$ are the weak charges of the atomic nucleus, given by
\begin{align}\label{e1:QW1}
Q_W^{(1)}(Z,N) &= -4g_A^e\klg{g_V^u(2Z + N) + g_V^d(Z+2N)}\ ,\\ \label{e1:QW2}
Q_W^{(2)}(Z,N) &= \phantom-4g_V^e\sum_q g_A^q \frac1I\Delta q(Z,N)\ .
\end{align}
The weak charge $Q_W^{(2)}(Z,N)$ exists, of course, only for nuclei with spin $I\neq 0$. The quantities $\Delta q(Z,N)$ are the
total polarisations of the nucleus carried by the quark species $q$. In the SM we have with (\ref{e1:gVA.e}) to (\ref{e1:gVA.ds})
\begin{align}\label{e1:QW1.SM}
Q_W^{(1)}(Z,N) &= \phantom-(1-4\sin^2\vartheta_W)Z - N\ ,\\ \label{e1:QW2.SM}
Q_W^{(2)}(Z,N) &= -\frac1I(1-4\sin^2\vartheta_W)\kle{\Delta u(Z,N)-\Delta d(Z,N)-\Delta s(Z,N)}\ .
\end{align}

It was pointed out recently in \cite{Bou05} that the measurement of $Q^{(1)}_W(Z,N)$ is still the main motivation for APV studies, since it allows for a determination of $\sin^2\vartheta_W$, which is complementary to high-energy physics. Such low-energy determinations of $\sin^2\vartheta_W$ have, of course, been done with neutrino and electron scattering experiments. For a review see section 10 in \cite{PDG06}. It is a challenge for experimentalists to see which method can give the best precision.
As is clear from (\ref{e1:QW2.SM}) APV measurements can also contribute to the study of the spin structure of light nuclei. For recent reviews of the available information on the spin structure of the nucleons see for instance \cite{Air07,Bas05,AsFl06,Bra06}.

\subsection{The $n=2$ states for hydrogen and deuterium}\label{s:Hydrogen}

We consider the subspace of atomic states with principal quantum number $n=2$ for hydrogen and deuterium. As discussed in section \extref{I.s:Generalities}, the effective Schr\"{o}dinger equation for these systems reads in the Wigner-Weisskopf approximation
\begin{align}
\I\frac{\partial}{\partial t}\rket t = \umat M(t)\rket{t}\ ,
\end{align}
where $\umat M(t)$ is the non-hermitian mass matrix and $\rket t$ the state vector of the undecayed atom in the $(n=2)$-subspace. Here and in the following we use, in contrast to the companion paper I,  the time $t$ instead of the reduced time $\tau$ (\extref{I.e2:red.time}) as variable for time dependent quantities. That is, we set $\tau_0=T$ in (\extref{I.e2:red.time}). As in paper I we suppose that $\umat M(t)$ is diagonalisable for all times $t$. Then we have for each time $t$ a complete set of right and left eigenvectors satisfying
\begin{align}\label{e2:EWG}
\begin{split}
\umat M(t)\rket{\alpha,t} &= E(\alpha,t)\rket{\alpha,t}\ ,\\
\lrbra{\alpha,t}\umat M(t) &= \lrbra{\alpha,t}E(\alpha,t)\ ,\\
E(\alpha,t) &= E_R(\alpha,t) - \frac\I2\Gamma(\alpha,t)\ ,\\
(\alpha &= 1,\ldots,N)\ .
\end{split}
\end{align}
Here $E(\alpha,t)$ are the complex energy eigenvalues which can be separated into a real part
$E_R(\alpha,t)$ and an imaginary part $-\frac12\Gamma(\alpha,t)$ leading to an exponential decay of the atomic state with decay rate $\Gamma(\alpha,t)$ (see section \extref{I.s:Generalities}).

For hydrogen with principal quantum number $n=2$ there are $N=16$ basis states, whereas for deuterium we have $N=24$ states. As a set of basis vectors we choose coupled states of nuclear spin $\ket{I,I_3}$, electron spin $\ket{\tfrac12,S_3}$ and electron orbital angular momentum $\ket{n,L,L_3}$. We denote these basis states by $\ket{n L_J,F,F_3}$, where $J$ is the total angular momentum of the electron and $F,F_3$ are the quantum numbers for the total angular momentum of the atom.

The mass matrix $\umat M(t)$ contains contributions from the external fields and the PV Hamiltonians (\ref{e1:HPV1}), (\ref{e1:HPV2}). The PV Hamiltonians mix the $2S$ and $2P$ states of hydrogen-like atoms in the $(n=2)$-subspace. In the Coulomb approximation for the wave functions we get the following matrix elements
\begin{align}\label{e2:HPV1}
\bra{2S_{1/2},F',F_3'}H^{(1)}_\PV\ket{2P_{1/2},F,F_3} 
&= -\I\delta_1(Z,N)L(Z,N)\,\delta_{F',F}\delta_{F'_3,F_3}\ ,\\ \label{e2:HPV2}
\begin{split}
\bra{2S_{1/2},F',F_3'}H_\PV^{(2)}\ket{2P_{1/2},F,F_3} 
&= -\I\delta_2(Z,N)L(Z,N)\kle{F(F+1)-I(I+1)-\tfrac34}\\ &\qquad\times \delta_{F',F}\delta_{F'_3,F_3}\ ,\\
\Big(|I-\tfrac12| \leq F, F' &\leq I+\tfrac12\Big)\ ,
\end{split} 
\end{align}
and
\begin{align}\label{e2:HPV.zeros}
\begin{split}
\bra{2S_{1/2},F',F_3'}H^{(1)}_\PV\ket{2P_{3/2},F,F_3} &= 0\ ,\\
\bra{2S_{1/2},F',F_3'}H^{(2)}_\PV\ket{2P_{3/2},F,F_3} &= 0\ ,\\
\Big(|I-\tfrac12| \leq F' \leq I+\tfrac12;\ |I-\tfrac32| \leq F &\leq I+\tfrac32\Big)\ .
\end{split}
\end{align}
In (\ref{e2:HPV1}), (\ref{e2:HPV2}) define the PV parameters
\begin{align}
\label{e2:deltaPV}
\begin{split}
\delta_i(Z,N) &= -\frac{\sqrt3\,G}{64\pi\sqrt2\,r_B^4(Z) m_e}\frac{Q_W^{(i)}(Z,N)}{L(Z,N)}\\
&= -2.68827(3)\cdot 10^{-17}\,\frac{Z^4\,Q_W^{(i)}(Z,N)}{L(Z,N)}\,\mathrm{eV}
\ ,\\
(i &= 1,2)\ .
\end{split}
\end{align}
Here $r_B(Z) = (Z\alpha m_e)^{-1} = Z^{-1}r_B(1)$ is the first Bohr radius and $L(Z,N)=E(2S_{1/2})-E(2P_{1/2})$ is the Lamb shift for a hydrogen-like atom with proton number $Z$ and neutron number $N$. The numerical factor in (\ref{e2:deltaPV}) was obtained using the values for $G$, $r_B(1)$ and $m_e$ from \cite{PDG06}. 

With the Lamb shifts listed in table \ref{t:values} of appendix \ref{s:Values} we get for ordinary hydrogen
\begin{align}\label{e2:deltaPV12H}
\delta_i(1,0) &= -6.14477(6)\cdot 10^{-12}\ Q_W^{(i)}(1,0)\ ,\quad (i=1,2)\ ,
\end{align}
and for deuterium
\begin{align}\label{e2:deltaPV12D}
\delta_i(1,1) &= -6.13671(6)\cdot 10^{-12}\ Q_W^{(i)}(1,1)\ ,\quad (i = 1,2)\ .
\end{align}
The numerical values of $\delta_{1,2}$ for hydrogen and deuterium can be found in table \ref{t:values} of appendix \ref{s:Values}. The mass matrix for these atoms in the $(n=2)$-subspace including PV contributions can now be written as
\begin{align}\label{e2:MM.full}
\umat M(t) = \umat M_0 - \uvec{D}\cdot\vmc E(t) - \uvec{\mu}\cdot\vmc B(t) + \delta_1\umat M_\PV^{(1)} + \delta_2\umat M_\PV^{(2)}\ ,
\end{align}
where we define for $i=1,2$
\begin{align}\label{e2:MPV}
\delta_i\umat M_\PV^{(i)} = \Big(\bra{2L'_{J'},F',F_3'}H^{(i)}_\PV\ket{2L_J,F,F_3}\Big)\ .
\end{align}
The mass matrix for zero external fields, 
\begin{align}\label{e2:MM.free}
\umat{\tilde M}_0(\delta_1,\delta_2) = \umat M_0 + \delta_1\umat M_\PV^{(1)} + \delta_2\umat M_\PV^{(2)}\ ,
\end{align}
and the matrix representations $\uvec D$ and $\uvec \mu$ of the electric and magnetic dipole operators are shown in tables \ref{t:H2.M0}, \ref{t:H2.D} and \ref{t:H2.Mu}
of appendix \ref{s:Values}, respectively.

At the end of this section, we introduce some notation. The atomic states for zero external fields, that is
the eigenstates of $\umat{\tilde M}_0(\delta_1,\delta_2)$, will be written either in the short-hand notation $\rket{\alpha,\delta_1,\delta_2}$ or, with explicit quantum numbers, in the form $\rket{2\hat L_J,F,F_3,\delta_1,\delta_2}$.
For non-zero electric and magnetic fields the eigenstates of the full mass matrix $\umat M(t)$ will be written in the forms $\rket{\alpha,t,\delta_1,\delta_2}$, $\rket{\alpha,\vmc E(t),\vmc B(t),\delta_1,\delta_2}$, and $\rket{2\hat L_J,F,F_3,\vmc E(t),\vmc B(t),\delta_1,\delta_2}$, depending on the context.

\section{Adiabatic limit and geometric phases}\label{s:AdiabaticLimit}

\begin{figure}
\includegraphics[width=13cm]{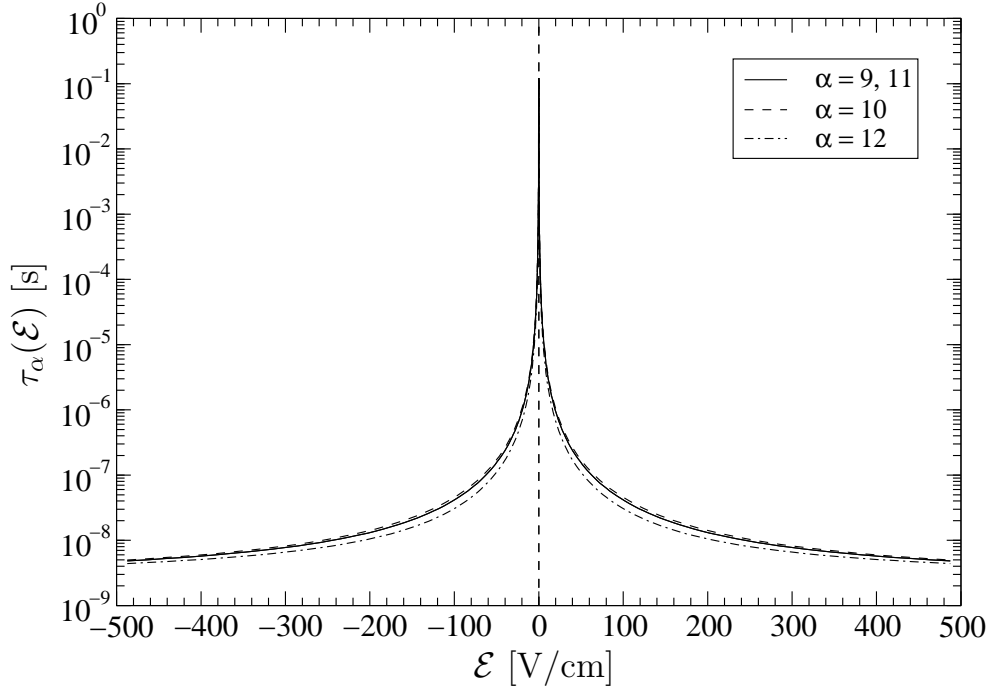}\\ {\bf (a)}\\[2mm]
\includegraphics[width=13cm]{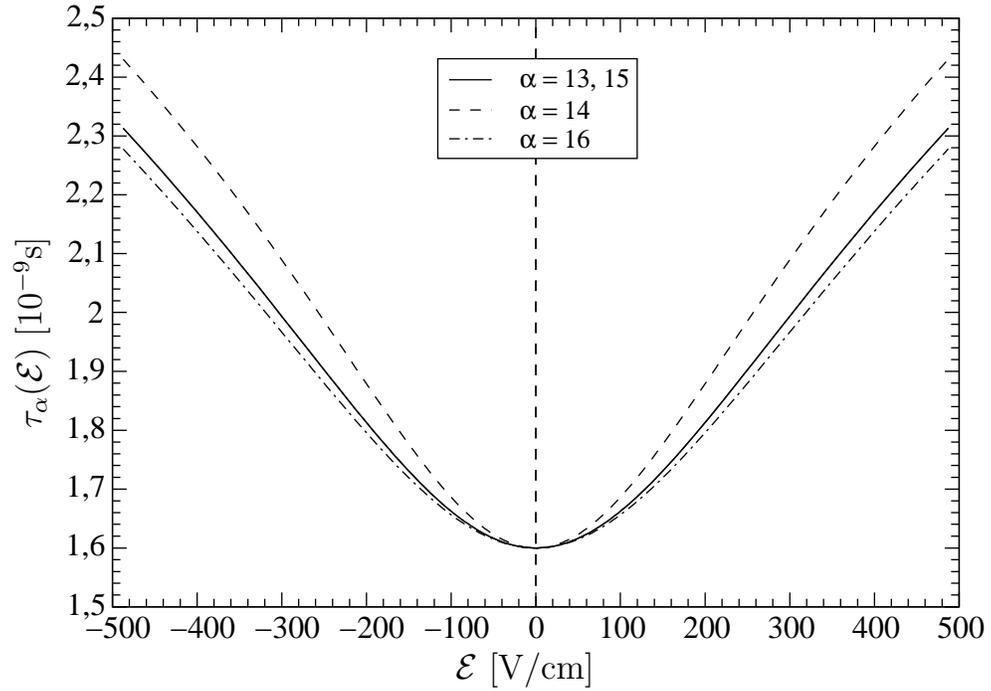}\\ {\bf (b)}
\caption{The lifetimes of the mixed $2S_{1/2}$ and $2P_{1/2}$ states of hydrogen in an external electric field $\vmc E = \mcal E\vec e_3$. Figure (a) shows the $2S_{1/2}$ states with labels $\alpha\in\klg{9,10,11,12}$, figure (b) shows the $2P_{1/2}$ states labelled by $\alpha\in\klg{13,14,15,16}$. The lifetimes of the $2P_{3/2}$ states with labels $\alpha\in\klg{1,\ldots,8}$ are not shown here. The variation of the $2P_{3/2}$ lifetimes is about two orders of magnitude smaller than that of the $2P_{1/2}$ lifetimes. For the numbering scheme see table \ref{t:state.labels} of appendix \ref{s:Values}.}\label{f:TauH}
\end{figure}

In the absence of external fields the $2S$ states are metastable, the $2P$ states decay fast. The theoretical lifetimes from recent QED calculations are
\begin{align}\label{e3:tauS}
\tau_S &= \Gamma_S^{-1} = 8.2207^{-1}\,\mathrm s = 0.1216\,\mathrm s\ ,\\ \label{e3:tauP}
\tau_P &= \Gamma_P^{-1} = (6.2649\cdot 10^8)^{-1}\,\mathrm s = 1.596\cdot 10^{-9}\,\mathrm s\ ,
\end{align}
see \cite{LaShSo05,Sap04}. An external electric field mixes the $2S$ and $2P$ states. The lifetimes of these mixed states are shown for hydrogen in figure \ref{f:TauH}, similar results can be obtained for deuterium. For the numbering scheme see table \ref{t:state.labels} of appendix \ref{s:Values}. Note that the metastable states are labelled in a different way as in I, where the numbering was $\alpha=1,\ldots,M$ for a total of $M$ metastable states.

In the electric field the $2S$ states, that is, the atomic states with $\alpha\in\klg{9,10,11,12}\equiv J^H_m$ for hydrogen and $\alpha\in\klg{13,14,15,16,17,18}\equiv J^D_m$ for deuterium quickly become more and more unstable with increasing $|\vmc E|$. Still, for
\begin{align}\label{e2:E.bound}
|\vmc E| \leq 250\,\mathrm{V/cm}
\end{align}
we have for the lifetimes of these states
\begin{align}\label{e2:tau.bound}
\tau_\alpha \gtrsim 10^{-8}\,\mathrm s\ ,\qquad (\alpha\in J_m^{H,D})\ ,
\end{align}
and
\begin{align}\label{e2:tau.ratio}
\tau_\alpha/\tau_\beta \gtrsim 5
\end{align}
for $\alpha\in J_m^{H,D}$ and $\beta$ any of the $2P$ states. The ratio of lifetimes (\ref{e2:tau.ratio}) should be sufficient in order to apply the results of I to the systems considered here.
In the following we thus have to keep in mind the restriction (\ref{e2:E.bound}) in the discussion of adiabatic limits and geometric phases. 

We suppose that our atom is subjected to slowly varying electric and magnetic fields. In I we have --- for mathematical convenience --- worked with the limit $T\to\infty$, making the variation of the fields smaller and smaller in real time. In reality we shall identify $T$ with the observation time being of the order of $\tau_S = \Gamma_S^{-1}$. We discuss the resulting adiabaticity condition in appendix \ref{s:Adiabaticity}, where based on the results of I we find that typically the rate of change of the external fields must satisfy
\begin{align}
\max_{t\in[0,T]}\frac{1}{\mcal E_0}\left|\frac{\partial\vmc E(t)}{\partial t}\right| &< \frac{1}{\tau_S}\ ,\\
\max_{t\in[0,T]}\frac{1}{\mcal B_0}\left|\frac{\partial\vmc B(t)}{\partial t}\right| &< \frac{1}{\tau_S}\ ,
\end{align}
where $\mcal E_0 = 477.3\,\mathrm{V/cm}$ and $\mcal B_0 = 45.65\,\mathrm{mT}$, see (\ref{B.7}), (\ref{B.13}), (\ref{B.16a}) and (\ref{B.18}). In the following we always suppose this to hold.

\begin{figure}[!Htbp]
\centering
\parbox{7cm}{\includegraphics[width=7cm]{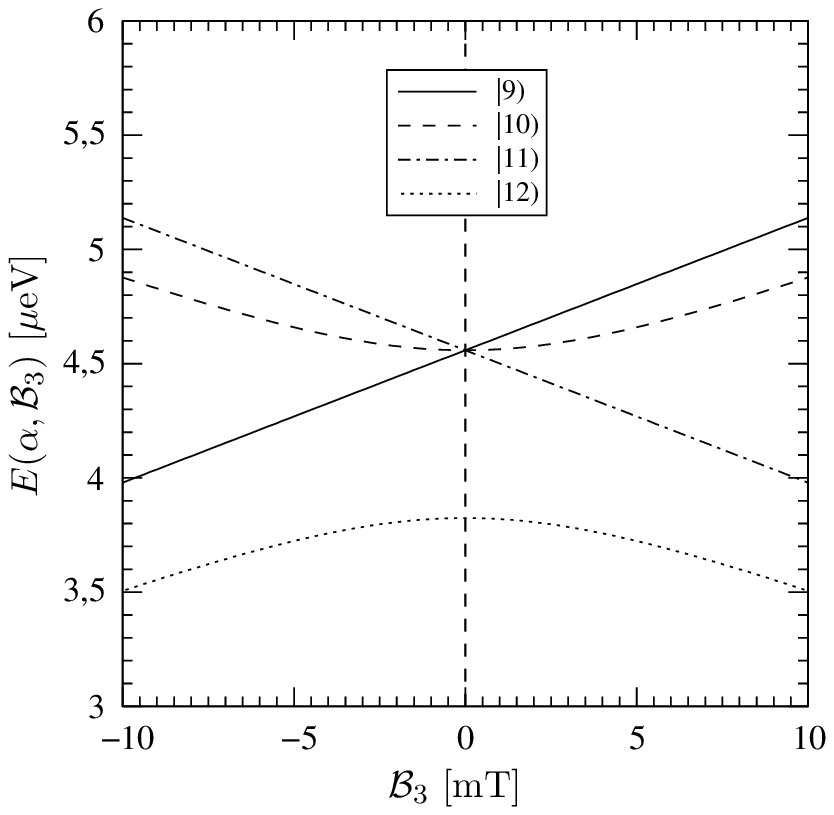}\\[-3mm] \textbf{(a)}}
\qquad
\parbox{7cm}{\includegraphics[width=7cm]{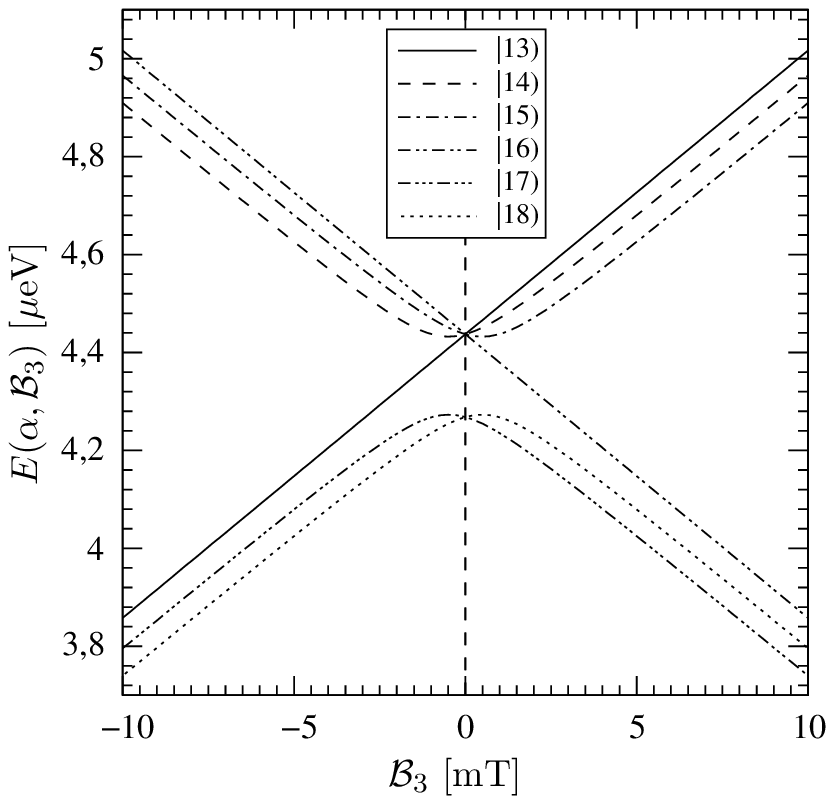}\\[-3mm] \textbf{(b)}}
\caption{The Breit-Rabi diagrams for the metastable states of hydrogen (a) and deuterium (b). The energy values are relative to the centre of zero-field energies of the $2P_{1/2}$ states.}\label{f:BreitRabi}
\end{figure}

Now we can apply the results of section \extref{I.s:MetastableStates}. We suppose that we start at time $t=0$ with metastable states only,
\begin{align}\label{e3:initial}
\psi_\alpha(0) = \begin{cases}
\chi_\alpha^{(0)}\qquad & \alpha\in J_m^{H,D}\\
0 & \text{otherwise}\ .
\end{cases}
\end{align}
The time evolution of the metastable states is then given by (\extref{I.e6:metastable.amplitudes.large.T}), neglecting non-adiabatic corrections,
\begin{align}\label{e4:nonad.corr}
\psi_\alpha(t) = \sum_{\beta\in J_m} U_{\alpha\beta}(t)\chi_\beta^{(0)}.
\end{align}
Here and in the following we use $J_m$ always understanding it as $J_m^H$ for hydrogen and $J_m^D$ for deuterium.
In the present paper we shall only discuss the case where, by suitable magnetic fields, the eigen\-energies of the metastable states are separated such that condition (\extref{I.e2:EW.non.deg}) holds. We have then, according to I, the case that the components of the metastable states evolve independently and no mixing occurs. In other words, the geometric phase matrix is diagonal (see (\extref{I.e6:abelian.amplitudes})-(\extref{I.e6:abelian.U.geom})) and we have (with $\tau_0=T$, $\tau=t$)
\begin{align}\label{e3:psidevelop}
\psi_\alpha(t) = \exp\kle{-\I\varphi_\alpha(t) + \I\gamma_{\alpha\alpha}(t)}\chi_\alpha^{(0)}
\end{align}
for $\alpha\in J_m$. Here $\varphi_\alpha(t)$ is the dynamical phase (see (\extref{I.e2:Dyn.Phase}))
\begin{align}\label{e3:dynphase}
\varphi_\alpha(t) = \int_0^t\mrmd t'\ E(\alpha,t')
\end{align}
and $\gamma_{\alpha\alpha}(t)$ is the geometric phase (\extref{I.e2:Berry.Phase}) given here by
\begin{align}\label{e3:BP}
\gamma_{\alpha\alpha}(t,\delta_1,\delta_2) = \int_0^t\mrmd t'\ \lrbra{\alpha,t',\delta_1,\delta_2}\I\frac{\partial}{\partial t'}\rket{\alpha,t',\delta_1,\delta_2}\ .
\end{align}

In detail we shall suppose that there is a constant magnetic field $\vmc B$ in 3-direction and that only the electric field $\vmc E(t)$ varies with time. The magnetic field alone will lead to the usual Breit-Rabi diagram of the energy levels. This is shown in figures \ref{f:BreitRabi}a and \ref{f:BreitRabi}b for the hydrogen and deuterium $2S$ states, respectively.

For a magnetic field $|\vmc B|\geq 1\,\mathrm{mT}$ the splittings in energy $\Delta E$ of the Zeeman levels satisfy
\begin{align}\label{e3:DeltaEbound}
\begin{split}
|\Delta E| &\gtrsim \Delta E_0 = 0.02\,\mu\mathrm{eV}\ ,\\
\Delta E_0/h &\cong 5\,\mathrm{MHz}\ .
\end{split}
\end{align}
Then the frequencies associated with the variation of the electric field should be much smaller than the one of (\ref{e3:DeltaEbound}). That is, we consider the Fourier analysis of the electric field strength:
\begin{align}
\vmc E(t) = \int_{-\infty}^\infty\frac{d\omega}{2\pi}\,e^{-\I\omega t}\vmc{\tilde E}(\omega)\ .
\end{align}
The requirement is that $|\vmc{\tilde E}(\omega)|\neq 0$ only for
\begin{align}
|\omega|\ll |\Delta E_0/\hbar| \cong 3\cdot 10^7\,\mathrm s^{-1}\ .
\end{align}
In other words, the variation of $\vmc{\tilde E}(t)$ should be negligible over time intervals of order $h/\Delta E_0 \cong 2\cdot 10^{-7}\mathrm s$. Then we will have a geometric phase matrix which is diagonal.

In \cite{BrGaNa99} it was shown that time reversal invariance forbids shifts linear in the P-violating parameters for the eigenenergies of states in a spatially constant electromagnetic field. This implies, in particular, that the energies $E(\alpha,t)$ and the dynamical phases $\varphi_\alpha(t)$ have no $\delta$-linear terms. They are insensitive to P-violation if we neglect terms of order $\delta^2$. The geometric phases, on the other hand, do have $\delta$-linear terms as we shall see explicitly below.

In the following it is convenient to discuss the geometric phases using the parameter space $\mcal R$, which in our case is the six dimensional space of electric and magnetic field strengths. The path $\mcal C\subset\mcal R$ along which the atom is led in $\mcal R$ is given by a time dependent parameter vector
\begin{align}\label{e3:R}
\vec R(t) = (\mcal E_1(t),\mcal E_2(t),\mcal E_3(t),\mcal B_1(t),\mcal B_2(t),\mcal B_3(t))\ .
\end{align}
The time dependence of the atomic basis states (\ref{e2:EWG}) is only through $\vec R$. That is, we can set
\begin{align}\label{e3:Notation.States}
\rket{\alpha,t,\delta_1,\delta_2} = \rket{\alpha,\vec R(t),\delta_1,\delta_2}
\end{align}
and thus
\begin{align}\label{e3:Kettenregel}
\frac{\partial}{\partial t}\rket{\alpha,t,\delta_1,\delta_2} = 
\vec\nabla_{\vec R}\rket{\alpha,\vec R,\delta_1,\delta_2}\Big\vert_{\vec R(t)}\frac{d\vec R(t)}{dt}\ .
\end{align}
Hence, the integral (\ref{e3:BP}) can be transformed into a line-integral in parameter space,
\begin{align}\label{e3:BP.PS}
\gamma_{\alpha\alpha}(\mcal C,\delta_1,\delta_2) = \int_{\mcal C}\mrmd\vec R\cdot\lrbra{\alpha,\vec R,\delta_1,\delta_2}\I\vec\nabla_{\vec R}\rket{\alpha,\vec R,\delta_1,\delta_2}\ .
\end{align}
In section \ref{s:Results} we shall consider closed paths $\mcal C$ and transform (\ref{e3:BP.PS}) into a surface-integral. From this we will derive the geometric flux densities that will turn out to be extremely useful for visualisation purposes of geometric phases.

However, numerical calculations of geometric phases for a given parametrised field confi\-guration can be performed quite easily by using (\ref{e3:BP}). Introducing the local matrix elements of the time derivative,
\begin{align}\label{e3:D}
\umat D_{\alpha\beta}(\vec R(t),\delta_1,\delta_2) = \lrbra{\alpha,\vec R(t),\delta_1,\delta_2}\frac{\partial}{\partial t}\rket{\beta,\vec R(t),\delta_1,\delta_2}\ ,
\end{align}
the geometric phase (\ref{e3:BP}) reads
\begin{align}\label{e3:BP.D}
\gamma_{\alpha\alpha}(t,\delta_1,\delta_2) = \I\int_0^t dt'\ \umat D_{\alpha\alpha}(\vec R(t'),\delta_1,\delta_2)\ .
\end{align}
(In the notation of I we have for $\tau_0 = T$\ \ $a_{\alpha\alpha} = \I\umat D_{\alpha\alpha}$, see (\extref{I.e2:Def.a})).

In order to study PC and PV contributions to the geometric phases separately, we use perturbation theory to expand the atomic states in powers of the PV parameters $\delta_{1,2}$. This is discussed in appendix \ref{s:GeometricFluxes}. We find with (\ref{eB:rket.PE})ff. for the matrix elements (\ref{e3:D}) for $\beta=\alpha$
\begin{align}\label{e3:D.PC.PV}
\umat D_{\alpha\alpha}(\vec R(t),\delta_1,\delta_2) = \umat D_{\PC,\alpha\alpha}(\vec R(t))
+ \delta_1\umat D^{(1)}_{\PV,\alpha\alpha}(\vec R(t)) + \delta_2\umat D^{(2)}_{\PV,\alpha\alpha}(\vec R(t))
+ \mcal O(\delta^2)\ ,
\end{align}
where $\mcal O(\delta^2)$ is the short-hand notation for $\mcal O(\delta_1^2,\delta_2^2,\delta_1\delta_2)$ and
\begin{align}\label{e3:D.PC}
\umat D_{\PC,\alpha\beta}(\vec R(t)) &= \lrbra{\alpha^{(0)},\vec R(t)}\frac\partial{\partial t}\rket{\beta^{(0)},\vec R(t)}\ ,\\
\label{e3:D.PV}
\begin{split}
\umat D_{\PV,\alpha\alpha}^{(i)}(\vec R(t)) &= \sum_{\gamma\neq\alpha}\Bigg(\frac{\umat M^{(i)}_{\PV,\alpha\gamma}(\vec R(t))\umat D_{\PC,\gamma\alpha}(\vec R(t))}{E(\alpha^{(0)},\vec R(t)) - E(\gamma^{(0)},\vec R(t))}\\
&\qquad\ + \frac{\umat D_{\PC,\alpha\gamma}(\vec R(t))\umat M^{(i)}_{\PV,\gamma\alpha}(\vec R(t))}{E(\alpha^{(0)},\vec R(t)) - E(\gamma^{(0)},\vec R(t))}\Bigg)\ ,
\end{split}\\
\label{e3:M.PV}
\umat M_{\PV,\alpha\gamma}^{(i)}(\vec R(t)) &= \lrbra{\alpha^{(0)},\vec R(t)}\umat M_\PV^{(i)}\rket{\gamma^{(0)},\vec R(t)}\ ,\\
\nonumber
(i &= 1,2)\ .
\end{align}
Here, $\rket{\alpha^{(0)},\vec R(t)}$ and $E(\alpha^{(0)},\vec R(t))$ are the eigenstates and eigenvalues of the mass matrix, obtained from (\ref{e2:MM.full}) by setting $\delta_1=\delta_2=0$,
\begin{align}
\umat M(\vec R(t),0) \equiv \umat M(\vec R(t),\delta_1,\delta_2)\big\vert_{\delta_1=\delta_2=0}
= \umat M_0 - \uvec D\cdot\vmc E(t) - \uvec\mu\cdot\vmc B(t)\ .
\end{align}
From (\ref{e3:BP.D})-(\ref{e3:M.PV}) we obtain the PC and PV contributions for the geometric phases,
\begin{align}\label{e3:BP.series}
\gamma_{\alpha\alpha}(t,\delta_1,\delta_2) &= \gamma_{\PC,\alpha\alpha}(t) + \delta_1\gamma^{(1)}_{\PV,\alpha\alpha}(t) + \delta_2\gamma^{(2)}_{\PV,\alpha\alpha}(t) 
+ \mcal O(\delta^2)\ ,\\ \label{e3:BP.PC}
\gamma_{\PC,\alpha\alpha}(t) &= \I\int_{0}^t\mrmd t'\ \umat D_{\PC,\alpha\alpha}(\vec R(t'))\ ,\\ \label{e3:BP.PV}
\gamma^{(i)}_{\PV,\alpha\alpha}(t) &= \I\int_{0}^t\mrmd t'\ \umat D^{(i)}_{\PV,\alpha\alpha}(\vec R(t'))\ ,\quad
(i=1,2)\ .
\end{align}

\section{Geometric flux densities}\label{s:Results}

In the following we always consider a closed path $\mcal C$ in parameter space.

The direct evaluation of (\ref{e3:BP.series})-(\ref{e3:BP.PV}) is the easiest way to obtain numerical values of geometric phases for a given path. The problem is to find suitable paths in parameter space that maximise PV or PC geometric phases while keeping the decay of metastable states to a minimum. In this section we will introduce geometric flux densities in order to get a visualisation and a better understanding of geometric phases. We will also discuss our numerical results for PV geometric phases at the end of the section.

By using the generalised Stokes' theorem the integral (\ref{e3:BP.PS}), for a closed path $\mcal C$, can be transformed into a surface-integral in parameter space. The detailed calculation can be found in appendix \ref{s:GeometricFluxes} and gives (see (\ref{e:BP.surface}))
\begin{align}\label{e4:BP.surface}
\gamma_{\alpha\alpha}(\mcal C) = 
  \int_{\mcal F}\vmc J^{(\vmc E)}_{\alpha\alpha}(\vec R)\cdot d\vec f^{(\vmc E)}
+ \int_{\mcal F}\vmc J^{(\vmc B)}_{\alpha\alpha}(\vec R)\cdot d\vec f^{(\vmc B)}
+ \int_{\mcal F}\mcal I^{(\vmc E,\vmc B)}_{\alpha\alpha}(\vec R)\ .
\end{align}
Here, $\mcal F$ is a two-dimensional surface in the parameter space $\mcal R$ with $\mcal C=\partial\mcal F$.

Thus, the geometric phases can be written as a sum of three integrals. The first integral is evaluated in the space of electric field strengths, the second integral in the space of magnetic field strengths and the third in the full six-dimensional parameter space. The integrands $\vmc J^{(\vmc E)}_{\alpha\alpha}(\vec R)$ and $\vmc J^{(\vmc B)}_{\alpha\alpha}(\vec R)$ are three-dimensional vector fields in the spaces of electric and magnetic field strengths, respectively. They can be interpreted as geometric flux densities and are a useful tool for visualisation and understanding of geometric phases as we will demonstrate below.

The geometric flux densities can be split into PC and PV parts by using perturbation theory. The details of this calculation can be found in appendix \ref{sB:Perturbation}, where we have used only one combined PV parameter $\delta$ (see (\ref{eB:Def.delta})),
\begin{align}
\delta = (\delta_1^2 + \delta_2^2)^{1/2}\ .
\end{align}
From the PV geometric flux density in electric-field space (\ref{eB:J.E.PV}) we can easily derive the nuclear spin independent and dependent flux densities. We obtain from (\ref{eB:J.E.PCPV})-(\ref{eB:J.E.PV}) the geometric flux densities in electric field space as
\begin{align}\label{e4:J.E.PCPV}
\mcal J^{(\vmc E)}_{\ell,\alpha\alpha}(\vec R,\delta_1,\delta_2) &= 
\mcal J^{(\vmc E,\PC)}_{\ell,\alpha\alpha}(\vec R) 
+ \delta_1\mcal J^{(\vmc E,\PV)}_{1,\ell,\alpha\alpha}(\vec R)
+ \delta_2\mcal J^{(\vmc E,\PV)}_{2,\ell,\alpha\alpha}(\vec R)\ ,\\ \label{e4:J.E.PC}
\mcal J^{(\vmc E,\PC)}_{\ell,\alpha\alpha}(\vec R) &= 
\frac\I2\sum_{i,j=1}^3\varepsilon_{ij\ell}\klr{
\sum_{\beta\neq\alpha}\frac{\underline D^\PC_{i,\alpha\beta}(\vec R)\underline D^\PC_{j,\beta\alpha}(\vec R) - (i\leftrightarrow j)}{\big(E(\alpha^{(0)},\vec R) - E(\beta^{(0)},\vec R)\big)^2}}\ ,\\ \label{e4:J.E.PV}
\begin{split}
\mcal J^{(\vmc E,\PV)}_{\ell,\varkappa,\alpha\alpha}(\vec R) &= \frac\I2\sum_{i,j=1}^3\varepsilon_{ij\ell}
\sum_{\beta\neq\alpha}\Bigg(
\frac{\underline D^\PC_{i,\alpha\beta}(\vec R)\underline D^\PV_{\varkappa,j,\beta\alpha}(\vec R) 
+ \underline D^\PV_{\varkappa,i,\alpha\beta}(\vec R)\underline D^\PC_{j,\beta\alpha}(\vec R) - (i\leftrightarrow j)
}{\big(E(\alpha^{(0)},\vec R) - E(\beta^{(0)},\vec R)\big)^2}\Bigg),
\end{split}
\end{align}
for  $\ell = 1,2,3$. Here the matrix elements of the electric dipole operator enter. For $\uvec{D}^{\PC}(\vec R)$ see (\ref{eB:Dip.PC}). The two PV contributions to the electric dipole operator follow from (\ref{eB:Dip.PV}) and read
\begin{align}\label{e4:Dip.PV}
\uvec D^{\PV}_{\varkappa,\alpha\beta}(\vec R) &= \sum_{\gamma\neq\alpha}
\frac{\umat M^{(\varkappa)}_{\PV,\alpha\gamma}(\vec R)\uvec D^\PC_{\gamma\beta}(\vec R)}{E(\alpha^{(0)},\vec R) - E(\gamma^{(0)},\vec R)}
+\sum_{\gamma\neq\beta}
\frac{\uvec D^\PC_{\alpha\gamma}(\vec R)\umat M^{(\varkappa)}_{\PV,\gamma\beta}(\vec R)}{E(\beta^{(0)},\vec R) - E(\gamma^{(0)},\vec R)}
\end{align}
for $\varkappa=1,2$.

Similar results follow for the geometric flux densities in magnetic field space by making the replacements $\vmc E\to\vmc B$ and $\uvec D\to\uvec\mu$ in (\ref{e4:J.E.PCPV})--(\ref{e4:Dip.PV}).

We will restrict ourselves in the following to the case of constant magnetic field. We are then dealing only with the three dimensional electric-field-strength parameter space. Correspondingly, the closed path $\mcal C$ and the surface $\mcal F$ with $\mcal C=\partial\mcal F$ refer to this space. The geometric phase (\ref{e4:BP.surface}) reads then
\begin{align}\label{e4:BPS.E}
\gamma_{\alpha\alpha}(\mcal C) = 
  \int_{\mcal F}\vmc J^{(\vmc E)}_{\alpha\alpha}(\vec R)\cdot d\vec f^{(\vmc E)}\qquad\text{(for $\vmc B=$ const)}\ .
\end{align}

\begin{turnpage}
\begin{figure}
\centering
\includegraphics[width=18cm]{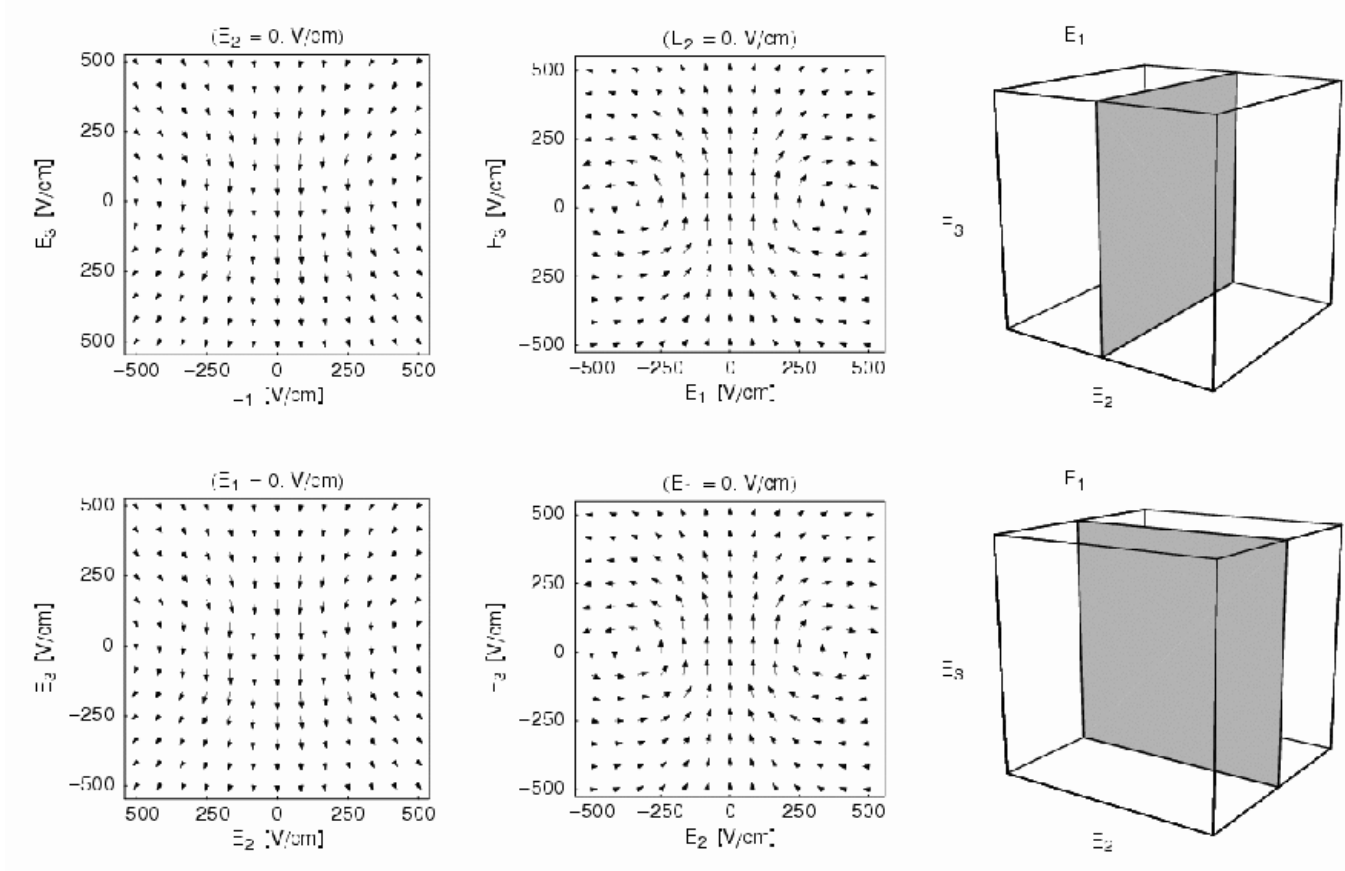}
\caption{The PC geometric flux density field of the hydrogen state $\rket{9} \equiv \rket{2\hat S_{1/2},1,1,\vmc E,\vmc B = 1\,\mathrm{mT}}$ in electric-field space. The diagram on the l.h.s. shows the real parts of the flux density vectors, the diagram in the middle shows the imaginary parts. The diagram on the r.h.s. indicates the position of the plotted plane in electric field space.
}\label{f:H2_PCa}
\end{figure}
\begin{figure}
\includegraphics[width=18cm]{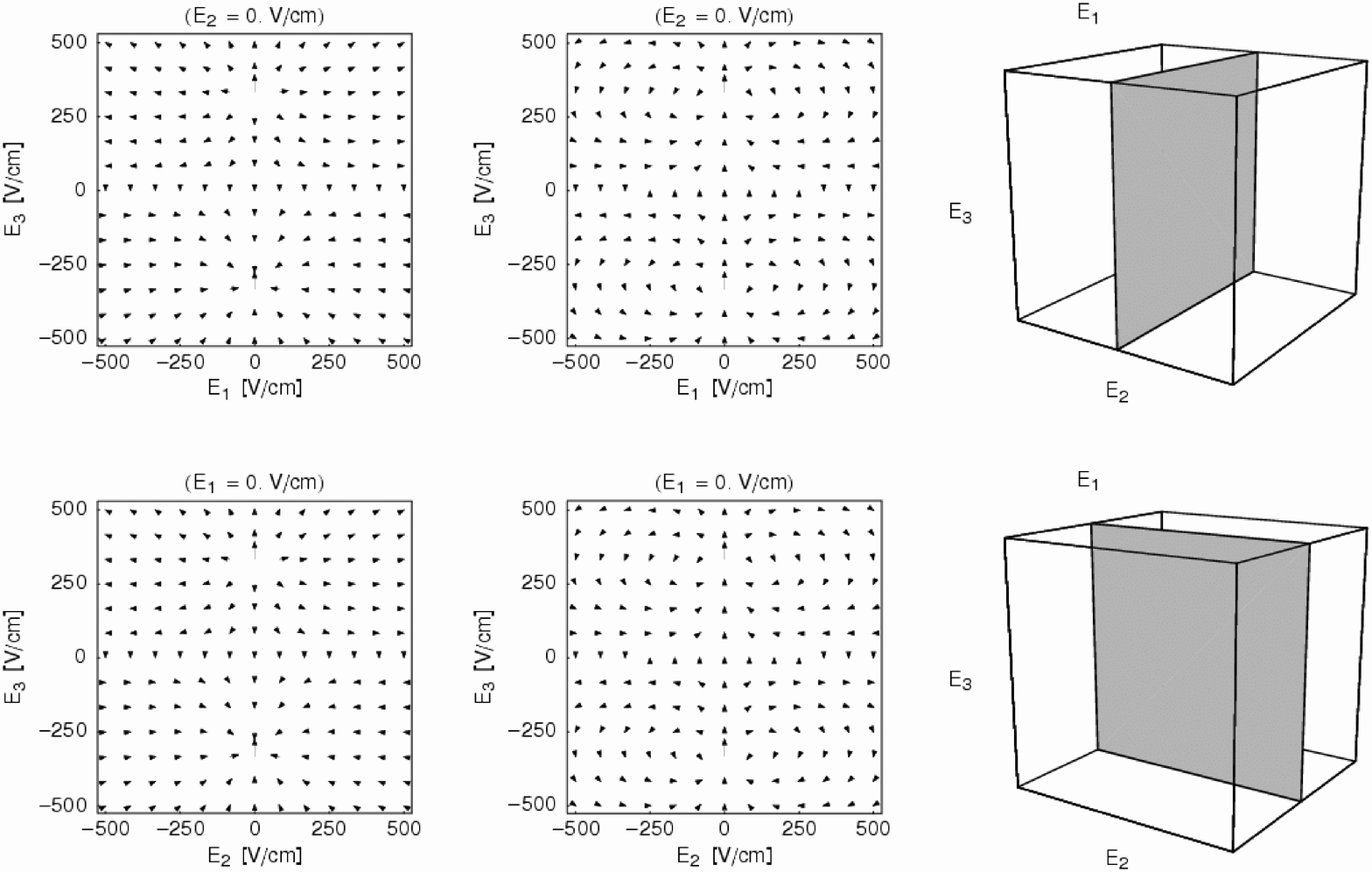}
\caption{The PC geometric flux density field of the hydrogen state $\rket{10} \equiv \rket{2\hat S_{1/2},1,0,\vmc E,\vmc B = 1\,\mathrm{mT}}$ in electric-field space. The diagram on the l.h.s. shows the real parts of the flux density vectors, the diagram in the middle shows the imaginary parts. The diagram on the r.h.s. indicates the position of the plotted plane in electric field space.}\label{f:H2_PCb}
\end{figure}
\end{turnpage}

We will now study $\vmc J^{(\vmc E)}_{\alpha\alpha}(\vec R)$ in some detail for the metastable states of hydrogen and deuterium. The numerical calculation of these vector fields is done for magnetic fields pointing in 3-direction. We show first some examples of flux density fields in electric-field space. In figures \ref{f:H2_PCa} and \ref{f:H2_PCb} the PC flux densities of the hydrogen states with $\alpha\in\klg{9,10}$ are shown. Rotational invariance of the fields around the $\mcal E_3$-axis is indicated here, as the plots of the $\mcal E_1$-$\mcal E_3$-plane are identical to the plots of the $\mcal E_2$-$\mcal E_3$-plane. In fig. \ref{f:H2_PCa} a toroidal structure of the flux density field shows up, whereas in fig. \ref{f:H2_PCb} a source and a sink can be seen. The corresponding plots for the other metastable hydrogen and deuterium states look very similar and can be found online \cite{Plots}.

\begin{turnpage}
\begin{figure}[!HTbp]
\centering
\includegraphics[width=23cm]{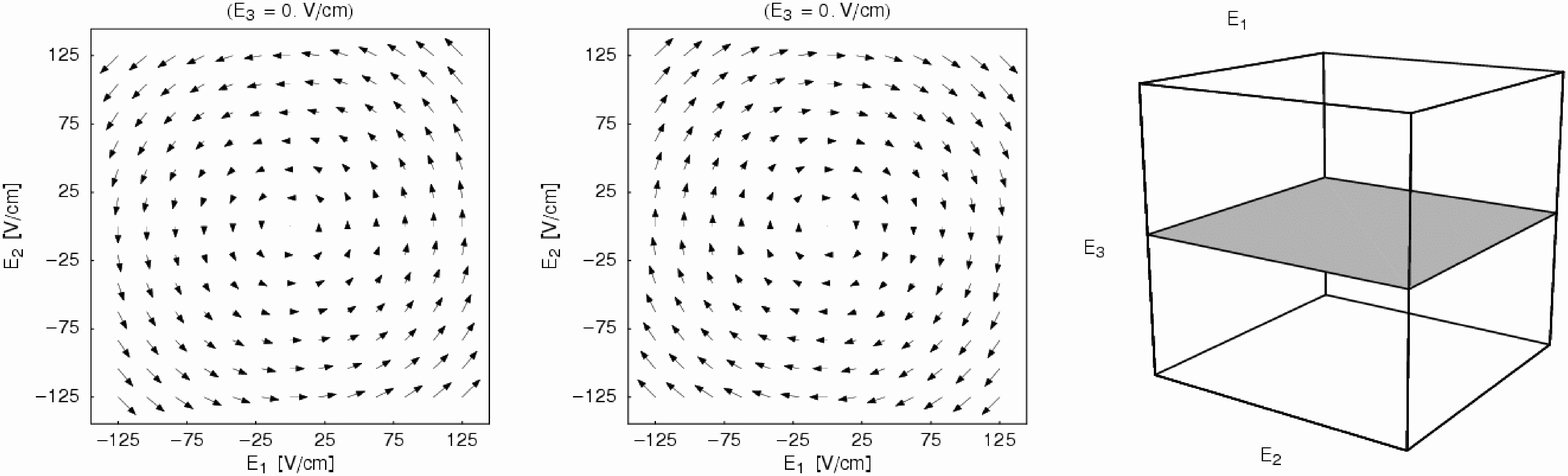}
\caption{The nuclear spin dependent PV flux density field of the hydrogen state $\rket{9} \equiv \rket{2\hat S_{1/2},1,1,\vmc E,\vmc B = 1\,\mathrm{mT}}$ in the $\mcal E_1$-$\mcal E_2$-plane with $\mcal E_3=0$. The figure shows from left to right the real and imaginary part of the flux density and the location of the plane in electric field space.}\label{f:H2_PV}
\end{figure}
\end{turnpage}

Figure \ref{f:H2_PV} shows the nuclear spin dependent PV flux density of the hydrogen state $\rket{9} \equiv \rket{2\hat S_{1/2},1,1,\vmc E,\vmc B}$ in the $\mcal E_1$-$\mcal E_2$-plane for $\mcal E_3=0$. The solenoidal character and the cylindrical symmetry of the flux density can be seen. Many similar plots for different values of $\mcal E_3$ and all other metastable hydrogen and deuterium states can also be found online \cite{Plots}.

\begin{figure}[!Htbp]
\centering
\parbox{5cm}{\includegraphics[width=5cm]{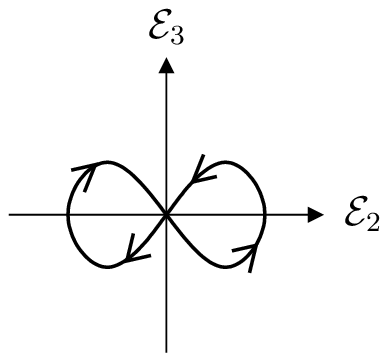}\\[3mm]{\bf (a)}}\qquad
\parbox{5cm}{\includegraphics[width=5cm]{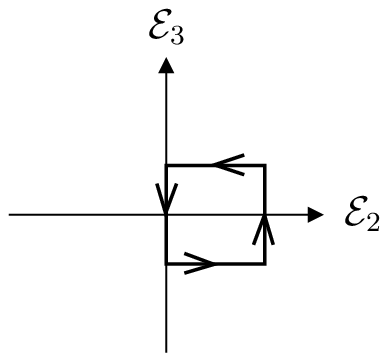}\\[3mm]{\bf (b)}}
\caption{Two example paths for a purely PV geometric phase, assuming a constant magnetic field in 3-direction.}\label{f:paths}
\end{figure}

From a study of many vector plots we find that $\vmc J_{\alpha\alpha}^{(\vmc E,\PC)}(\vmc E,\vmc B)$ has only radial and 3-components whereas $\vmc J_{\varkappa,\alpha\alpha}^{(\vmc E,\PV)}(\vmc E,\vmc B)$ ($\varkappa=1,2$) has only an azimuthal component. At least, this holds to the accuracy of the numerical calculations in all cases studied so far.

Furthermore, the vector plots directly confirm the PC and PV character of the vector fields $\vmc J^{(\vmc E,\PC)}_{\alpha\alpha}(\vmc E,\vmc B)$ and $\vmc J^{(\vmc E,\PV)}_{\varkappa,\alpha\alpha}(\vmc E,\vmc B)$, $(\varkappa=1,2)$. This is discussed in appendix \ref{sB:Perturbation}, see (\ref{e:P.J.E.PC}) and (\ref{e:P.J.E.PV}), respectively. The plots also show that in all cases studied so far the PV vector fields in electric field space are source free, that is we have, to our numerical accuracy,
\begin{align}
\vec\nabla_{\vmc E}\cdot\vmc J^{(\vmc E,\PV)}_{\varkappa,\alpha\alpha}(\vmc E,\vmc B) &= 0
\end{align}
for $\varkappa=1,2$, whereas the PC flux fields can have sources and sinks.

These vector field plots show that the PC and PV flux densities are orthogonal to each other and that they, furthermore, make it easy to choose closed paths in parameter space which maximise PC or PV geometric phases, respectively. Two examples of closed paths in parameter space that will give purely PV geometric phases are shown in fig. \ref{f:paths}. The path (a) --- consisting of two-point symmetric loops --- gives a purely PV geometric phase for an arbitrary constant magnetic field since the PC contributions of the two loops cancel each other due to (\ref{e:P.J.E.PC}), whereas due to (\ref{e:P.J.E.PV}) the PV contributions will be the same for each loop and add up.

The PV geometric phases of all metastable hydrogen and deuterium states have been calculated numerically by evaluating the surface-integral (\ref{e4:BPS.E}) over the PV geometric flux densities (\ref{e4:J.E.PV}) for a rectangular surface as shown in fig. \ref{f:paths}b. In detail, we define the surface $\mcal F$ by
\begin{align}\label{e4:Def.F}
\mcal F = \klg{\vmc E;\ \mcal E_1 = 0,\,\mcal E_2\in\kle{0,250\,\mathrm{V/cm}},\,\mcal E_3\in\kle{-125\,\mathrm{V/cm},125\,\mathrm{V/cm}}}
\end{align}
and use a constant magnetic field
\begin{align}\label{e4:B.example}
\vmc B = (0,0,1\,\mathrm{mT})\ .
\end{align}
We have also evaluated the line-integrals (\ref{e3:BP.PV}) along $\partial\mcal F$ numerically in order to cross-check our results which are listed in table \ref{t:PV_12}. We find that the results from both methods match nicely and give an indication of the accuracy of the numerical calculations. The total PV geometric phases are (see (\ref{e3:BP.series}))
\begin{align}\label{e4:PVtotal}
\gamma_{\PV,\alpha\alpha} = \delta_1\gamma_{\PV,\alpha\alpha}^{(1)} + \delta_2\gamma_{\PV,\alpha\alpha}^{(2)},
\end{align}
with the PV parameters $\delta_{1,2}$ taken from table \ref{t:values}. The $\gamma_{\PV,\alpha\alpha}$ are listed in table \ref{t:PV_all}.

{
\begin{table}
\centering
\begin{tabular}{|c|l||r|r||r|r|} \hline
& & \multicolumn{2}{c||}{$\gamma^{(1)}_{\PV,\alpha\alpha}$} & \multicolumn{2}{c|}{$\gamma^{(2)}_{\PV,\alpha\alpha}$}\\
$\alpha$ & $\rket{2\hat S_{1/2},F,F_3}$ & \multicolumn{1}{c|}{surface} & \multicolumn{1}{c||}{line} & \multicolumn{1}{c|}{surface} & \multicolumn{1}{c|}{line}\\ \hline
9 & $\rket{2\hat S_{1/2},1,1}$   & $-0.1966 + 0.0290\I$ & $-0.1918 + 0.0287\I$ & $-3.0547 + 0.3101\I$ & $-3.0548 + 0.3099\I$\\
10 & $\rket{2\hat S_{1/2},1,0}$  & $0.1438 - 0.0194\I$ & $0.1438 - 0.0194\I$ & $4.2083 - 0.3872\I$ & $4.2132 - 0.3873\I$ \\
11 & $\rket{2\hat S_{1/2},1,-1}$ & $0.0394 - 0.0078\I$ & $0.0352 - 0.0074\I$ & $-0.9889 + 0.0584\I$ & $-0.9936 + 0.0586\I$ \\ \hline
12 & $\rket{2\hat S_{1/2},0,0}$  & $0.0339 - 0.0055\I$ & $0.0330 - 0.0054\I$ & $-0.1610 + 0.0179\I$ & $-0.1610 + 0.0180\I$ \\ \hline
\end{tabular}\\[3mm]
\textbf{(a)}\\[5mm]
\begin{tabular}{|c|l||r|r||r|r|} \hline
& & \multicolumn{2}{c||}{$\gamma^{(1)}_{\PV,\alpha\alpha}$} & \multicolumn{2}{c|}{$\gamma^{(2)}_{\PV,\alpha\alpha}$}\\
$\alpha$ & $\rket{2\hat S_{1/2},F,F_3}$ & \multicolumn{1}{c|}{surface} & \multicolumn{1}{c||}{line} & \multicolumn{1}{c|}{surface} & \multicolumn{1}{c|}{line}\\ \hline
13 & $\rket{2\hat S_{1/2},\tfrac32,\tfrac32}$    & $-0.1382 + 0.0218\I$ & $-0.1333 + 0.0214\I$ & $-20.004 + 2.247\I$ & $-20.033 + 2.249\I$ \\
14 & $\rket{2\hat S_{1/2},\tfrac32,\tfrac12}$    & $-0.0767 + 0.0119\I$ & $-0.0733 + 0.0116\I$ & $ 11.642 - 1.296\I$ & $ 11.665 - 1.297\I$ \\
15 & $\rket{2\hat S_{1/2},\tfrac32,-\tfrac12}$   & $-0.0123 + 0.0019\I$ & $-0.0108 + 0.0018\I$ & $ 13.441 - 1.456\I$ & $ 13.460 - 1.458\I$ \\
16 & $\rket{2\hat S_{1/2},\tfrac32,-\tfrac32}$   & $ 0.0924 - 0.0149\I$ & $ 0.0876 - 0.0146\I$ & $- 0.235 - 0.040\I$ & $ -0.242 - 0.040\I$ \\ \hline
17 & $\rket{2\hat S_{1/2},\tfrac12,\tfrac12}$    & $ 0.1284 - 0.0200\I$ & $ 0.1246 - 0.0197\I$ & $-11.545 + 1.346\I$ & $-11.564 + 1.348\I$ \\
18 & $\rket{2\hat S_{1/2},\tfrac12,-\tfrac12}$   & $ 0.0365 - 0.0059\I$ & $ 0.0348 - 0.0058\I$ & $  6.703 - 0.801\I$ & $  6.715 - 0.802\I$ \\ \hline
\end{tabular}\\[3mm]
\textbf{(b)}
\caption{Numerical results of PV geometric phases for metastable hydrogen (a) and deuterium (b).}\label{t:PV_12}
\end{table}}

{
\begin{table}
\centering
\begin{tabular}{|c|l||r|r|} \hline
& & \multicolumn{2}{c|}{$\gamma_{\PV,\alpha\alpha}\ [10^{-13}\,\mathrm{rad}]$} \\
$\alpha$ & $\rket{2\hat S_{1/2},F,F_3}$ & \multicolumn{1}{c|}{surface} & \multicolumn{1}{c|}{line}\\ \hline
9 & $\rket{2\hat S_{1/2},1,1}$   & $-20.96(59) + 2.10(6)\I$ & $-20.97(59) + 2.10(6)\I$ \\
10 & $\rket{2\hat S_{1/2},1,0}$  & $29.23(81) - 2.67(7)\I$ & $29.26(81) - 2.67(7)\I$  \\
11 & $\rket{2\hat S_{1/2},1,-1}$ & $-7.07(19) + 0.41(1)\I$ & $-7.09(19) + 0.43(1)\I$  \\ \hline
12 & $\rket{2\hat S_{1/2},0,0}$  & $-1.23(3) + 0.141(3)\I$ & $-1.23(3) + 0.141(3)\I$ \\ \hline
\end{tabular}\\[3mm]
\textbf{(a)}\\[5mm]
\begin{tabular}{|c|l||r|r|} \hline
& & \multicolumn{2}{c|}{$\gamma_{\PV,\alpha\alpha}\ [10^{-13}\,\mathrm{rad}]$} \\
$\alpha$ & $\rket{2\hat S_{1/2},F,F_3}$ & \multicolumn{1}{c|}{surface} & \multicolumn{1}{c|}{line}\\ \hline
13 & $\rket{2\hat S_{1/2},\tfrac32,\tfrac32}$    & $-8.7(3.4) + 1.34(38)\I$ & $-8.4(3.4) + 1.32(38)\I$ \\
14 & $\rket{2\hat S_{1/2},\tfrac32,\tfrac12}$    & $-4.1(2.0) + 0.66(22)\I$ & $-3.9(2.0) + 0.64(22)\I$ \\
15 & $\rket{2\hat S_{1/2},\tfrac32,-\tfrac12}$   & $-0.32(2.3) + 0.07(24)\I$ & $-0.23(2.3) + 0.06(25)\I$ \\
16 & $\rket{2\hat S_{1/2},\tfrac32,-\tfrac32}$   & $5.404(36) - 0.876(8)\I$ & $5.128(38) - 0.856(7)\I$ \\ \hline
17 & $\rket{2\hat S_{1/2},\tfrac12,\tfrac12}$    & $7.2(2.0) - 1.13(23)\I$ & $7.0(2.0) - 1.12(23)\I$ \\
18 & $\rket{2\hat S_{1/2},\tfrac12,-\tfrac12}$   & $2.3(1.1) - 0.37(13)\I$ & $2.2(1.1) - 0.36(14)\I$ \\ \hline
\end{tabular}\\[3mm]
\textbf{(b)}
\caption{The total PV geometric phases (\ref{e4:PVtotal}) for the metastable states of hydrogen (a) and deuterium (b). The errors quoted are discussed in the text.}\label{t:PV_all}
\end{table}

}

\begin{figure}
\centering
\parbox{7cm}{\includegraphics[width=7cm]{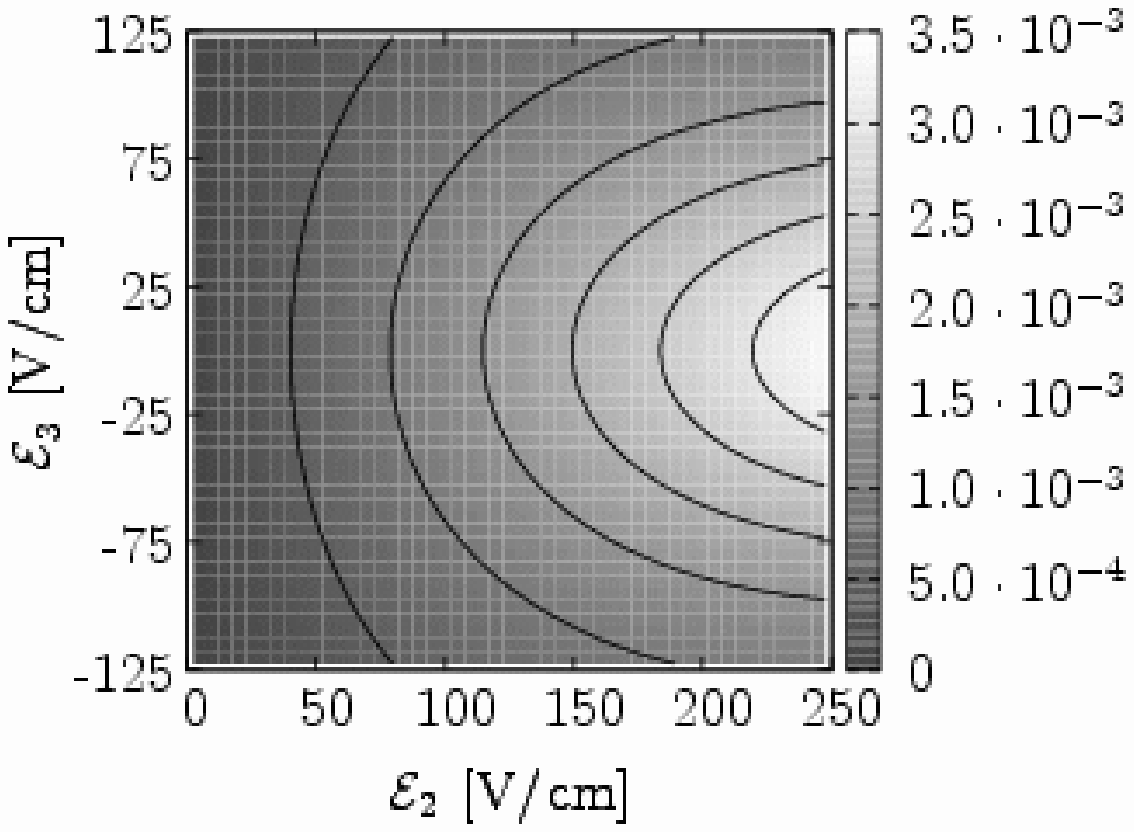}\\[2mm]\textbf{(a)}}
\qquad
\parbox{7cm}{\includegraphics[width=7cm]{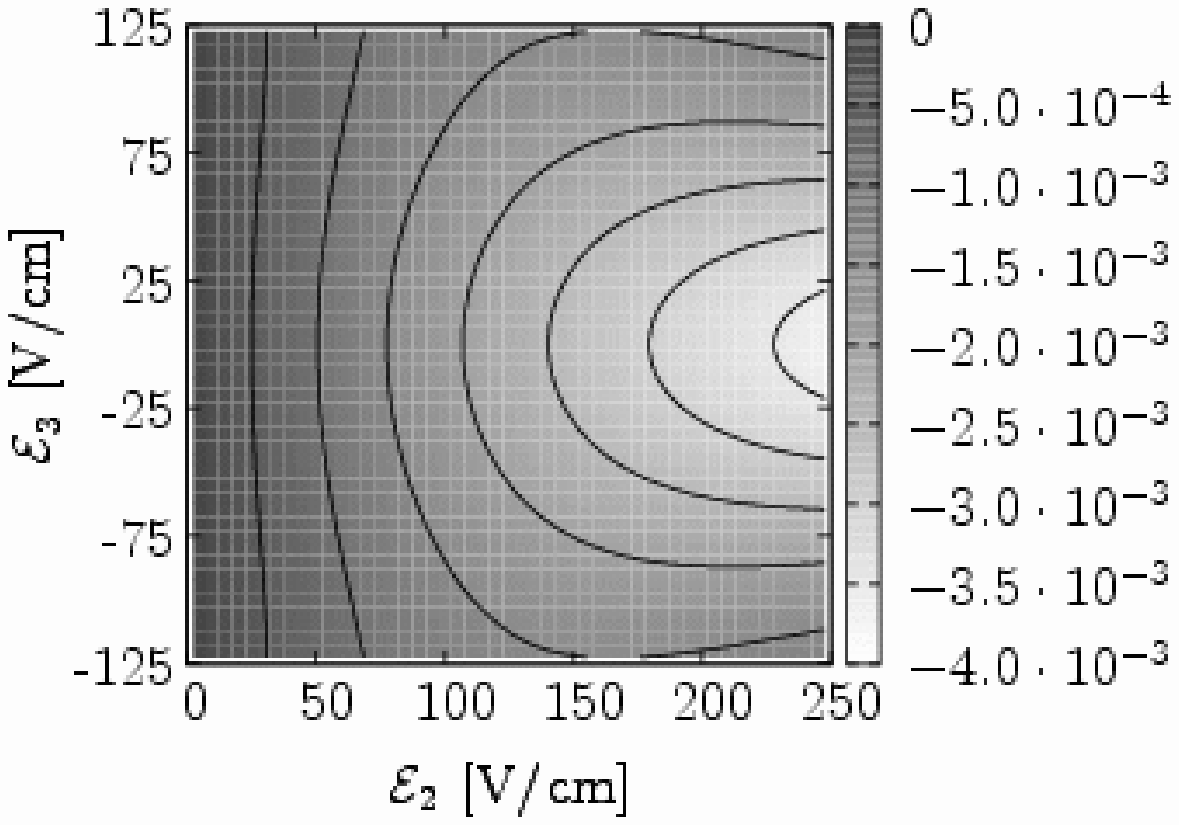}\\[2mm]\textbf{(b)}}
\caption{The real parts of the projected, nuclear spin dependent flux densities $\vmc J^{(\vmc E,\PV)}_{2,\alpha\alpha}(\vmc E,\vmc B)\cdot d\vec f^{(\vmc E)}/|d\vec f^{(\vmc E)}|$ of the metastable hydrogen states $\rket{9}$ (a) and $\rket{10}$ (b). Brighter areas correspond to a larger absolute value of the geometric flux. The solid black lines are contour lines for the levels shown on the right hand sides of the density plots.}\label{f:H2_RePV}
\end{figure}

We will now discuss the results shown in table \ref{t:PV_all}. First of all, the metastable hydrogen states $\rket{9}$ and $\rket{10}$ have the largest PV geometric phases. In figure \ref{f:H2_RePV}, the corresponding real parts of the nuclear spin dependent fluxes $\vmc J^{(\vmc E,\PV)}_{2,\alpha\alpha}(\vmc E,\vmc B)\cdot d\vec f^{(\vmc E)}/|d\vec f^{(\vmc E)}|$ are shown. They represent the dominant contributions to $\gamma_{\PV,\alpha\alpha}$ for these states. The areas of the largest contributions to the flux are centred around $\mcal E_3=0$ and extend to higher values of $\mcal E_2$. In fact, the PV geometric flux can be increased by one order of magnitude for a surface with $\mcal E_2\in\kle{0,1000\,\mathrm{V/cm}}$, but then, of course, (\ref{e2:E.bound}) no longer holds for every point on that surface and the mixed $2S$ states can in general no longer be considered as metastable, see appendix \ref{s:Adiabaticity}.

The quoted uncertainties in the results of table \ref{t:PV_all} only take into account the errors of the PV parameters $\delta_{1,2}$ given in table \ref{t:values}. The numerical uncertainties have been neglected but can be estimated roughly by comparison of the results for the surface and the line-integration and seem to be around a few percent. The only PV geometric phase for which the surface- and the line-integration results do not coincide within $1\sigma$ is the result for the metastable deuterium state $\rket{16} = \rket{2\hat S_{1/2},\tfrac32,-\tfrac32,\vmc E,\vmc B}$. The small uncertainty due to $\delta_{1,2}$ in $\gamma_{\PV,16\,16}$ is due to its small nuclear spin dependent contribution $\gamma_{\PV,16\,16}^{(2)}$ (see table \ref{t:PV_12}). For all other hydrogen and deuterium states the contribution from $\gamma_{\PV,\alpha\alpha}^{(2)}$ dominates over that from $\gamma_{\PV,\alpha\alpha}^{(1)}$ in $\gamma_{\PV,\alpha\alpha}$. This, together with the fact that $\delta_2$ for deuterium is consistent with zero and has a statistical error larger than $500\%$ (see table \ref{t:values}), is the reason for the large uncertainties due to $\delta_{1,2}$ (up to $50\%$) of all metastable deuterium states with $\alpha\neq16$.

\begin{figure}
\centering
\parbox{7cm}{\includegraphics[width=7cm]{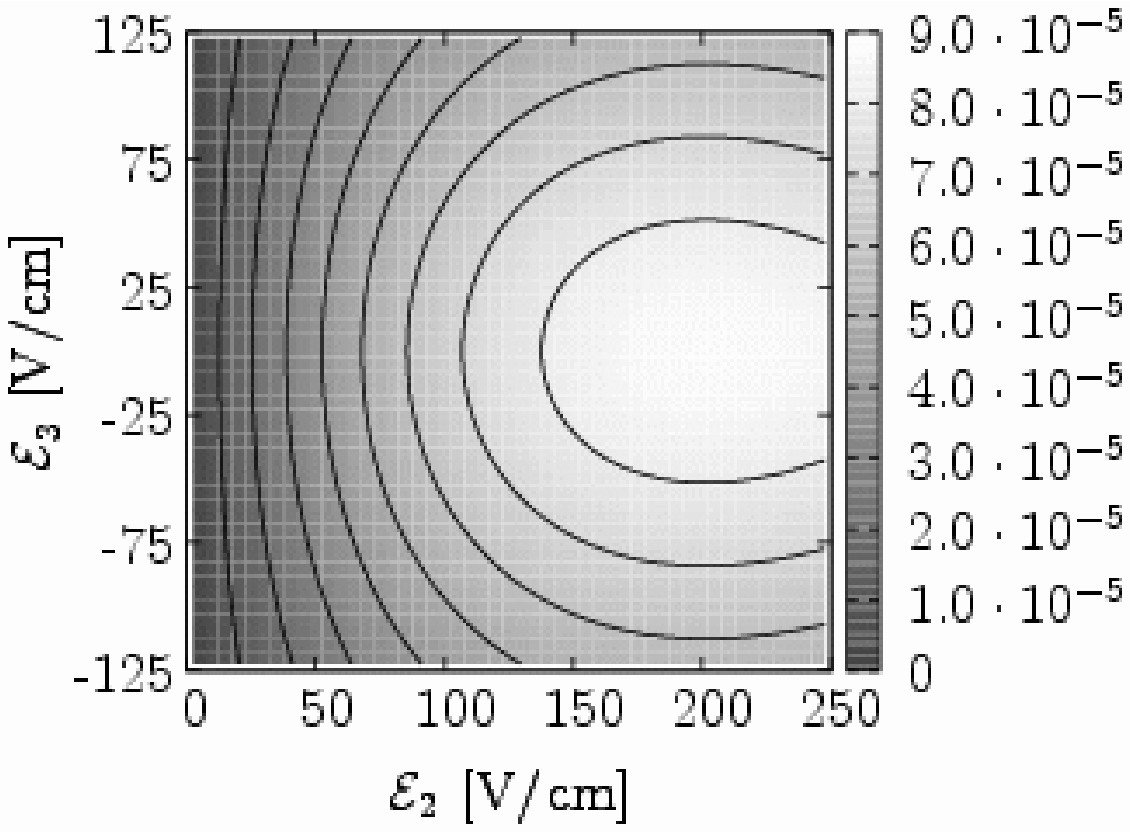}\\[-3mm] \textbf{(a)}}
\qquad
\parbox{7cm}{\includegraphics[width=7cm]{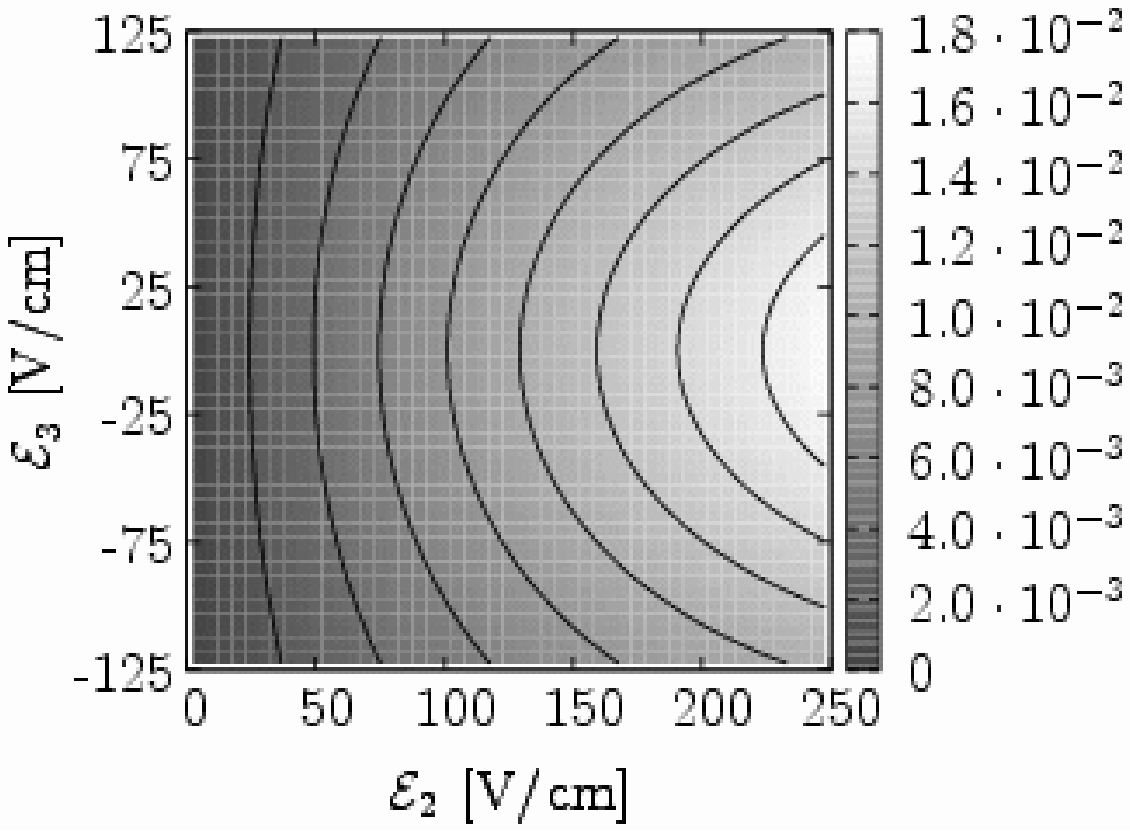}\\[-3mm] \textbf{(b)}}\\[3mm]
\parbox{7cm}{\includegraphics[width=7cm]{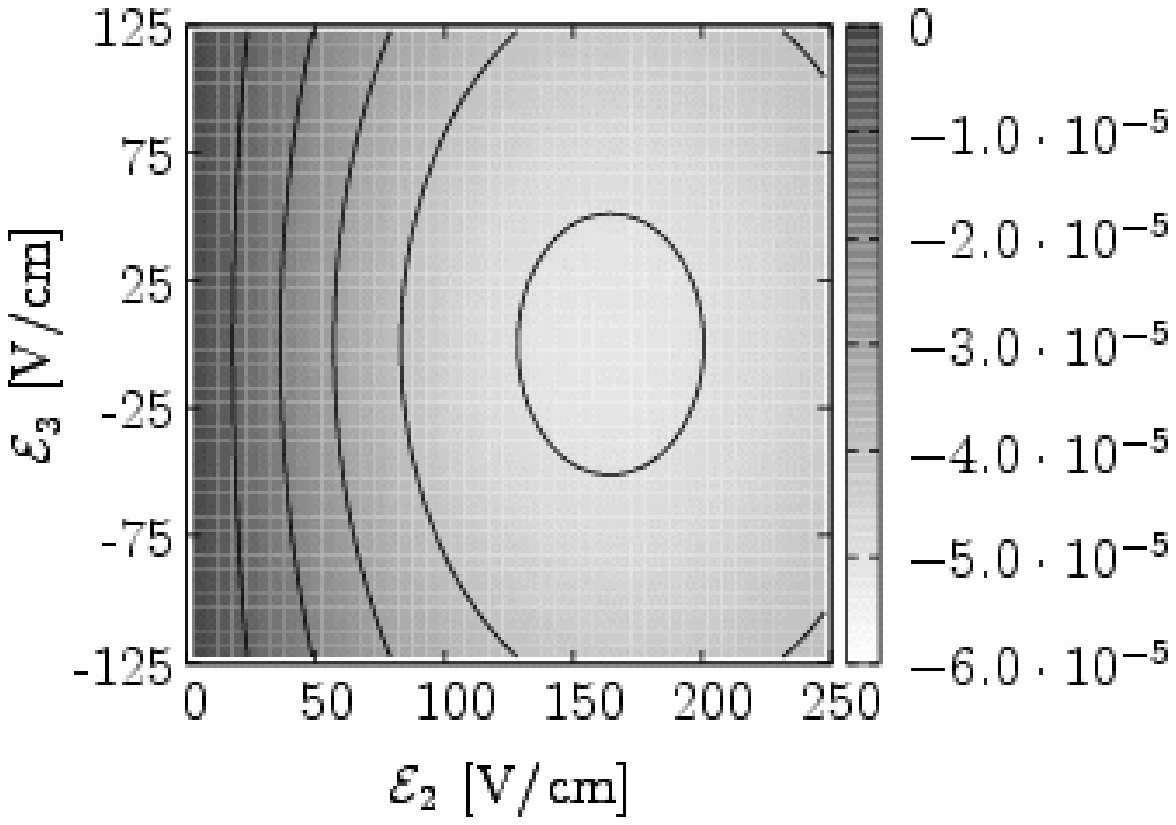}\\[-3mm] \textbf{(c)}}
\qquad
\parbox{7cm}{\includegraphics[width=7cm]{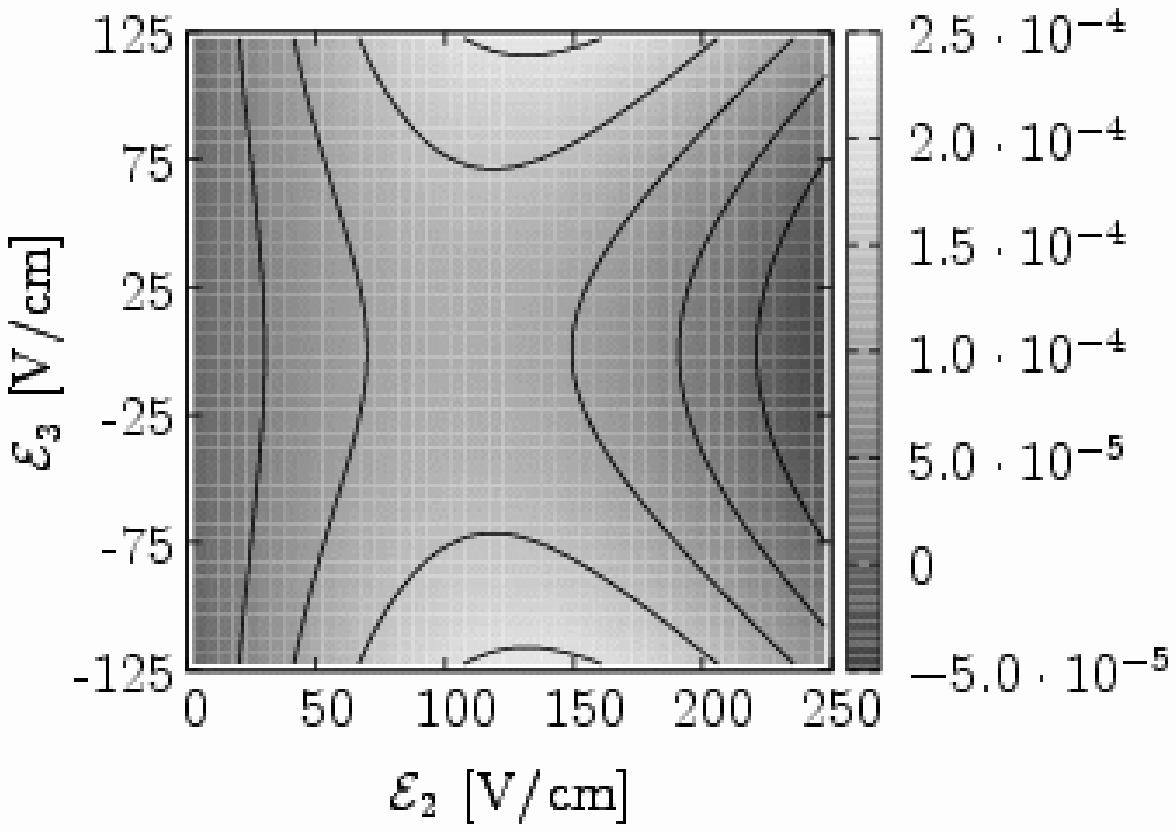}\\[-3mm] \textbf{(d)}}
\caption{The real parts of the two PV contributions to the geometric fluxes of the metastable deuterium states $\rket{13}=\rket{2\hat S_{1/2},\tfrac32,\tfrac32,\vmc E,\vmc B}$ (figures (a), (b)) and $\rket{16}=\rket{2\hat S_{1/2},\tfrac32,-\tfrac32,\vmc E,\vmc B}$ (figures (c), (d)). In (a) and (c) the nuclear spin independent, whereas in (b) and (d) the nuclear spin dependent contributions are shown.}\label{f:D2_RePV}
\end{figure}

In fig. \ref{f:D2_RePV} we show the real parts of the PV flux densities for the deuterium states $\rket{13}=\rket{2\hat S_{1/2},\tfrac32,\tfrac32,\vmc E,\vmc B}$ and $\rket{16}=\rket{2\hat S_{1/2},\tfrac32,-\tfrac32,\vmc E,\vmc B}$.
Figures \ref{f:D2_RePV}a-c are similar to the corresponding hydrogen flux densities shown in fig. \ref{f:H2_RePV}. Only fig. \ref{f:D2_RePV}d, which shows the nuclear spin dependent PV flux density of $\rket{16}$, looks different. Compared to fig. \ref{f:D2_RePV}b one can see a difference of two orders of magnitude between both flux densities. Furthermore, fig. \ref{f:D2_RePV}d shows a change of sign between the maximum and minimum of the plotted flux density values. The geometric flux density also increases both in positive and negative $\mcal E_3$-direction.

\section{Conclusions and Outlook}\label{s:Conclusions}

In this work we have shown that PV geometric phases in the lightest metastable atoms, $2S$ hydrogen and deuterium, exist. We have considered adiabatic changes of the electric field along suitable closed paths $\mcal C$ with a constant magnetic field in order to avoid degeneracies of atomic energy levels. The metastable states of hydrogen and deuterium indeed pick up non-zero PV geometric phases in this way.

The geometric flux densities provide us with a good way to visualise geometric phases as fluxes through areas $\mcal F$ in parameter space with $\mcal C=\partial F$. In this way we could see how to choose the path $\mcal C$ in order to get only a PC or only a PV geometric phase.

We have then studied numerical values of geometric phases for a closed rectangular loop in electric-field space (see fig. \ref{f:paths}b) where only PV and no PC phases occur. The PV phases have been calculated numerically both from the surface and from the line-integrals, see (\ref{e4:BPS.E}) and (\ref{e3:BP.PV}), respectively. The results agreed within the numerical uncertainty. The resulting PV geometric phases are listed in tables \ref{t:PV_12} and \ref{t:PV_all}. The geometric contributions $\gamma_{\PV,\alpha\alpha}^{(1,2)}$ can be large, see table \ref{t:PV_12}. The smallness of the results for the total PV geometric phase in table \ref{t:PV_all} is solely due to the smallness of the PV parameters $\delta_{1,2}$. It remains to be seen if such small effects can be enhanced by suitable iterations or other means in order to make them large enough to be measurable.

Even if the results for PV geometric phases presented in this paper are numerically small we find that they have an interest of principle. They show that geometric phases can carry information from parity violation in atoms. This was first discussed in \cite{DissTG} and has now been investigated in detail for the $2S$ states of hydrogen and deuterium.

There are numerous things left to be studied in the future. So far, we have only investigated abelian geometric phases for constant magnetic fields. The investigation of abelian geometric phases for varying magnetic field strengths is a challenging task for further studies. Furthermore, non-abelian geometric phases could reveal interesting new features of PV geometric phases.

\section*{Acknowledgements}

The authors thank M. DeKieviet and D. Dubbers for many useful discussions. This work was supported by Deutsche Forschungsgemeinschaft under project No. NA\,296/3-1.


\appendix
\titleformat{name=\section}[block]{\large\bfseries\sffamily}{Appendix \thesection}{1ex}{}
\titleformat{name=\subsection}{\bfseries\sffamily}{\Alph{section}.\arabic{subsection}.}{1ex}{}
\renewcommand{\thesection}{\Alph{section}}
\renewcommand{\thesubsection}{\!.\arabic{subsection}}
\renewcommand{\theequation}{\Alph{section}.\arabic{equation}}

\section{Values for quantities related to $n=2$ hydrogen and deuterium}\label{s:Values}

In this appendix we collect the numerical values for the quantities entering our calculations for hydrogen and deuterium states with principal quantum number $n=2$. We specify our numbering scheme for these states and give the expressions for the mass matrices at zero external fields, and for the electric and the magnetic dipole operators.

In table \ref{t:values} we present the numerical values for the weak charges $Q_W^{(\varkappa)}$, $\varkappa=1,2$, the quantities $\Delta q$, the Lamb shift $L = E(2S_{1/2})-E(2P_{1/2})$, and the fine structure splitting $\Delta = E(2P_{3/2}) - E(2P_{1/2})$. The ground state hyperfine splitting energy is denoted by $\mcal A$. We have $\mcal A = E(1S_{1/2},F=1)-E(1S_{1/2},F=0)$ for hydrogen and $\mcal A = E(1S_{1/2},F=3/2)-E(1S_{1/2},F=1/2)$ for deuterium.
The $n=2$ states of hydrogen and deuterium in the absence of P-violation and for zero external field are denoted by $\rket{2L_J,F,F_3}$, where $L$, $J$, $F$ and $F_3$ are the quantum numbers of the electron's orbital angular momentum, its total angular momentum, the total atomic angular momentum and its third component, respectively. The quantum numbers $S$ for the electron spin and $I$ for the nuclear spin are omitted, since these are fixed quantities for each atomic species. In the following the ordering of the atomic states in the matrix representations of operators is according to decreasing $F_3$, $F$, $J$ and $L$.

\begin{table}[tbhp]
\begin{tabular}{c||c|c|c}
 & ${}_1^1$H &${}^2_1$D & Ref.\\ \toprule
$Z$ & 1 & 1 &\\
$N$ & 0 & 1 &\\
$I$ & $\tfrac12$ & 1 &\\  \hline
$Q_W^{(1)}(Z,N)$ & 0.04532(64) & -0.95468(64) & (\ref{e1:QW1.SM})\\ 
$\delta_1(Z,N)$ & $-2.78(4)\cdot 10^{-13}$ & $58.59(4)\cdot 10^{-13}$ & (\ref{e2:deltaPV12H}),(\ref{e2:deltaPV12D}) \\  \hline
$\Delta u(Z,N) - \Delta d(Z,N)$ & $1.2695(29)$ & 0 & \cite{PDG06}\\
$\Delta s(Z,N)$ & $0.006(29)(7)$ & $0.012(58)(14)$ & \cite{Jac06} \\
$Q_W^{(2)}(Z,N)$ & $-0.1145(31)$ & $0.0005(26)$ & (\ref{e1:QW2.SM})\\
$\delta_2(Z,N)$ & $7.04(19)\cdot 10^{-13}$ & $0.03(17)\cdot 10^{-13}$ & (\ref{e2:deltaPV12H}),(\ref{e2:deltaPV12D}) \\  \hline
$L(Z,N)/h$ & \multicolumn{1}{r|}{1057.8440(24) MHz} & \multicolumn{1}{r|}{1059.2330(26) MHz} & \cite{NISTData}\\
$\Delta(Z,N)/h$ & \multicolumn{1}{r|}{10969.0416(48) MHz} & \multicolumn{1}{r|}{10972.0355(48) MHz} & \cite{NISTData}\\
$\mcal A(Z,N)/h$ & \multicolumn{1}{r|}{\ 1420.405751768(1) MHz} & \multicolumn{1}{r|}{\ 327.384352522(2) MHz} & \cite{Kar05}\\ \toprule
\end{tabular}
\caption{Values of parameters for numerical calculations. The weak mixing angle in the low energy limit, $\sin^2\vartheta_W = 0.23867(16)$, was taken from \cite{ErRa05}. The uncertainty in $\delta_1$ is dominated by the uncertainty of $\sin^2\vartheta_W$. The values of $Q_W^{(2)}$ and $\delta_2$ for deuterium are consistent with zero, according to the value of $\Delta s$. The uncertainties in $Q_W^{(2)}$ and $\delta_2$ for hydrogen are resulting from the errors of the weak mixing angle and $\Delta s$ in equal shares.}
\label{t:values}
\end{table}

In table \ref{t:state.labels} we give the numbering scheme for the states which we consider. For electric field $\vmc E$ and magnetic field $\vmc B$ equal to zero we have the free $2S$ and $2P$ states. We write $\hat L$, $\hat P$, $\hat S$ since these states include the parity mixing due to $H_{\PV}$ (\ref{e1:eff.HPV}).

Consider first atoms in a constant $\vmc B$-field pointing in positive 3-direction,
\begin{align}\label{eA:B3}
\vmc B = \mcal B\vec e_3\ ,\quad \mcal B\geq 0\ .
\end{align}
The corresponding states $\rket{2\hat L_J, F,F_3,0,\mcal B\vec e_3}$ are obtained from those at $\mcal B=0$ by continuously turning on $\vmc B$ in the form (\ref{eA:B3}). Of course, for $|\vmc B| \neq 0$, $F$ is no longer a good quantum number. Here it is merely a label for the states.

We now choose a reference field $\vmc B_\mathrm{ref}=\mcal B_\mathrm{ref}\vec e_3$, $\mcal B_\mathrm{ref}>0$, below the first crossings in the Breit-Rabi diagrams, for instance $\mcal B_{\mathrm{ref}} = 0.05\,\mathrm{mT}$. We define the states $\rket{2\hat L_J,F,F_3,\vmc E,\vmc B}$ for arbitrary $\vmc B$ fields in the neighbourhood of $\vmc B_\mathrm{ref}$ as the states obtained continuously from $\rket{2\hat L_J,F,F_3,0,\vmc B_\mathrm{ref}}$ by turning on $\vmc E$ and $(\vmc B-\vmc B_\mathrm{ref})$ as $\lambda\vmc E$ and $\lambda(\vmc B-\vmc B_\mathrm{ref})$, respectively, with $\lambda\in[0,1]$. Then both $F$ and $F_3$ are only labels for these states and no longer good quantum numbers.

\newcommand{\phm}{\phantom-}
{
\begin{table}[htbp]
\caption{The numbering scheme for the atomic states of hydrogen and deuterium.}
\label{t:state.labels}\vskip10pt
\begin{tabular}{c|lp{7mm}c|l}
\multicolumn{2}{c}{hydrogen} && \multicolumn{2}{c}{deuterium}\\ \cline{1-2}\cline{4-5}
$\alpha$ & $\rket{2\hat L_J,F,F_3,\vmc E,\vmc B}$ && $\alpha$ & $\rket{2\hat L_J,F,F_3,\vmc E,\vmc B}$ \\ 
\cline{1-2}\cline{4-5}
1 & $\rket{2\hat P_{3/2},2,\phm2,\vmc E,\vmc B}$  &&  1 & $\rket{2\hat P_{3/2},\tfrac52,\phm\tfrac52,\vmc E,\vmc B}$\\
2 & $\rket{2\hat P_{3/2},2,\phm1,\vmc E,\vmc B}$  &&  2 & $\rket{2\hat P_{3/2},\tfrac52,\phm\tfrac32,\vmc E,\vmc B}$\\
3 & $\rket{2\hat P_{3/2},2,\phm0,\vmc E,\vmc B}$  &&  3 & $\rket{2\hat P_{3/2},\tfrac52,\phm\tfrac12,\vmc E,\vmc B}$\\
4 & $\rket{2\hat P_{3/2},2,-1,\vmc E,\vmc B}$     &&  4 & $\rket{2\hat P_{3/2},\tfrac52,-\tfrac12,\vmc E,\vmc B}$\\
5 & $\rket{2\hat P_{3/2},2,-2,\vmc E,\vmc B}$     &&  5 & $\rket{2\hat P_{3/2},\tfrac52,-\tfrac32,\vmc E,\vmc B}$\\
6 & $\rket{2\hat P_{3/2},1,\phm1,\vmc E,\vmc B}$  &&  6 & $\rket{2\hat P_{3/2},\tfrac52,-\tfrac52,\vmc E,\vmc B}$\\
7 & $\rket{2\hat P_{3/2},1,\phm0,\vmc E,\vmc B}$  &&  7 & $\rket{2\hat P_{3/2},\tfrac32,\phm\tfrac32,\vmc E,\vmc B}$\\
8 & $\rket{2\hat P_{3/2},1,-1,\vmc E,\vmc B}$     &&  8 & $\rket{2\hat P_{3/2},\tfrac32,\phm\tfrac12,\vmc E,\vmc B}$\\ \cline{1-2}
9 & $\rket{2\hat S_{1/2},1,\phm1,\vmc E,\vmc B}$  &&  9 & $\rket{2\hat P_{3/2},\tfrac32,-\tfrac12,\vmc E,\vmc B}$\\
10 & $\rket{2\hat S_{1/2},1,\phm0,\vmc E,\vmc B}$ && 10 & $\rket{2\hat P_{3/2},\tfrac32,-\tfrac32,\vmc E,\vmc B}$\\
11 & $\rket{2\hat S_{1/2},1,-1,\vmc E,\vmc B}$    && 11 & $\rket{2\hat P_{3/2},\tfrac12,\phm\tfrac12,\vmc E,\vmc B}$\\
12 & $\rket{2\hat S_{1/2},0,\phm0,\vmc E,\vmc B}$ && 12 & $\rket{2\hat P_{3/2},\tfrac12,-\tfrac12,\vmc E,\vmc B}$\\ \cline{1-2}\cline{4-5}
13 & $\rket{2\hat P_{1/2},1,\phm1,\vmc E,\vmc B}$ && 13 & $\rket{2\hat S_{1/2},\tfrac32,\phm\tfrac32,\vmc E,\vmc B}$\\
14 & $\rket{2\hat P_{1/2},1,\phm0,\vmc E,\vmc B}$ && 14 & $\rket{2\hat S_{1/2},\tfrac32,\phm\tfrac12,\vmc E,\vmc B}$\\
15 & $\rket{2\hat P_{1/2},1,-1,\vmc E,\vmc B}$    && 15 & $\rket{2\hat S_{1/2},\tfrac32,-\tfrac12,\vmc E,\vmc B}$\\
16 & $\rket{2\hat P_{1/2},0,\phm0,\vmc E,\vmc B}$ && 16 & $\rket{2\hat S_{1/2},\tfrac32,-\tfrac32,\vmc E,\vmc B}$\\
\multicolumn{3}{c}{} & 17 & $\rket{2\hat S_{1/2},\tfrac12,\phm\tfrac12,\vmc E,\vmc B}$\\ 
\multicolumn{3}{c}{} & 18 & $\rket{2\hat S_{1/2},\tfrac12,-\tfrac12,\vmc E,\vmc B}$\\ \cline{4-5}
\multicolumn{3}{c}{} & 19 & $\rket{2\hat P_{1/2},\tfrac32,\phm\tfrac32,\vmc E,\vmc B}$\\ 
\multicolumn{3}{c}{} & 20 & $\rket{2\hat P_{1/2},\tfrac32,\phm\tfrac12,\vmc E,\vmc B}$\\ 
\multicolumn{3}{c}{} & 21 & $\rket{2\hat P_{1/2},\tfrac32,-\tfrac12,\vmc E,\vmc B}$\\ 
\multicolumn{3}{c}{} & 22 & $\rket{2\hat P_{1/2},\tfrac32,-\tfrac32,\vmc E,\vmc B}$\\ 
\multicolumn{3}{c}{} & 23 & $\rket{2\hat P_{1/2},\tfrac12,\phm\tfrac12,\vmc E,\vmc B}$\\ 
\multicolumn{3}{c}{} & 24 & $\rket{2\hat P_{1/2},\tfrac12,-\tfrac12,\vmc E,\vmc B}$\\ 
\end{tabular}
\end{table}}
\clearpage

Tables \ref{t:H2.M0}, \ref{t:H2.D} and \ref{t:H2.Mu} show the non-zero parts of the mass matrix $\umat{\tilde M}_0$ for zero external fields, of the electric dipole operator $\uvec{D}$ and of the magnetic dipole operator $\uvec{\mu}$ for the $n=2$ states of hydrogen.

In tables \ref{t:H2.D} and \ref{t:H2.Mu} we use the spherical unit vectors, which are defined as
\begin{align}
\vec e_0 = \vec e_3\ ,\qquad\vec e_\pm = \mp\frac1{\sqrt2}\klr{\vec e_1 \pm \I\vec e_2}\ ,
\end{align}
where $\vec e_i$ $(i=1,2,3)$ are the Cartesian unit vectors. For $\vec e_\pm$, the following relation holds:
\begin{align}
\vec e_\pm^* = -\vec e_\mp\ .
\end{align}

{
\begin{table}[htbp]
\caption{The mass matrix $\umat{\tilde M}_0(\delta_1,\delta_2)$ (\ref{e2:MM.free}) for the $n=2$ states of hydrogen for the case of zero external fields. For the explanation of the variables $\Delta$, $L$ and $\mcal A$ and their numerical values see the introduction of this appendix and table \ref{t:values}. The PV parameters $\delta_{1,2}$ can also be found in table \ref{t:values}, the decay rates $\Gamma_{P.S}$ are given in (\ref{e3:tauS}) and (\ref{e3:tauP}).
}\label{t:H2.M0}
\vskip10pt
\begin{tabular}{l||c|cccc|}
& \parbox{18mm}{\vskip5pt$\ 2P_{3/2},2,2\ $\vskip4pt} & $\ 2P_{3/2},2,1\ $ & $\ 2P_{3/2},1,1\ $ & $\ 2P_{1/2},1,1\ $ & $\ 2S_{1/2},1,1\ $\\  \hline\hline
$\ 2P_{3/2},2,2\ $ & \mycell{18mm}{$\FineStructure+\frac{\HyperFineSplitting}{160}$\\[-1.5ex]$-\tfrac{\I }{2}\Gamma_P$}
& 0 & 0 & 0 & 0\\ \hline
$\ 2P_{3/2},2,1\ $ & 0 & 
\mycell{18mm}{${\FineStructure}+\frac{\HyperFineSplitting}{160}$\\[-1.5ex]$-\frac{\I}{2}\,{\Gamma_P}$}
& 0 & 0 & 0\\  
$\ 2P_{3/2},1,1\ $ & 0 & 0 & 
\mycell{18mm}{${\FineStructure}-\frac{\HyperFineSplitting}{96}$\\[-1.5ex]$-\frac{\I}{2}\,{\Gamma_P}$} & 
$-\frac{\HyperFineSplitting}{192\,{\sqrt{2}}}$ & 0\\  
$\ 2P_{1/2},1,1\ $ & 0 & 0 & 
$-\frac{\HyperFineSplitting}{192\,{\sqrt{2}}}$ & 
$\frac{\HyperFineSplitting}{96}-\frac\I2{\Gamma_P}$ & 
\mycell{18mm}{$\I{\delta_1}{\LambShift}$\\[-1.5ex]$+\frac\I2\delta_2 \LambShift$} \\ 
$\ 2S_{1/2},1,1\ $ & 0 & 0 & 0 & 
\mycell{18mm}{$-\I{\delta_1}{\LambShift}$\\[-1.5ex]$-\frac\I2 {\delta_2}{\LambShift}$} &
\mycell{18mm}{${\LambShift}+\frac{\HyperFineSplitting}{32}$\\[-1.5ex]$-\frac{\I}{2}\,{\Gamma_S}$}\\  \hline
\end{tabular}\\[15pt]
\centerline{\bf (Table \ref{t:H2.M0}a)}
\end{table}
\begin{table}
\begin{tabular}{l||cccccc|}
& \parbox{18mm}{\vskip5pt$\ 2P_{3/2},2,0\ $\vskip4pt} & $\ 2P_{3/2},1,0\ $ & $\ 2P_{1/2},1,0\ $ & $\ 2S_{1/2},1,0\ $ & $\ 2P_{1/2},0,0\ $ & $\ 2P_{3/2},0,0\ $\\ \hline\hline
$\ 2P_{3/2},2,0\ $ & 
\mycell{18mm}{${\FineStructure}+\frac{\HyperFineSplitting}{160}$\\[-1.5ex]$-\frac{\I}{2}{\Gamma_P}$} & 0 & 0 & 0
& 0 & 0\\  
$\ 2P_{3/2},1,0\ $ & 0 & 
\mycell{18mm}{${\FineStructure}-\frac{\HyperFineSplitting}{96}$\\[-1.5ex]$-\frac{\I}{2}{\Gamma_P}$} & $-\frac{\HyperFineSplitting}{192{\sqrt{2}}}$ & 0 & 0 & 0 \\  
$\ 2P_{1/2},1,0\ $ & 0 & 
$-\frac{\HyperFineSplitting}{192{\sqrt{2}}}$ & $\frac{\HyperFineSplitting}{96} - \frac\I2{\Gamma_P}$ & 
\mycell{18mm}{$\I\delta_1{\LambShift}$\\[-1.5ex]$+\frac\I2\delta_2{\LambShift}$} & 0 & 0 \\ 
$\ 2S_{1/2},1,0\ $ & 0 & 0 & 
\mycell{18mm}{$-\I{\delta_1}{\LambShift}$\\[-1.5ex]$-\frac\I2{\delta_2}{\LambShift}$} & 
\mycell{18mm}{${\LambShift}+\frac{\HyperFineSplitting}{32}$\\[-1.5ex]$-\frac{\I}{2}{\Gamma_S}$} & 0 & 0 \\ 
$\ 2P_{1/2},0,0\ $ & 0 & 0 & 0 & 0 & 
$-\frac{\HyperFineSplitting}{32} - \frac\I2{\Gamma_P}$ & 
\mycell{18mm}{$\I{\delta_1}{\LambShift}$\\[-1.5ex]$+\frac{3}{2}\I{\delta_2}{\LambShift}$} \\
$\ 2S_{1/2},0,0\ $ & 0 & 0 & 0 & 0 & 
\mycell{18mm}{$-\I{\delta_1}{\LambShift}$\\[-1.5ex]$-\frac{3}{2}\I{\delta_2}{\LambShift}$} & 
\mycell{18mm}{${\LambShift}-\frac{3\HyperFineSplitting}{32}$\\[-1.5ex]$-\frac{\I}{2}{\Gamma_S}$} \\ \hline
\end{tabular}\\[15pt]
\centerline{\bf (Table \ref{t:H2.M0}b)}
\end{table}
\begin{table}
\begin{tabular}{l||cccc|c|}
& \parbox{20mm}{\vskip5pt$\ 2P_{3/2},2,-1\ $\vskip4pt} & $\ 2P_{3/2},1,-1\ $ & $\ 2P_{1/2},1,-1\ $ & $\ 2S_{1/2},1,-1\ $ & $\ 2P_{3/2},2,-2\ $\\ \hline\hline
$\ 2P_{3/2},2,-1\ $ & 
\mycell{18mm}{${\FineStructure}+\frac{\HyperFineSplitting}{160}$\\[-1.5ex]$-\frac{\I}{2}{\Gamma_P}$} 
& 0 & 0 & 0 & 0 \\  
$\ 2P_{3/2},1,-1\ $ & 0 & 
\mycell{18mm}{${\FineStructure}-\frac{\HyperFineSplitting}{96}$\\[-1.5ex]$-\frac{\I}{2}{\Gamma_P}$} & 
$-\frac{\HyperFineSplitting}{192{\sqrt{2}}}$ & 0 & 0 \\ 
$\ 2P_{1/2},1,-1\ $ & 0 & 
$-\frac{\HyperFineSplitting}{192{\sqrt{2}}}$ & 
$\frac{\HyperFineSplitting}{96} - \frac\I2{\Gamma_P}$ & 
\mycell{18mm}{$\I{\delta_1}{\LambShift}$\\[-1.5ex]$+\frac\I2{\delta_2}{\LambShift}$} & 0 \\ 
$\ 2S_{1/2},1,-1\ $ & 0 & 0 & 
\mycell{18mm}{$-\I{\delta_1}{\LambShift}$\\[-1.5ex]$-\frac\I2{\delta_2}{\LambShift}$} & 
\mycell{18mm}{${\LambShift}+\frac{\HyperFineSplitting}{32}$\\[-1.5ex]$-\frac{\I}{2}{\Gamma_S}$} & 0 \\  \hline
$\ 2P_{3/2},2,-2\ $ & 0 & 0 & 0 & 0 & 
\mycell{18mm}{${\FineStructure}+\frac{\HyperFineSplitting}{160}$\\[-1.5ex]$-\frac{\I}{2}{\Gamma_P}$} \\ \hline
\end{tabular}\\[15pt]
\centerline{\bf (Table \ref{t:H2.M0}c)}
\end{table}}

\begin{turnpage}
{
\begin{table}
\caption{The suitably normalised electric dipole operator $\uvec{D}/(e\,r_B(1))$ for the $n=2$ states of hydrogen.}\label{t:H2.D}
\vskip10pt
\begin{tabular}{l||c|cccc|}
& $\ 2P_{3/2},2,2\ $ & $\ 2P_{3/2},2,1\ $ & $\ 2P_{3/2},1,1\ $ & $\ 2P_{1/2},1,1\ $ & $\ 2S_{1/2},1,1\ $\\ 
\hline\hline
$\ 2P_{3/2},2,2\ $ & 0 & 0 & 0 & 0 & $-3\sem$\\ \hline
$\ 2P_{3/2},2,1\ $ & 0 & 0 & 0 & 0 & 
$\frac{3}{{\sqrt{2}}}\sez$\\
$\ 2P_{3/2},1,1\ $ & 0 & 0 & 0 & 0 & 
$-{\sqrt{\frac{3}{2}}}\sez$\\
$\ 2P_{1/2},1,1\ $ & 0 & 0 & 0 & 0 & 
$-{\sqrt{3}}\sez$\\
$\ 2S_{1/2},1,1\ $ & $3\sep$ & 
$\frac{3}{{\sqrt{2}}}\sez$ & 
$-{\sqrt{\frac{3}{2}}}\sez$ & 
$-{\sqrt{3}}\sez$ & 0\\ \hline
\end{tabular}\\[15pt]
\centerline{\bf (Table \ref{t:H2.D}a)}
\vskip15pt
\begin{tabular}{l||cccccc|}
& $\ 2P_{3/2},2,0\ $ & $\ 2P_{3/2},1,0\ $ & $\ 2P_{1/2},1,0\ $ & $\ 2S_{1/2},1,0\ $ & $\ 2P_{1/2},0,0\ $ & $\ 2S_{1/2},0,0\ $\\ \hline\hline
$\ 2P_{3/2},2,1\ $ & 0 & 0 & 0 & 
$-\frac{3}{\sqrt2}\sem$ & 0 & 0\\ 
$\ 2P_{3/2},1,1\ $ & 0 & 0 & 0 & 
$-\sqrt{\frac{3}{2}}\sem$ & 0 & 
$-\sqrt{6}\sem$\\ 
$\ 2P_{1/2},1,1\ $ & 0 & 0 & 0 & 
$-\sqrt{3}\sem$ & 0 & 
$\sqrt{3}\sem$\\ 
$\ 2S_{1/2},1,1\ $ & 
$\sqrt{\frac{3}{2}}\sem$ & 
$-\sqrt{\frac{3}{2}}\sem$ & 
$-\sqrt{3}\sem$ & 0 & 
$\sqrt{3}\sem$ & 0\\ \hline
\end{tabular}\\[15pt]
\centerline{\bf (Table \ref{t:H2.D}b)}
\end{table}
\begin{table}
\begin{tabular}{l||cccc}
& $\ 2P_{3/2},2,1\ $ & $\ 2P_{3/2},1,1\ $ & $\ 2P_{1/2},1,1\ $ & $\ 2S_{1/2},1,1\ $\\ \hline\hline
$\ 2P_{3/2},2,0\ $ & 0 & 0 & 0 & $-\sqrt{\frac32}\sep$ \\ 
$\ 2P_{3/2},1,0\ $ & 0 & 0 & 0 & $\sqrt{\frac32}\sep$ \\ 
$\ 2P_{1/2},1,0\ $ & 0 & 0 & 0 & $\sqrt3\sep$ \\ 
$\ 2S_{1/2},1,0\ $ & $\frac3{\sqrt2}\sep$ & $\sqrt{\frac32}\sep$ & $\sqrt3\sep$ & 0 \\ 
$\ 2P_{1/2},0,0\ $ & 0 & 0 & 0 & $-\sqrt3\sep$ \\ 
$\ 2S_{1/2},0,0\ $ & 0 & $\sqrt6\sep$ & $-\sqrt3\sep$ & 0 \\ \hline
\end{tabular}\\[15pt]
\centerline{\bf (Table \ref{t:H2.D}c)}
\vskip15pt
\begin{tabular}{l||cccccc|}
& $\ 2P_{3/2},2,0\ $ & $\ 2P_{3/2},1,0\ $ & $\ 2P_{1/2},1,0\ $ & $\ 2S_{1/2},1,0\ $ & $\ 2P_{1/2},0,0\ $ & $\ 2S_{1/2},0,0\ $\\ \hline\hline
$\ 2P_{3/2},2,0\ $ & 0 & 0 & 0 & 
${\sqrt{6}}\sez$ & 0 & 0\\
$\ 2P_{3/2},1,0\ $ & 0 & 0 & 0 & 0 & 0 & 
${\sqrt{6}}\sez$\\
$\ 2P_{1/2},1,0\ $ & 0 & 0 & 0 & 0 & 0 & 
$-{\sqrt{3}}\sez$\\
$\ 2S_{1/2},1,0\ $ & 
${\sqrt{6}}\sez$ & 0 & 0 & 0 & 
$-{\sqrt{3}}\sez$ & 0\\
$\ 2P_{1/2},0,0\ $ & 0 & 0 & 0 & 
$-{\sqrt{3}}\sez$ & 0 & 0\\
$\ 2S_{1/2},0,0\ $ & 0 & 
${\sqrt{6}}\sez$ & 
$-{\sqrt{3}}\sez$ & 0 & 0 & 0\\ \hline
\end{tabular}\\[15pt]
\centerline{\bf (Table \ref{t:H2.D}d)}
\end{table}
\begin{table}
\begin{tabular}{l||cccc|}
& $\ 2P_{3/2},2,-1\ $ & $\ 2P_{3/2},1,-1\ $ & $\ 2P_{1/2},1,-1\ $ & $\ 2S_{1/2},1,-1\ $\\ \hline\hline
$\ 2P_{3/2},2,0\ $ & 0 & 0 & 0 & 
$-\sqrt{\frac{3}{2}}\sem$\\ 
$\ 2P_{3/2},1,0\ $ & 0 & 0 & 0 & 
$-\sqrt{\frac{3}{2}}\sem$\\
$\ 2P_{1/2},1,0\ $ & 0 & 0 & 0 & 
$-\sqrt{3}\sem$\\
$\ 2S_{1/2},1,0\ $ & 
$\frac{3}{\sqrt2}\sem$ & 
$-\sqrt{\frac{3}{2}}\sem$ & 
$-\sqrt{3}\sem$ & 0\\
$\ 2P_{1/2},0,0\ $ & 0 & 0 & 0 & 
$-\sqrt{3}\sem$\\
$\ 2S_{1/2},0,0\ $ & 0 & 
$\sqrt{6}\sem$ & 
$-\sqrt{3}\sem$ & 0\\ \hline
\end{tabular}\\[15pt]
\centerline{\bf (Table \ref{t:H2.D}e)}\vskip15pt
\begin{tabular}{l||cccccc|}
& $\ 2P_{3/2},2,0\ $ & $\ 2P_{3/2},1,0\ $ & $\ 2P_{1/2},1,0\ $ & $\ 2S_{1/2},1,0\ $ & $\ 2P_{1/2},0,0\ $ & $\ 2S_{1/2},0,0\ $\\ \hline\hline
$\ 2P_{3/2},2,-1\ $ & 0 & 0 & 0 & $-\frac3{\sqrt2}\sep$ & 0 & 0 \\ 
$\ 2P_{3/2},1,-1\ $ & 0 & 0 & 0 & $\sqrt{\frac32}\sep$ & 0 & $-\sqrt6\sep$ \\ 
$\ 2P_{1/2},1,-1\ $ & 0 & 0 & 0 & $\sqrt3\sep$ & 0 & $\sqrt3\sep$ \\ 
$\ 2S_{1/2},1,-1\ $ & $\sqrt{\frac32}\sep$ & $\sqrt{\frac32}\sep$ & $\sqrt3\sep$ & 0 & $\sqrt3\sep$ & 0 \\ \hline
\end{tabular}\\[15pt]
\centerline{\bf (Table \ref{t:H2.D}f)}
\end{table}
\begin{table}
\begin{tabular}{l||cccc|c|}
& $\ 2P_{3/2},2,-1\ $ & $\ 2P_{3/2},1,-1\ $ & $\ 2P_{1/2},1,-1\ $ & $\ 2S_{1/2},1,-1\ $ & $\ 2P_{3/2},2,-2\ $\\ \hline\hline
$\ 2P_{3/2},2,-1\ $ & 0 & 0 & 0 & 
$\frac{3}{{\sqrt{2}}}\sez$ & 0\\  
$\ 2P_{3/2},1,-1\ $ & 0 & 0 & 0 & 
${\sqrt{\frac{3}{2}}}\sez$ & 0\\  
$\ 2P_{1/2},1,-1\ $ & 0 & 0 & 0 & 
${\sqrt{3}}\sez$ & 0\\  
$\ 2S_{1/2},1,-1\ $ & 
$\frac{3}{{\sqrt{2}}}\sez$ & 
${\sqrt{\frac{3}{2}}}\sez$ & 
${\sqrt{3}}\sez$ & 0 & $3\sem$\\ \hline
$\ 2P_{3/2},2,-2\ $ & 0 & 0 & 0 & $-3\sep$ & 0\\ \hline
\end{tabular}\\[15pt]
\centerline{\bf (Table \ref{t:H2.D}g)}
\end{table}}

{
\begin{table}
\caption{The suitably normalised magnetic dipole operator $\uvec{\mu}/\mu_B$ for the $n=2$ states of hydrogen, where $\mu_B = e\hbar/(2m_e)$ is the Bohr magneton and $g=2.002319304(76)$ is the Land\'{e} factor of the electron \cite{Moh05}.}\label{t:H2.Mu}
\vskip10pt
\begin{tabular}{l||c|cccc|}
& $\ 2P_{3/2},2,2\ $ & $\ 2P_{3/2},2,1\ $ & $\ 2P_{3/2},1,1\ $ & $\ 2P_{1/2},1,1\ $ & $\ 2S_{1/2},1,1\ $\\ 
\hline\hline
$\ 2P_{3/2},2,2\ $ & $-\frac{g+2}{2}\sez$ & $-\frac{\sqrt2(g+2)}{4}\sem$ & $\frac{\sqrt2(g+2)}{4\sqrt3}\sem$ & $-\frac{g-1}{\sqrt3}\sem$ & 0\\ \hline
$\ 2P_{3/2},2,1\ $ & $\frac{\sqrt2(g+2)}{4}\sep$ & $-\frac{g+2}{4}\sez$ & $-\frac{g+2}{4\sqrt3}\sez$ & $-\frac{g-1}{\sqrt6}\sez$ & 0\\
$\ 2P_{3/2},1,1\ $ & $-\frac{\sqrt2(g+2)}{4\sqrt3}\sep$ & $-\frac{g+2}{4\sqrt3}\sez$ & $-\frac{5(g+2)}{12}\sez$ & $\frac{g-1}{3\sqrt2}\sez$ & 0\\
$\ 2P_{1/2},1,1\ $ & $\frac{g-1}{\sqrt3}\sep$ & $-\frac{g-1}{\sqrt6}\sem$ & $\frac{g-1}{3\sqrt2}\sez$ & $\frac{g-4}6\sez$ & 0\\
$\ 2S_{1/2},1,1\ $ & 0 & 0 & 0 & 0 & $-\frac{g}2\sez$\\ \hline
\end{tabular}\\[15pt]
\centerline{\bf (Table \ref{t:H2.Mu}a)}\vskip15pt
\begin{tabular}{l||cccccc|}
& $\ 2P_{3/2},2,0\ $ & $\ 2P_{3/2},1,0\ $ & $\ 2P_{1/2},1,0\ $ & $\ 2S_{1/2},1,0\ $ & $\ 2P_{1/2},0,0\ $ & $\ 2S_{1/2},0,0\ $\\ \hline\hline
$\ 2P_{3/2},2,1\ $ & $-\frac{\sqrt3(g+2)}4\sem$ & $\frac{g+2}{4\sqrt3}\sem$ & $-\frac{g-1}{\sqrt6}\sem$ & 0 & 0 & 0\\ 
$\ 2P_{3/2},1,1\ $ & $-\frac{g+2}{12}\sem$ & $-\frac{5(g+2)}{12}\sem$ & $-\frac{\sqrt2(g-1)}{6}\sem$ & 0 & $-\frac{\sqrt2(g-1)}{3}\sem$ & 0\\ 
$\ 2P_{1/2},1,1\ $ & $\frac{\sqrt2(g-1)}{6}\sem$ & $-\frac{\sqrt2(g-1)}{6}\sem$ & $\frac{g-4}{6}\sem$ & 0 & $-\frac{g-4}{6}\sem$ & 0\\ 
$\ 2S_{1/2},1,1\ $ & 0 & 0 & 0 & $-\frac{g}2\sem$ & 0 & $\frac{g}2\sem$\\ \hline
\end{tabular}\\[15pt]
\centerline{\bf (Table \ref{t:H2.Mu}b)}
\end{table}
\begin{table}
\begin{tabular}{l||cccc|}
& $\ 2P_{3/2},2,1\ $ & $\ 2P_{3/2},1,1\ $ & $\ 2P_{1/2},1,1\ $ & $\ 2S_{1/2},1,1\ $\\ \hline\hline
$\ 2P_{3/2},2,0\ $ & $\frac{\sqrt3(g+2)}4\sep$ & $\frac{g+2}{12}\sep$ & $-\frac{\sqrt2(g-1)}{6}\sep$ & 0 \\ 
$\ 2P_{3/2},1,0\ $ & $-\frac{g+2}{4\sqrt3}\sep$ & $\frac{5(g+2)}{12}\sep$ & $\frac{\sqrt2(g-1)}{6}\sep$ & 0 \\ 
$\ 2P_{1/2},1,0\ $ & $\frac{g-1}{\sqrt6}\sep$ & $\frac{\sqrt2(g-1)}{6}\sep$ & $-\frac{g-4}{6}\sep$ & 0 \\ 
$\ 2S_{1/2},1,0\ $ & 0 & 0 & 0 & $\frac{g}2\sep$ \\ 
$\ 2P_{1/2},0,0\ $ & 0 & $\frac{\sqrt2(g-1)}{3}\sep$ & $\frac{g-4}6\sep$ & 0 \\ 
$\ 2S_{1/2},0,0\ $ & 0 & 0 & 0 & $-\frac{g}2\sep$ \\ \hline
\end{tabular}\\[15pt]
\centerline{\bf (Table \ref{t:H2.Mu}c)}\vskip15pt
\begin{tabular}{l||cccccc|}
& $\ 2P_{3/2},2,0\ $ & $\ 2P_{3/2},1,0\ $ & $\ 2P_{1/2},1,0\ $ & $\ 2S_{1/2},1,0\ $ & $\ 2P_{1/2},0,0\ $ & $\ 2S_{1/2},0,0\ $\\ \hline\hline
$\ 2P_{3/2},2,0\ $ & 0 & 
$-\frac{g+2}{6}\sez$ & 
$-\frac{\sqrt2(g-1)}{3}\sez$ & 0 & 0 & 0\\
$\ 2P_{3/2},1,0\ $ & 
$-\frac{g+2}{6}\sez$ & 0 & 0 & 0 & 
$-\frac{{\sqrt{2}}(g-1)}{3}\sez$ & 0\\ 
$\ 2P_{1/2},1,0\ $ & 
$-\frac{{\sqrt{2}}(g-1)}{3}\sez$ & 0 & 0 & 0 & 
$\frac{g-4}{6}\sez$ & 0\\ 
$\ 2S_{1/2},1,0\ $ & 0 & 0 & 0 & 0 & 0 & 
$-\frac{g}{2}\sez$\\ 
$\ 2P_{1/2},0,0\ $ & 0 & 
$-\frac{{\sqrt{2}}(g-1)}{3}\sez$ & 
$\frac{g-4}{6}\sez$ & 0 & 0 & 0\\
$\ 2S_{1/2},0,0\ $ & 
0 & 0 & 0 & 
$-\frac{g}{2}\sez$ & 
0 & 0\\ \hline
\end{tabular}\\[15pt]
\centerline{\bf (Table \ref{t:H2.Mu}d)}
\end{table}
\begin{table}
\begin{tabular}{l||cccc|}
& $\ 2P_{3/2},2,-1\ $ & $\ 2P_{3/2},1,-1\ $ & $\ 2P_{1/2},1,-1\ $ & $\ 2S_{1/2},1,-1\ $\\ \hline\hline
$\ 2P_{3/2},2,0\ $ & 
$-\frac{\sqrt3(g+2)}{4}\sem$ & 
$\frac{g+2}{12}\sem$ & 
$-\frac{\sqrt2(g-1)}{6}\sem$ & 0\\
$\ 2P_{3/2},1,0\ $ & 
$-\frac{g+2}{4 \sqrt{3}}\sem$ & 
$-\frac{5 (g+2)}{12}\sem$ & 
$-\frac{\sqrt2(g-1)}{6}\sem$ & 0\\ 
$\ 2P_{1/2},1,0\ $ & 
$\frac{g-1}{\sqrt{6}}\sem$ & 
$-\frac{\sqrt2(g-1)}{6}\sem$ & 
$\frac{g-4}{6}\sem$ & 0\\ 
$\ 2S_{1/2},1,0\ $ & 0 & 0 & 0 & 
$-\frac{g}{2}\sem$\\ 
$\ 2P_{1/2},0,0\ $ & 0 & 
$\frac{\sqrt2(g-1)}{3}\sem$ & 
$\frac{g-4}{6}\sem$ & 0\\ 
$\ 2S_{1/2},0,0\ $ & 0 & 0 & 0 & 
$-\frac{g}{2}\sem$\\ \hline
\end{tabular}\\[15pt]
\centerline{\bf (Table \ref{t:H2.Mu}e)}\vskip15pt
\begin{tabular}{l||cccccc|}
& $\ 2P_{3/2},2,0\ $ & $\ 2P_{3/2},1,0\ $ & $\ 2P_{1/2},1,0\ $ & $\ 2S_{1/2},1,0\ $ & $\ 2P_{1/2},0,0\ $ & $\ 2S_{1/2},0,0\ $\\ \hline\hline
$\ 2P_{3/2},2,-1\ $ & 
$\frac{\sqrt3(g+2)}{4}\sep$ & 
$\frac{g+2}{4 \sqrt{3}}\sep$ & 
$-\frac{g-1}{\sqrt{6}}\sep$ & 0 & 0 & 0 \\ 
$\ 2P_{3/2},1,-1\ $ & 
$-\frac{g+2}{12}\sep$ & 
$\frac{5 (g+2)}{12}\sep$ & 
$\frac{\sqrt2(g-1)}{6}\sep$ & 0 & 
$-\frac{\sqrt2(g-1)}{3}\sep$ & 0 \\ 
$\ 2P_{1/2},1,-1\ $ & 
$\frac{\sqrt2(g-1)}{6}\sep$ & 
$\frac{\sqrt2(g-1)}{6}\sep$ & 
$-\frac{g-4}{6}\sep$ & 0 & 
$-\frac{g-4}{6}\sep$ & 0 \\ 
$\ 2S_{1/2},1,-1\ $ & 0 & 0 & 0 & 
$\frac{g}{2}\sep$ & 0 & $\frac{g}{2}\sep$ \\ \hline
\end{tabular}\\[15pt]
\centerline{\bf (Table \ref{t:H2.Mu}f)}
\end{table}
\begin{table}
\begin{tabular}{l||cccc|c|}
& $\ 2P_{3/2},2,-1\ $ & $\ 2P_{3/2},1,-1\ $ & $\ 2P_{1/2},1,-1\ $ & $\ 2S_{1/2},1,-1\ $ & $\ 2P_{3/2},2,-2\ $\\ \hline\hline
$\ 2P_{3/2},2,-1\ $ & 
$\frac{g+2}{4}\sez$ & 
$-\frac{g+2}{4{\sqrt{3}}}\sez$ & 
$-\frac{g-1}{{\sqrt{6}}}\sez$ & 
0 & $-\frac{\sqrt2(g+2)}{4}\sem$\\
$\ 2P_{3/2},1,-1\ $ & 
$-\frac{g+2}{4{\sqrt{3}}\sez}$ & 
$\frac{5(g+2)}{12}\sez$ & 
$-\frac{g-1}{3{\sqrt{2}}}\sez$ & 
0 & $-\frac{\sqrt2(g+2)}{4\sqrt3}\sem$\\ 
$\ 2P_{1/2},1,-1\ $ & 
$-\frac{g-1}{{\sqrt{6}}}\sez$ & 
$-\frac{g-1}{3{\sqrt{2}}}\sez$ & 
$-\frac{g-4}{6}\sez$ & 
0 & $\frac{g-1}{\sqrt3}\sem$\\ 
$\ 2S_{1/2},1,-1\ $ & 
0 & 0 & 0 & 
$\frac{g}{2}\sez$ & 0\\ \hline
$\ 2P_{3/2},2,-2\ $ & $\frac{\sqrt2(g+2)}{4}\sep$ & $\frac{\sqrt2(g+2)}{4\sqrt3}\sep$ & $-\frac{g-1}{\sqrt3}\sep$ & 0 & 
$\frac{2 + g}{2}\sez$\\ \hline
\end{tabular}\\[15pt]
\centerline{\bf (Table \ref{t:H2.Mu}g)}
\end{table}}
\vfill
\end{turnpage}

\clearpage
Tables \ref{t:D2.M0}, \ref{t:D2.D} and \ref{t:D2.Mu} show the mass matrix $\umat{\tilde M}_0$ for zero external fields, the electric dipole operator $\uvec D$ and the magnetic dipole operator $\uvec\mu$ for the $n=2$ states of deuterium.

\begin{table}[hbp]
\caption{The mass matrix $\umat{\tilde M}_0(\delta_1,\delta_2)$ (\ref{e2:MM.free}) for the $n=2$ states of deuterium for the case of zero external fields. For the explanation of the variables $\Delta$, $L$ and $\mcal A$ and their numerical values see the introduction of this appendix and table \ref{t:values}. The PV parameters $\delta_{1,2}$ can also be found in table \ref{t:values}, the decay rates $\Gamma_{P.S}$ are given in (\ref{e3:tauS}) and (\ref{e3:tauP}).
}\label{t:D2.M0}
\vskip10pt
\begin{tabular}{l||c|cccc|}
& \parbox{18mm}{\vskip5pt$\ 2P_{3/2},\tfrac{5}{2},\tfrac{5}{2}\ $\vskip4pt} & $\ 2P_{3/2},\tfrac{5}{2},\tfrac{3}{2}\ $ & $\ 2P_{3/2},\tfrac{3}{2},\tfrac{3}{2}\ $ & $\ 2P_{1/2},\tfrac{3}{2},\tfrac{3}{2}\ $ & $\ 2S_{1/2},\tfrac{3}{2},\tfrac{3}{2}\ $\\ \hline\hline
$\ 2P_{3/2},\tfrac{5}{2},\tfrac{5}{2}\ $ & 
\mycell{18mm}{${\FineStructure}+\frac{\HyperFineSplitting}{120}$\\[-1.5ex]$ - \frac{\I}{2}{\Gamma_P}$} & 0 & 0 & 0 & 0\\ \hline
$\ 2P_{3/2},\tfrac{5}{2},\tfrac{3}{2}\ $ & 0 & 
\mycell{18mm}{${\FineStructure}+\frac{\HyperFineSplitting}{120}$\\[-1.5ex]$ - \frac{\I}{2}{\Gamma_P}$} & 0 & 0 & 0\\
$\ 2P_{3/2},\tfrac{3}{2},\tfrac{3}{2}\ $ & 0 & 0 & 
\mycell{18mm}{${\FineStructure}-\frac{\HyperFineSplitting}{180}$\\[-1.5ex]$ - \frac{\I}{2}{\Gamma_P}$} & 
$-\frac{{\sqrt{5}}\HyperFineSplitting}{576}$ & 0\\ 
$\ 2P_{1/2},\tfrac{3}{2},\tfrac{3}{2}\ $ & 0 & 0 & 
$-\frac{{\sqrt{5}}\HyperFineSplitting}{576}$ & 
$\frac{\HyperFineSplitting}{72} - \frac\I2{\Gamma_P}$ & 
\mycell{18mm}{$\phm\I{\delta_1}{\LambShift}$\\[-1.5ex]$ + \I{\delta_2}{\LambShift}$}\\ 
$\ 2S_{1/2},\tfrac{3}{2},\tfrac{3}{2}\ $ & 0 & 0 & 0 & 
\mycell{18mm}{$-\I{\delta_1}\LambShift$\\[-1.5ex]$ - \I{\delta_2}{\LambShift}$} &
\mycell{18mm}{${\LambShift}+\frac{\HyperFineSplitting}{24}$\\[-1.5ex]$ - \frac{\I}{2}{\Gamma_S}$}\\ \hline
\end{tabular}\\[15pt]
\centerline{\bf (Table \ref{t:D2.M0}a)}
\end{table}
\begin{turnpage}
{
\begin{table}
\begin{tabular}{l||ccccccc|}
& \parbox{18mm}{\vskip5pt$\ 2P_{3/2},\tfrac{5}{2},\tfrac{1}{2}\ $\vskip4pt} & $\ 2P_{3/2},\tfrac{3}{2},\tfrac{1}{2}\ $ & $\ 2P_{1/2},\tfrac{3}{2},\tfrac{1}{2}\ $ & $\ 2S_{1/2},\tfrac{3}{2},\tfrac{1}{2}\ $ & $\ 2P_{3/2},\tfrac{1}{2},\tfrac{1}{2}\ $ & $\ 2P_{1/2},\tfrac{1}{2},\tfrac{1}{2}\ $ & $\ 2S_{1/2},\tfrac{1}{2},\tfrac{1}{2}\ $\\ \hline\hline
$\ 2P_{3/2},\tfrac{5}{2},\tfrac{1}{2}\ $ & 
\mycell{18mm}{${\FineStructure}+\frac{\HyperFineSplitting}{120}$\\[-1.5ex]$ - \frac{\I}{2}{\Gamma_P}$} & 0 & 0 & 0 & 0 & 0 & 0\\
$2P_{3/2},\tfrac{3}{2},\tfrac{1}{2}\ $ & 0 & 
\mycell{18mm}{${\FineStructure}-\frac{\HyperFineSplitting}{180}$\\[-1.5ex]$ - \frac{\I}{2}{\Gamma_P}$} & 
$-\frac{\sqrt{5}\HyperFineSplitting}{576}$ & 0 & 0 & 0 & 0\\ 
$\ 2P_{1/2},\tfrac{3}{2},\tfrac{1}{2}\ $ & 0 & 
$-\frac{{\sqrt{5}}\HyperFineSplitting}{576}$ & 
$\frac{\HyperFineSplitting}{72} - \frac\I2{\Gamma_P}$ & 
\mycell{18mm}{$\phm\I{\delta_1}\LambShift$\\[-1.5ex]$ + \I{\delta_2}{\LambShift}$} & 0 & 0 & 0\\
$\ 2S_{1/2},\tfrac{3}{2},\tfrac{1}{2}\ $ & 0 & 0 & 
\mycell{18mm}{$-\I{\delta_1}\LambShift$\\[-1.5ex]$ - \I{\delta_2}{\LambShift}$} & 
\mycell{18mm}{${\LambShift}+\frac{\HyperFineSplitting}{24}$\\[-1.5ex]$ - \frac{\I}{2}{\Gamma_S}$} & 0 & 0 & 0\\ 
$\ 2P_{3/2},\tfrac{1}{2},\tfrac{1}{2}\ $ & 0 & 0 & 0 & 0 & 
\mycell{18mm}{${\FineStructure}-\frac{\HyperFineSplitting}{72}$\\[-1.5ex]$ - \frac{\I}{2}{\Gamma_P}$} & 
$-\frac{\HyperFineSplitting}{288{\sqrt{2}}}$ & 0\\ 
$\ 2P_{1/2},\tfrac{1}{2},\tfrac{1}{2}\ $ & 0 & 0 & 0 & 0 & 
$-\frac{\HyperFineSplitting}{288{\sqrt{2}}}$ & 
\mycell{18mm}{$-\frac{\HyperFineSplitting}{36}$\\[-1.5ex]$ - \frac\I2{\Gamma_P}$} & 
\mycell{18mm}{$\phantom{-2}\I{\delta_1}\LambShift$\\[-1.5ex]$ - 2\I{\delta_2} {\LambShift}$}\\ 
$\ 2S_{1/2},\tfrac{1}{2},\tfrac{1}{2}\ $ & 0 & 0 & 0 & 0 & 0 & 
\mycell{18mm}{$-\phantom2\I{\delta_1}{\LambShift}$\\[-1.5ex]$ + 2\I{\delta_2} {\LambShift}$} & 
\mycell{18mm}{${\LambShift}-\frac{\HyperFineSplitting}{12}$\\[-1.5ex]$ - \frac{\I}{2}{\Gamma_S}$}\\ \hline
\end{tabular}\\[15pt]
\centerline{\bf (Table \ref{t:D2.M0}b)}
\end{table}
\begin{table}
\begin{tabular}{l||ccccccc|}
& \parbox{22mm}{\vskip5pt$\ 2P_{3/2},\tfrac{5}{2},-\tfrac{1}{2}\ $\vskip4pt} & $\ 2P_{3/2},\tfrac{3}{2},-\tfrac{1}{2}\ $ & $\ 2P_{1/2},\tfrac{3}{2},-\tfrac{1}{2}\ $ & $\ 2S_{1/2},\tfrac{3}{2},-\tfrac{1}{2}\ $ & $\ 2P_{3/2},\tfrac{1}{2},-\tfrac{1}{2}\ $ & $\ 2P_{1/2},\tfrac{1}{2},-\tfrac{1}{2}\ $ & $\ 2S_{1/2},\tfrac{1}{2},-\tfrac{1}{2}\ $\\ \hline\hline
$\ 2P_{3/2},\tfrac{5}{2},-\tfrac{1}{2}\ $ & 
\mycell{18mm}{${\FineStructure}+\frac{\HyperFineSplitting}{120}$\\[-1.5ex]$ - \frac{\I}{2}{\Gamma_P}$} 
& 0 & 0 & 0 & 0 & 0 & 0\\
$\ 2P_{3/2},\tfrac{3}{2},-\tfrac{1}{2}\ $ & 0 & 
\mycell{18mm}{${\FineStructure}-\frac{\HyperFineSplitting}{180}$\\[-1.5ex]$ - \frac{\I}{2}{\Gamma_P}$} & 
$-\frac{{\sqrt{5}}\HyperFineSplitting}{576}$ & 0 & 0 & 0 & 0\\ 
$\ 2P_{1/2},\tfrac{3}{2},-\tfrac{1}{2}\ $ & 0 & 
$-\frac{{\sqrt{5}}\HyperFineSplitting}{576}$ & 
$\frac{\HyperFineSplitting}{72} - \frac\I2{\Gamma_P}$ & 
\mycell{18mm}{$\phm\I{\delta_1}\LambShift$\\[-1.5ex]$ + \I{\delta_2} {\LambShift}$} & 0 & 0 & 0\\ 
$\ 2S_{1/2},\tfrac{3}{2},-\tfrac{1}{2}\ $ & 0 & 0 & 
\mycell{18mm}{$-\I {\delta_1}\LambShift$\\[-1.5ex]$ - \I{\delta_2} {\LambShift}$} & 
\mycell{18mm}{${\LambShift}+\frac{\HyperFineSplitting}{24}$\\[-1.5ex]$ - \frac{\I}{2}{\Gamma_S}$} 
& 0 & 0 & 0\\ 
$\ 2P_{3/2},\tfrac{1}{2},-\tfrac{1}{2}\ $ & 0 & 0 & 0 & 0 & 
\mycell{18mm}{${\FineStructure}-\frac{\HyperFineSplitting}{72}$\\[-1.5ex]$ - \frac{\I}{2}{\Gamma_P}$} & 
$-\frac{\HyperFineSplitting}{288{\sqrt{2}}}$ & 0\\ 
$\ 2P_{1/2},\tfrac{1}{2},-\tfrac{1}{2}\ $ & 0 & 0 & 0 & 0 & 
$-\frac{\HyperFineSplitting}{288{\sqrt{2}}}$ & 
\mycell{18mm}{$-\frac{\HyperFineSplitting}{36}$\\[-1.5ex]$ - \frac\I2{\Gamma_P}$} & 
\mycell{18mm}{$\phantom{-2}\I{\delta_1}\LambShift$\\[-1.5ex]$ - 2\I{\delta_2} {\LambShift}$}\\
$\ 2S_{1/2},\tfrac{1}{2},-\tfrac{1}{2}\ $ & 0 & 0 & 0 & 0 & 0 & 
\mycell{18mm}{$-\phantom2\I{\delta_1}\LambShift$\\[-1.5ex]$ + 2\I{\delta_2}{\LambShift}$} & 
\mycell{18mm}{${\LambShift}-\frac{\HyperFineSplitting}{12}$\\[-1.5ex]$ - \frac{\I}{2}{\Gamma_S}$}\\ \hline
\end{tabular}\\[15pt]
\centerline{\bf (Table \ref{t:D2.M0}c)}
\end{table}
\begin{table}
\begin{tabular}{l||cccc|c|}
& \parbox{22mm}{\vskip5pt$\ 2P_{3/2},\tfrac{5}{2},-\tfrac{3}{2}\ $\vskip4pt} & $\ 2P_{3/2},\tfrac{3}{2},-\tfrac{3}{2}\ $ &
$\ 2P_{1/2},\tfrac{3}{2},-\tfrac{3}{2}\ $ & $\ 2S_{1/2},\tfrac{3}{2},-\tfrac{3}{2}\ $ & 
$\ 2P_{3/2},\tfrac{5}{2},-\tfrac{5}{2}\ $\\ \hline\hline
$\ 2P_{3/2},\tfrac{5}{2},-\tfrac{3}{2}\ $ & 
\mycell{18mm}{${\FineStructure}+\frac{\HyperFineSplitting}{120}$\\[-1.5ex]$ - \frac{\I}{2}{\Gamma_P}$} & 0 & 0 & 0 & 0\\ 
$\ 2P_{3/2},\tfrac{3}{2},-\tfrac{3}{2}\ $ & 0 & 
\mycell{18mm}{${\FineStructure}-\frac{\HyperFineSplitting}{180}$\\[-1.5ex]$ - \frac{\I}{2}{\Gamma_P}$} & 
$-\frac{{\sqrt{5}}\HyperFineSplitting}{576}$ & 0 & 0\\
$\ 2P_{1/2},\tfrac{3}{2},-\tfrac{3}{2}\ $ & 0 & 
$-\frac{{\sqrt{5}}\HyperFineSplitting}{576}$ & 
$\frac{\HyperFineSplitting}{72} - \frac\I2 {\Gamma_P}$ & 
\mycell{18mm}{$\phm\I{\delta_1}{\LambShift}$\\[-1.5ex]$ + \I{\delta_2}{\LambShift}$} & 0\\ 
$\ 2S_{1/2},\tfrac{3}{2},-\tfrac{3}{2}\ $ & 0 & 0 & 
\mycell{18mm}{$-\I{\delta_1}{\LambShift}$\\[-1.5ex]$ - \I{\delta_2}{\LambShift}$} &
\mycell{18mm}{${\LambShift}+\frac{\HyperFineSplitting}{24}$\\[-1.5ex]$ - \frac{\I}{2}{\Gamma_S}$} & 0\\ \hline
$\ 2P_{3/2},\tfrac{5}{2},-\tfrac{5}{2}\ $ & 0 & 0 & 0 & 0 & 
\mycell{18mm}{${\FineStructure}+\frac{\HyperFineSplitting}{120}$\\[-1.5ex]$ - \frac{\I}{2}{\Gamma_P}$}\\ \hline
\end{tabular}\\[15pt]
\centerline{\bf (Table \ref{t:D2.M0}d)}
\end{table}}

{
\begin{table}
\caption{The suitably normalised electric dipole operator $\uvec{D}/(e\,r_B(1))$ for the $n=2$ states of deuterium.}\label{t:D2.D}
\vskip10pt
\begin{tabular}{l||c|cccc|}
& \parbox{18mm}{\vskip5pt$\ 2P_{3/2},\tfrac{5}{2},\tfrac{5}{2}\ $\vskip4pt} & $\ 2P_{3/2},\tfrac{5}{2},\tfrac{3}{2}\ $ & $\ 2P_{3/2},\tfrac{3}{2},\tfrac{3}{2}\ $ & $\ 2P_{1/2},\tfrac{3}{2},\tfrac{3}{2}\ $ & $\ 2S_{1/2},\tfrac{3}{2},\tfrac{3}{2}\ $\\ \hline\hline
$\ 2P_{3/2},\tfrac{5}{2},\tfrac{5}{2}\ $ & 0 & 0 & 0 & 0 & $-3\sem$\\ \hline
$\ 2P_{3/2},\tfrac{5}{2},\tfrac{3}{2}\ $ & 0 & 0 & 0 & 0 & $3\,{\sqrt{\frac{2}{5}}}\sez$\\
$\ 2P_{3/2},\tfrac{3}{2},\tfrac{3}{2}\ $ & 0 & 0 & 0 & 0 & $-2\,{\sqrt{\frac{3}{5}}}\sez$\\  
$\ 2P_{1/2},\tfrac{3}{2},\tfrac{3}{2}\ $ & 0 & 0 & 0 & 0 & $-{\sqrt{3}}\sez$\\  
$\ 2S_{1/2},\tfrac{3}{2},\tfrac{3}{2}\ $ & $3\sep$ & $3\,{\sqrt{\frac{2}{5}}}\sez$ & $-2\,{\sqrt{\frac{3}{5}}}\sez$ & $-{\sqrt{3}}\sez$ & 0\\ \hline
\end{tabular}\\[15pt]
\centerline{\bf (Table \ref{t:D2.D}a)}\vskip15pt
\begin{tabular}{l||ccccccc|}
& \parbox{20mm}{\vskip5pt$\ 2P_{3/2},\tfrac{5}{2},\tfrac{1}{2}\ $\vskip4pt} & $\ 2P_{3/2},\tfrac{3}{2},\tfrac{1}{2}\ $ & $\ 2P_{1/2},\tfrac{3}{2},\tfrac{1}{2}\ $ & $\ 2S_{1/2},\tfrac{3}{2},\tfrac{1}{2}\ $ & $\ 2P_{3/2},\tfrac{1}{2},\tfrac{1}{2}\ $ & $\ 2P_{1/2},\tfrac{1}{2},\tfrac{1}{2}\ $ & $\ 2S_{1/2},\tfrac{1}{2},\tfrac{1}{2}\ $\\ \hline\hline
$\ 2P_{3/2},\tfrac{5}{2},\tfrac{3}{2}\ $ & 0 & 0 & 0 & $-3 \sqrt{\frac{3}{5}}\sem$ & 0 & 0 & 0\\
$\ 2P_{3/2},\tfrac{3}{2},\tfrac{3}{2}\ $ & 0 & 0 & 0 & $-\sqrt{\frac{8}{5}}\sem$ & 0 & 0 & $-\sqrt{5}\sem$\\  
$\ 2P_{1/2},\tfrac{3}{2},\tfrac{3}{2}\ $ & 0 & 0 & 0 & $-\sqrt2\sem$ & 0 & 0 & $2\sem$\\  
$\ 2S_{1/2},\tfrac{3}{2},\tfrac{3}{2}\ $ & $\frac{3}{\sqrt{10}}\sem$ & $-\sqrt{\frac{8}{5}}\sem$ & $-\sqrt2\sem$ & 0 & $\frac{1}{\sqrt2}\sem$ & $2\sem$ & 0\\ \hline
\end{tabular}\\[15pt]
\centerline{\bf (Table \ref{t:D2.D}b)}
\end{table}
\begin{table}
\begin{tabular}{l||cccc|}
& \parbox{20mm}{\vskip5pt$\ 2P_{3/2},\tfrac{5}{2},\tfrac{3}{2}\ $\vskip4pt} & $\ 2P_{3/2},\tfrac{3}{2},\tfrac{3}{2}\ $ & $\ 2P_{1/2},\tfrac{3}{2},\tfrac{3}{2}\ $ & $\ 2S_{1/2},\tfrac{3}{2},\tfrac{3}{2}\ $\\ \hline\hline
$\ 2P_{3/2},\tfrac{5}{2},\tfrac{1}{2}\ $ & 0 & 0 & 0 & $-\frac{3}{\sqrt{10}}\sep$\\
$\ 2P_{3/2},\tfrac{3}{2},\tfrac{1}{2}\ $ & 0 & 0 & 0 & $\sqrt{\frac{8}{5}}\sep$ \\
$\ 2P_{1/2},\tfrac{3}{2},\tfrac{1}{2}\ $ & 0 & 0 & 0 & $\sqrt2\sep$ \\
$\ 2S_{1/2},\tfrac{3}{2},\tfrac{1}{2}\ $ & $3 \sqrt{\frac{3}{5}}\sep$ & $\sqrt{\frac{8}{5}}\sep$ & $\sqrt2\sep$ & 0 \\
$\ 2P_{3/2},\tfrac{1}{2},\tfrac{1}{2}\ $ & 0 & 0 & 0 & $-\frac{1}{\sqrt2}\sep$ \\
$\ 2P_{1/2},\tfrac{1}{2},\tfrac{1}{2}\ $ & 0 & 0 & 0 & $-2\sep$ \\
$\ 2S_{1/2},\tfrac{1}{2},\tfrac{1}{2}\ $ & 0 & $\sqrt{5}\sep$ & $-2\sep$ & 0 \\ \hline
\end{tabular}\\[15pt]
\centerline{\bf (Table \ref{t:D2.D}c)}\vskip15pt
\begin{tabular}{l||ccccccc|}
& \parbox{20mm}{\vskip5pt$\ 2P_{3/2},\tfrac{5}{2},\tfrac{1}{2}\ $\vskip4pt} & $\ 2P_{3/2},\tfrac{3}{2},\tfrac{1}{2}\ $ & $\ 2P_{1/2},\tfrac{3}{2},\tfrac{1}{2}\ $ & $\ 2S_{1/2},\tfrac{3}{2},\tfrac{1}{2}\ $ & $\ 2P_{3/2},\tfrac{1}{2},\tfrac{1}{2}\ $ & $\ 2P_{1/2},\tfrac{1}{2},\tfrac{1}{2}\ $ & $\ 2S_{1/2},\tfrac{1}{2},\tfrac{1}{2}\ $\\ \hline\hline
$\ 2P_{3/2},\tfrac{5}{2},\tfrac{1}{2}\ $ & 0 & 0 & 0 & $3\sqrt{\frac35}\sez$ & 0 & 0 & 0\\
$\ 2P_{3/2},\tfrac{3}{2},\tfrac{1}{2}\ $ & 0 & 0 & 0 & $-\frac2{\sqrt{15}}\sez$ & 0 & 0 & $\sqrt{\frac{10}3}\sez$\\ 
$\ 2P_{1/2},\tfrac{3}{2},\tfrac{1}{2}\ $ & 0 & 0 & 0 & $-\frac1{\sqrt3}\sez$ & 0 & 0 & $-2\sqrt{\frac23}\sez$\\
$\ 2S_{1/2},\tfrac{3}{2},\tfrac{1}{2}\ $ & $3\sqrt{\frac35}\sez$ & $-\frac2{\sqrt{15}}\sez$ & $-\frac1{\sqrt3}\sez$ & 0 & $-\frac1{\sqrt3}\sez$ & $-2\sqrt{\frac23}\sez$ & 0\\
$\ 2P_{3/2},\tfrac{1}{2},\tfrac{1}{2}\ $ & 0 & 0 & 0 & $-\frac1{\sqrt3}\sez$ & 0 & 0 & $-2\sqrt{\frac23}\sez$\\
$\ 2P_{1/2},\tfrac{1}{2},\tfrac{1}{2}\ $ & 0 & 0 & 0 & $-2\sqrt{\frac23}\sez$ & 0 & 0 & $\frac1{\sqrt3}\sez$\\
$\ 2S_{1/2},\tfrac{1}{2},\tfrac{1}{2}\ $ & 0 & $\sqrt{\frac{10}3}\sez$ & $-2\sqrt{\frac23}\sez$ & 0 & $-2\sqrt{\frac23}\sez$ & $\frac1{\sqrt3}\sez$ & 0\\ \hline
\end{tabular}\\[15pt]
\centerline{\bf (Table \ref{t:D2.D}d)}
\end{table}
\begin{table}
\begin{tabular}{l||ccccccc|}
& \parbox{22mm}{\vskip5pt$\ 2P_{3/2},\tfrac{5}{2},-\tfrac{1}{2}\ $\vskip4pt} & $\ 2P_{3/2},\tfrac{3}{2},-\tfrac{1}{2}\ $ & $\ 2P_{1/2},\tfrac{3}{2},-\tfrac{1}{2}\ $ & $\ 2S_{1/2},\tfrac{3}{2},-\tfrac{1}{2}\ $ & $\ 2P_{3/2},\tfrac{1}{2},-\tfrac{1}{2}\ $ & $\ 2P_{1/2},\tfrac{1}{2},-\tfrac{1}{2}\ $ & $\ 2S_{1/2},\tfrac{1}{2},-\tfrac{1}{2}\ $\\ \hline\hline
$\ 2P_{3/2},\tfrac{5}{2},\tfrac{1}{2}\ $ & 0 & 0 & 0 & $-3\sqrt{\frac3{10}}\sem$ & 0 & 0 & 0\\
$\ 2P_{3/2},\tfrac{3}{2},\tfrac{1}{2}\ $ & 0 & 0 & 0 & $-4\sqrt{\frac{2}{15}}\sem$ & 0 & 0 & $-\sqrt{\frac53}\sem$\\ 
$\ 2P_{1/2},\tfrac{3}{2},\tfrac{1}{2}\ $ & 0 & 0 & 0 & $-\sqrt{\frac83}\sem$ & 0 & 0 & $\sqrt{\frac43}\sem$\\
$\ 2S_{1/2},\tfrac{3}{2},\tfrac{1}{2}\ $ & $3\sqrt{\frac3{10}}\sem$ & $-4\sqrt{\frac{2}{15}}\sem$ & $-\sqrt{\frac83}\sem$ & 0 & $\frac1{\sqrt6}\sem$ & $\sqrt{\frac43}\sem$ & 0\\
$\ 2P_{3/2},\tfrac{1}{2},\tfrac{1}{2}\ $ & 0 & 0 & 0 & $-\frac1{\sqrt6}\sem$ & 0 & 0 & $-2\sqrt{\frac43}\sem$\\
$\ 2P_{1/2},\tfrac{1}{2},\tfrac{1}{2}\ $ & 0 & 0 & 0 & $-\sqrt{\frac43}\sem$ & 0 & 0 & $\sqrt{\frac23}\sem$\\
$\ 2S_{1/2},\tfrac{1}{2},\tfrac{1}{2}\ $ & 0 & $\sqrt{\frac53}\sem$ & $-\sqrt{\frac43}\sem$ & 0 & $-2\sqrt{\frac43}\sem$ & $\sqrt{\frac23}\sem$ & 0\\ \hline
\end{tabular}\\[15pt]
\centerline{\bf (Table \ref{t:D2.D}e)}\vskip15pt
\begin{tabular}{l||ccccccc|}
& \parbox{20mm}{\vskip5pt$\ 2P_{3/2},\tfrac{5}{2},\tfrac{1}{2}\ $\vskip4pt} & $\ 2P_{3/2},\tfrac{3}{2},\tfrac{1}{2}\ $ & $\ 2P_{1/2},\tfrac{3}{2},\tfrac{1}{2}\ $ & $\ 2S_{1/2},\tfrac{3}{2},\tfrac{1}{2}\ $ & $\ 2P_{3/2},\tfrac{1}{2},\tfrac{1}{2}\ $ & $\ 2P_{1/2},\tfrac{1}{2},\tfrac{1}{2}\ $ & $\ 2S_{1/2},\tfrac{1}{2},\tfrac{1}{2}\ $\\ \hline\hline
$\ 2P_{3/2},\tfrac{5}{2},-\tfrac{1}{2}\ $ & 0 & 0 & 0 & $-3\sqrt{\frac3{10}}\sep$ & 0 & 0 & 0\\
$\ 2P_{3/2},\tfrac{3}{2},-\tfrac{1}{2}\ $ & 0 & 0 & 0 & $4\sqrt{\frac{2}{15}}\sep$ & 0 & 0 & $-\sqrt{\frac53}\sep$\\ 
$\ 2P_{1/2},\tfrac{3}{2},-\tfrac{1}{2}\ $ & 0 & 0 & 0 & $\sqrt{\frac83}\sep$ & 0 & 0 & $\sqrt{\frac43}\sep$\\
$\ 2S_{1/2},\tfrac{3}{2},-\tfrac{1}{2}\ $ & $3\sqrt{\frac3{10}}\sep$ & $4\sqrt{\frac{2}{15}}\sep$ & $\sqrt{\frac83}\sep$ & 0 & $\frac1{\sqrt6}\sep$ & $\sqrt{\frac43}\sep$ & 0\\
$\ 2P_{3/2},\tfrac{1}{2},-\tfrac{1}{2}\ $ & 0 & 0 & 0 & $-\frac1{\sqrt6}\sep$ & 0 & 0 & $2\sqrt{\frac43}\sep$\\
$\ 2P_{1/2},\tfrac{1}{2},-\tfrac{1}{2}\ $ & 0 & 0 & 0 & $-\sqrt{\frac43}\sep$ & 0 & 0 & $-\sqrt{\frac23}\sep$\\
$\ 2S_{1/2},\tfrac{1}{2},-\tfrac{1}{2}\ $ & 0 & $\sqrt{\frac53}\sep$ & $-\sqrt{\frac43}\sep$ & 0 & $2\sqrt{\frac43}\sep$ & $-\sqrt{\frac23}\sep$ & 0\\ \hline
\end{tabular}\\[15pt]
\centerline{\bf (Table \ref{t:D2.D}f)}
\end{table}
\begin{table}
\begin{tabular}{l||ccccccc|}
& \parbox{22mm}{\vskip5pt$\ 2P_{3/2},\tfrac{5}{2},-\tfrac{1}{2}\ $\vskip4pt} & $\ 2P_{3/2},\tfrac{3}{2},-\tfrac{1}{2}\ $ & $\ 2P_{1/2},\tfrac{3}{2},-\tfrac{1}{2}\ $ & $\ 2S_{1/2},\tfrac{3}{2},-\tfrac{1}{2}\ $ & $\ 2P_{3/2},\tfrac{1}{2},-\tfrac{1}{2}\ $ & $\ 2P_{1/2},\tfrac{1}{2},-\tfrac{1}{2}\ $ & $\ 2S_{1/2},\tfrac{1}{2},-\tfrac{1}{2}\ $\\ \hline\hline
$\ 2P_{3/2},\tfrac{5}{2},-\tfrac{1}{2}\ $ & 0 & 0 & 0 & $3\sqrt{\frac3{5}}\sez$ & 0 & 0 & 0\\
$\ 2P_{3/2},\tfrac{3}{2},-\tfrac{1}{2}\ $ & 0 & 0 & 0 & $\frac{2}{\sqrt{15}}\sez$ & 0 & 0 & $\sqrt{\frac{10}3}\sez$\\ 
$\ 2P_{1/2},\tfrac{3}{2},-\tfrac{1}{2}\ $ & 0 & 0 & 0 & $\frac1{\sqrt{3}}\sez$ & 0 & 0 & $-2\sqrt{\frac23}\sez$\\
$\ 2S_{1/2},\tfrac{3}{2},-\tfrac{1}{2}\ $ & $3\sqrt{\frac3{5}}\sez$ & $\frac2{\sqrt{15}}\sez$ & $\frac1{\sqrt{3}}\sez$ & 0 & $-\frac1{\sqrt3}\sez$ & $-2\sqrt{\frac23}\sez$ & 0\\
$\ 2P_{3/2},\tfrac{1}{2},-\tfrac{1}{2}\ $ & 0 & 0 & 0 & $-\frac1{\sqrt3}\sez$ & 0 & 0 & $2\sqrt{\frac23}\sez$\\
$\ 2P_{1/2},\tfrac{1}{2},-\tfrac{1}{2}\ $ & 0 & 0 & 0 & $-2\sqrt{\frac23}\sez$ & 0 & 0 & $-\frac{1}{\sqrt{3}}\sez$\\
$\ 2S_{1/2},\tfrac{1}{2},-\tfrac{1}{2}\ $ & 0 & $\sqrt{\frac{10}3}\sez$ & $-2\sqrt{\frac23}\sez$ & 0 & $2\sqrt{\frac23}\sez$ & $-\frac1{\sqrt3}\sez$ & 0\\ \hline
\end{tabular}\\[15pt]
\centerline{\bf (Table \ref{t:D2.D}g)}\vskip15pt
\begin{tabular}{l||cccc|}
& \parbox{22mm}{\vskip5pt$\ 2P_{3/2},\tfrac{5}{2},-\tfrac{3}{2}\ $\vskip4pt} & $\ 2P_{3/2},\tfrac{3}{2},-\tfrac{3}{2}\ $ & $\ 2P_{1/2},\tfrac{3}{2},-\tfrac{3}{2}\ $ & $\ 2S_{1/2},\tfrac{3}{2},-\tfrac{3}{2}\ $\\ \hline\hline
$\ 2P_{3/2},\tfrac{5}{2},-\tfrac{1}{2}\ $ & 0 & 0 & 0 & $-\frac{3}{\sqrt{10}}\sem$\\
$\ 2P_{3/2},\tfrac{3}{2},-\tfrac{1}{2}\ $ & 0 & 0 & 0 & $-\sqrt{\frac85}\sem$\\ 
$\ 2P_{1/2},\tfrac{3}{2},-\tfrac{1}{2}\ $ & 0 & 0 & 0 & $-\sqrt2\sem$\\
$\ 2S_{1/2},\tfrac{3}{2},-\tfrac{1}{2}\ $ & $3\sqrt{\frac35}\sem$ & $-\sqrt{\frac85}\sem$ & $-\sqrt2\sem$ & 0\\
$\ 2P_{3/2},\tfrac{1}{2},-\tfrac{1}{2}\ $ & 0 & 0 & 0 & $-\frac1{\sqrt2}\sem$\\
$\ 2P_{1/2},\tfrac{1}{2},-\tfrac{1}{2}\ $ & 0 & 0 & 0 & $-2\sem$\\
$\ 2S_{1/2},\tfrac{1}{2},-\tfrac{1}{2}\ $ & 0 & $\sqrt5\sem$ & $-2\sem$ & 0\\ \hline
\end{tabular}\\[15pt]
\centerline{\bf (Table \ref{t:D2.D}h)}
\end{table}
\begin{table}
\begin{tabular}{l||ccccccc|}
& \parbox{22mm}{\vskip5pt$\ 2P_{3/2},\tfrac{5}{2},-\tfrac{1}{2}\ $\vskip4pt} & $\ 2P_{3/2},\tfrac{3}{2},-\tfrac{1}{2}\ $ & $\ 2P_{1/2},\tfrac{3}{2},-\tfrac{1}{2}\ $ & $\ 2S_{1/2},\tfrac{3}{2},-\tfrac{1}{2}\ $ & $\ 2P_{3/2},\tfrac{1}{2},-\tfrac{1}{2}\ $ & $\ 2P_{1/2},\tfrac{1}{2},-\tfrac{1}{2}\ $ & $\ 2S_{1/2},\tfrac{1}{2},-\tfrac{1}{2}\ $\\ \hline\hline
$\ 2P_{3/2},\tfrac{5}{2},-\tfrac{3}{2}\ $ & 0 & 0 & 0 & $-3 \sqrt{\frac{3}{5}}\sep$ & 0 & 0 & 0\\
$\ 2P_{3/2},\tfrac{3}{2},-\tfrac{3}{2}\ $ & 0 & 0 & 0 & $\sqrt{\frac{8}{5}}\sep$ & 0 & 0 & $-\sqrt{5}\sep$\\  
$\ 2P_{1/2},\tfrac{3}{2},-\tfrac{3}{2}\ $ & 0 & 0 & 0 & $\sqrt2\sep$ & 0 & 0 & $2\sep$\\  
$\ 2S_{1/2},\tfrac{3}{2},-\tfrac{3}{2}\ $ & $\frac{3}{\sqrt{10}}\sep$ & $\sqrt{\frac{8}{5}}\sep$ & $\sqrt2\sep$ & 0 & $\frac{1}{\sqrt2}\sep$ & $2\sep$ & 0\\ \hline
\end{tabular}\\[15pt]
\centerline{\bf (Table \ref{t:D2.D}i)}\vskip15pt
\begin{tabular}{l||cccc|c|}
& $\ 2P_{3/2},\tfrac{5}{2},-\tfrac{3}{2}\ $ & $\ 2P_{3/2},\tfrac{3}{2},-\tfrac{3}{2}\ $ & $\ 2P_{1/2},\tfrac{3}{2},-\tfrac{3}{2}\ $ & $\ 2S_{1/2},\tfrac{3}{2},-\tfrac{3}{2}\ $ & \parbox{20mm}{\vskip5pt$\ 2P_{3/2},\tfrac{5}{2},-\tfrac{5}{2}\ $\vskip4pt}\\ \hline\hline
$\ 2P_{3/2},\tfrac{5}{2},-\tfrac{3}{2}\ $ & 0 & 0 & 0 & $3\,{\sqrt{\frac{2}{5}}}\sez$ & 0\\
$\ 2P_{3/2},\tfrac{3}{2},-\tfrac{3}{2}\ $ & 0 & 0 & 0 & $2\,{\sqrt{\frac{3}{5}}}\sez$ & 0\\  
$\ 2P_{1/2},\tfrac{3}{2},-\tfrac{3}{2}\ $ & 0 & 0 & 0 & ${\sqrt{3}}\sez$ & 0\\  
$\ 2S_{1/2},\tfrac{3}{2},-\tfrac{3}{2}\ $ & $3\,{\sqrt{\frac{2}{5}}}\sez$ & $2\,{\sqrt{\frac{3}{5}}}\sez$ & ${\sqrt{3}}\sez$ & 0 & $3\sep$\\ \hline
$\ 2P_{3/2},\tfrac{5}{2},-\tfrac{5}{2}\ $ & 0 & 0 & 0 & $-3\sem$ & 0\\ \hline
\end{tabular}\\[15pt]
\centerline{\bf (Table \ref{t:D2.D}j)}
\end{table}}
\clearpage

{
\begin{table}
\caption{The suitably normalised magnetic dipole operator $\uvec{\mu}/\mu_B$ for the $n=2$ states of deuterium, where $\mu_B = e\hbar/(2m_e)$ is the Bohr magneton and $g=2.002319304(76)$ is the Land\'{e} factor of the electron \cite{Moh05}.}\label{t:D2.Mu}
\vskip10pt
\begin{tabular}{l||c|cccc|}
& \parbox{20mm}{\vskip5pt$\ 2P_{3/2},\tfrac{5}{2},\tfrac{5}{2}\ $\vskip4pt} & $\ 2P_{3/2},\tfrac{5}{2},\tfrac{3}{2}\ $ & $\ 2P_{3/2},\tfrac{3}{2},\tfrac{3}{2}\ $ & $\ 2P_{1/2},\tfrac{3}{2},\tfrac{3}{2}\ $ & $\ 2S_{1/2},\tfrac{3}{2},\tfrac{3}{2}\ $\\ \hline\hline
$\ 2P_{3/2},\tfrac{5}{2},\tfrac{5}{2}\ $ & $-\frac{2+g}2\sez$ & $-\frac{g+2}{\sqrt{10}}\sem$ & $\frac{g+2}{\sqrt{15}}\sem$ & $-\frac{g-1}{\sqrt3}\sem$ & 0 \\ \hline
$\ 2P_{3/2},\tfrac{5}{2},\tfrac{3}{2}\ $ & $\frac{g+2}{\sqrt{10}}\sep$ & $-\frac{3(g+2)}{10}\sez$ & $-\frac{\sqrt2(g+2)}{5\sqrt3}\sez$ & $\frac{\sqrt2(g-1)}{\sqrt{15}}\sez$ & 0\\
$\ 2P_{3/2},\tfrac{3}{2},\tfrac{3}{2}\ $ & $-\frac{g+2}{\sqrt{15}}\sep$ & $-\frac{\sqrt2(g+2)}{5\sqrt3}\sez$ & $-\frac{11(g+2)}{30}\sez$ & $-\frac{2(g-1)}{3\sqrt5}\sez$ & 0\\  
$\ 2P_{1/2},\tfrac{3}{2},\tfrac{3}{2}\ $ & $\frac{g-1}{\sqrt3}\sep$ & $\frac{\sqrt2(g-1)}{\sqrt{15}}\sez$ & $-\frac{2(g-1)}{3\sqrt5}\sez$ & $\frac{g-4}6\sez$ & 0\\  
$\ 2S_{1/2},\tfrac{3}{2},\tfrac{3}{2}\ $ & 0 & 0 & 0 & 0 & $-\frac g2\sez$\\ \hline
\end{tabular}\\[15pt]
\centerline{\bf (Table \ref{t:D2.Mu}a)}\vskip15pt
\begin{tabular}{l||ccccccc|}
& \parbox{20mm}{\vskip5pt$\ 2P_{3/2},\tfrac{5}{2},\tfrac{1}{2}\ $\vskip4pt} & $\ 2P_{3/2},\tfrac{3}{2},\tfrac{1}{2}\ $ & $\ 2P_{1/2},\tfrac{3}{2},\tfrac{1}{2}\ $ & $\ 2S_{1/2},\tfrac{3}{2},\tfrac{1}{2}\ $ & $\ 2P_{3/2},\tfrac{1}{2},\tfrac{1}{2}\ $ & $\ 2P_{1/2},\tfrac{1}{2},\tfrac{1}{2}\ $ & $\ 2S_{1/2},\tfrac{1}{2},\tfrac{1}{2}\ $\\ \hline\hline
$\ 2P_{3/2},\tfrac{5}{2},\tfrac{3}{2}\ $ & $-\frac{2(g+2)}{5}\sem$ & $\frac{g+2}{5}\sem$ & $-\frac{g-1}{\sqrt{5}}\sem$ & 0 & 0 & 0 & 0\\
$\ 2P_{3/2},\tfrac{3}{2},\tfrac{3}{2}\ $ & $-\frac{\sqrt2(g+2)}{10 \sqrt{3}}\sem$ & $-\frac{11\sqrt2(g+2)}{30 \sqrt{3}}\sem$ & $-\frac{\sqrt8(g-1)}{3 \sqrt{15}}\sem$ & 0 & $\frac{\sqrt{10}(g+2)}{6\sqrt3}\sem$ & $-\frac{\sqrt5(g-1)}{3\sqrt3}\sem$ & 0\\
$\ 2P_{1/2},\tfrac{3}{2},\tfrac{3}{2}\ $ & $\frac{g-1}{\sqrt{30}}\sem$ & $-\frac{\sqrt8(g-1)}{3 \sqrt{15}}\sem$ & $\frac{\sqrt2(g-4)}{6 \sqrt{3}}\sem$ & 0 & $\frac{\sqrt2(g-1)}{6 \sqrt{3}}\sem$ & $-\frac{g-4}{3 \sqrt{3}}\sem$ & 0\\
$\ 2S_{1/2},\tfrac{3}{2},\tfrac{3}{2}\ $ & 0 & 0 & 0 & $-\frac{g}{\sqrt{6}}\sem$ & 0 & 0 & $\frac{g}{\sqrt{3}}\sem$\\ \hline
\end{tabular}\\[15pt]
\centerline{\bf (Table \ref{t:D2.Mu}b)}
\end{table}
\begin{table}
\begin{tabular}{l||cccc|}
& \parbox{20mm}{\vskip5pt$\ 2P_{3/2},\tfrac{5}{2},\tfrac{3}{2}\ $\vskip4pt} & $\ 2P_{3/2},\tfrac{3}{2},\tfrac{3}{2}\ $ & $\ 2P_{1/2},\tfrac{3}{2},\tfrac{3}{2}\ $ & $\ 2S_{1/2},\tfrac{3}{2},\tfrac{3}{2}\ $\\ \hline\hline
$\ 2P_{3/2},\tfrac{5}{2},\tfrac{1}{2}\ $ & $\frac{2(g+2)}{5}\sep$ & $\frac{\sqrt2(g+2)}{10 \sqrt{3}}\sep$ & $-\frac{g-1}{\sqrt{30}}\sep$ & 0 \\
$\ 2P_{3/2},\tfrac{3}{2},\tfrac{1}{2}\ $ & $-\frac{g+2}{5}\sep$ & $\frac{11\sqrt2(g+2)}{30 \sqrt{3}}\sep$ & $\frac{\sqrt8(g-1)}{3 \sqrt{15}}\sep$ & 0 \\
$\ 2P_{1/2},\tfrac{3}{2},\tfrac{1}{2}\ $ & $\frac{g-1}{\sqrt{5}}\sep$ & $\frac{\sqrt8(g-1)}{3 \sqrt{15}}\sep$ & $-\frac{\sqrt2(g-4)}{6 \sqrt{3}}\sep$ & 0 \\
$\ 2S_{1/2},\tfrac{3}{2},\tfrac{1}{2}\ $ & 0 & 0 & 0 & $\frac{g}{\sqrt{6}}\sep$ \\
$\ 2P_{3/2},\tfrac{1}{2},\tfrac{1}{2}\ $ & 0 & $-\frac{\sqrt{10}(g+2)}{6\sqrt3}\sep$ & $-\frac{\sqrt2(g-1)}{6 \sqrt{3}}\sep$ & 0 \\
$\ 2P_{1/2},\tfrac{1}{2},\tfrac{1}{2}\ $ & 0 & $\frac{\sqrt5(g-1)}{3\sqrt3}\sep$ & $\frac{g-4}{3 \sqrt{3}}\sep$ & 0 \\
$\ 2S_{1/2},\tfrac{1}{2},\tfrac{1}{2}\ $ & 0 & 0 & 0 & $-\frac{g}{\sqrt{3}}\sep$ \\ \hline
\end{tabular}\\[15pt]
\centerline{\bf (Table \ref{t:D2.Mu}c)}\vskip15pt
\begin{tabular}{l||ccccccc|}
& \parbox{20mm}{\vskip5pt$\ 2P_{3/2},\tfrac{5}{2},\tfrac{1}{2}\ $\vskip4pt} & $\ 2P_{3/2},\tfrac{3}{2},\tfrac{1}{2}\ $ & $\ 2P_{1/2},\tfrac{3}{2},\tfrac{1}{2}\ $ & $\ 2S_{1/2},\tfrac{3}{2},\tfrac{1}{2}\ $ & $\ 2P_{3/2},\tfrac{1}{2},\tfrac{1}{2}\ $ & $\ 2P_{1/2},\tfrac{1}{2},\tfrac{1}{2}\ $ & $\ 2S_{1/2},\tfrac{1}{2},\tfrac{1}{2}\ $\\ \hline\hline
$\ 2P_{3/2},\tfrac{5}{2},\tfrac{1}{2}\ $ & $-\frac{g+2}{10}\sez$ & $-\frac{g+2}{5}\sez$ & $\frac{g-1}{{\sqrt{5}}}\sez$ & 0 & 0 & 0 & 0\\ 
$\ 2P_{3/2},\tfrac{3}{2},\tfrac{1}{2}\ $ & $-\frac{g+2}{5}\sez$ & $-\frac{11(g+2)}{90}\sez$ & $-\frac{2(g-1)}{9{\sqrt{5}}}\sez$ & 0 & $-\frac{{\sqrt{5}}(g+2)}{9}\sez$ & $\frac{{\sqrt{10}}(g-1)}{9}\sez$ & 0\\ 
$\ 2P_{1/2},\tfrac{3}{2},\tfrac{1}{2}\ $ & $\frac{g-1}{{\sqrt{5}}}\sez$ & $-\frac{2(g-1)}{9{\sqrt{5}}}\sez$ & $\frac{g-4}{18}\sez$ & 0 & $-\frac{g-1}{9}\sez$ & $\frac{{\sqrt{2}}(g-4)}{9}\sez$ & 0\\ 
$\ 2S_{1/2},\tfrac{3}{2},\tfrac{1}{2}\ $ & 0 & 0 & 0 & $-\frac{g}{6}\sez$ & 0 & 0 & $-\frac{{\sqrt{2}}g}{3}\sez$ \\ 
$\ 2P_{3/2},\tfrac{1}{2},\tfrac{1}{2}\ $ & 0 & $-\frac{{\sqrt{5}}(g+2)}{9}\sez$ & $-\frac{g-1}{9}\sez$ & 0 & $-\frac{5(g+2)}{18}\sez$ & $-\frac{2{\sqrt{2}}(g-1)}{9}\sez$ & 0\\ 
$\ 2P_{1/2},\tfrac{1}{2},\tfrac{1}{2}\ $ & 0 & $\frac{{\sqrt{10}}(g-1)}{9}\sez$ & $\frac{{\sqrt{2}}(g-4)}{9}\sez$ & 0 & $-\frac{2{\sqrt{2}}(g-1)}{9}\sez$ & $-\frac{g-4}{18}\sez$ & 0\\ 
$\ 2S_{1/2},\tfrac{1}{2},\tfrac{1}{2}\ $ & 0 & 0 & 0 & $-\frac{{\sqrt{2}}g}{3}\sez$ & 0 & 0 & $\frac{g}{6}\sez$ \\ \hline
\end{tabular}\\[15pt]
\centerline{\bf (Table \ref{t:D2.Mu}d)}
\end{table}
\begin{table}
\begin{tabular}{l||ccccccc|}
& \parbox{22mm}{\vskip5pt$\ 2P_{3/2},\tfrac{5}{2},-\tfrac{1}{2}\ $\vskip4pt} & $\ 2P_{3/2},\tfrac{3}{2},-\tfrac{1}{2}\ $ & $\ 2P_{1/2},\tfrac{3}{2},-\tfrac{1}{2}\ $ & $\ 2S_{1/2},\tfrac{3}{2},-\tfrac{1}{2}\ $ & $\ 2P_{3/2},\tfrac{1}{2},-\tfrac{1}{2}\ $ & $\ 2P_{1/2},\tfrac{1}{2},-\tfrac{1}{2}\ $ & $\ 2S_{1/2},\tfrac{1}{2},-\tfrac{1}{2}\ $\\ \hline\hline
$\ 2P_{3/2},\tfrac{5}{2},\tfrac{1}{2}\ $ & $-\frac{3\sqrt2(g+2)}{10}\sem$ & $\frac{\sqrt2(g+2)}{10}\sem$ & $-\frac{g-1}{{\sqrt{10}}}\sem$ & 0 & 0 & 0 & 0\\ 
$\ 2P_{3/2},\tfrac{3}{2},\tfrac{1}{2}\ $ & $-\frac{\sqrt2(g+2)}{10}\sem$ & $-\frac{11\sqrt2(g+2)}{45}\sem$ & $-\frac{4\sqrt2(g-1)}{9{\sqrt{5}}}\sem$ & 0 & $\frac{{\sqrt{10}}(g+2)}{18}\sem$ & $-\frac{{\sqrt{5}}(g-1)}{9}\sem$ & 0\\ 
$\ 2P_{1/2},\tfrac{3}{2},\tfrac{1}{2}\ $ & $\frac{g-1}{{\sqrt{10}}}\sem$ & $-\frac{4\sqrt2(g-1)}{9{\sqrt{5}}}\sem$ & $\frac{\sqrt2(g-4)}{9}\sem$ & 0 & $\frac{\sqrt2(g-1)}{18}\sem$ & $-\frac{(g-4)}{9}\sem$ & 0\\ 
$\ 2S_{1/2},\tfrac{3}{2},\tfrac{1}{2}\ $ & 0 & 0 & 0 & $-\frac{\sqrt2\,g}{3}\sem$ & 0 & 0 & $\frac{g}{3}\sem$ \\ 
$\ 2P_{3/2},\tfrac{1}{2},\tfrac{1}{2}\ $ & 0 & $-\frac{{\sqrt{10}}(g+2)}{18}\sem$ & $-\frac{\sqrt2(g-1)}{18}\sem$ & 0 & $-\frac{5\sqrt2(g+2)}{18}\sem$ & $-\frac{4(g-1)}{9}\sem$ & 0\\ 
$\ 2P_{1/2},\tfrac{1}{2},\tfrac{1}{2}\ $ & 0 & $\frac{{\sqrt{5}}(g-1)}{9}\sem$ & $\frac{(g-4)}{9}\sem$ & 0 & $-\frac{4(g-1)}{9}\sem$ & $-\frac{\sqrt2(g-4)}{18}\sem$ & 0\\ 
$\ 2S_{1/2},\tfrac{1}{2},\tfrac{1}{2}\ $ & 0 & 0 & 0 & $-\frac{g}{3}\sem$ & 0 & 0 & $\frac{\sqrt2\,g}{6}\sem$ \\ \hline
\end{tabular}\\[15pt]
\centerline{\bf (Table \ref{t:D2.Mu}e)}\vskip15pt
\begin{tabular}{l||ccccccc|}
& \parbox{20mm}{\vskip5pt$\ 2P_{3/2},\tfrac{5}{2},\tfrac{1}{2}\ $\vskip4pt} & $\ 2P_{3/2},\tfrac{3}{2},\tfrac{1}{2}\ $ & $\ 2P_{1/2},\tfrac{3}{2},\tfrac{1}{2}\ $ & $\ 2S_{1/2},\tfrac{3}{2},\tfrac{1}{2}\ $ & $\ 2P_{3/2},\tfrac{1}{2},\tfrac{1}{2}\ $ & $\ 2P_{1/2},\tfrac{1}{2},\tfrac{1}{2}\ $ & $\ 2S_{1/2},\tfrac{1}{2},\tfrac{1}{2}\ $\\ \hline\hline
$\ 2P_{3/2},\tfrac{5}{2},-\tfrac{1}{2}\ $ & $\frac{3\sqrt2(g+2)}{10}\sep$ & $\frac{\sqrt2(g+2)}{10}\sep$ & $-\frac{g-1}{{\sqrt{10}}}\sep$ & 0 & 0 & 0 & 0\\ 
$\ 2P_{3/2},\tfrac{3}{2},-\tfrac{1}{2}\ $ & $-\frac{\sqrt2(g+2)}{10}\sep$ & $\frac{11\sqrt2(g+2)}{45}\sep$ & $\frac{4\sqrt2(g-1)}{9{\sqrt{5}}}\sep$ & 0 & $\frac{{\sqrt{10}}(g+2)}{18}\sep$ & $-\frac{{\sqrt{5}}(g-1)}{9}\sep$ & 0\\ 
$\ 2P_{1/2},\tfrac{3}{2},-\tfrac{1}{2}\ $ & $\frac{g-1}{{\sqrt{10}}}\sep$ & $\frac{4\sqrt2(g-1)}{9{\sqrt{5}}}\sep$ & $-\frac{\sqrt2(g-4)}{9}\sep$ & 0 & $\frac{\sqrt2(g-1)}{18}\sep$ & $-\frac{(g-4)}{9}\sep$ & 0\\ 
$\ 2S_{1/2},\tfrac{3}{2},-\tfrac{1}{2}\ $ & 0 & 0 & 0 & $\frac{\sqrt2\,g}{3}\sep$ & 0 & 0 & $\frac{g}{3}\sep$ \\ 
$\ 2P_{3/2},\tfrac{1}{2},-\tfrac{1}{2}\ $ & 0 & $-\frac{{\sqrt{10}}(g+2)}{18}\sep$ & $-\frac{\sqrt2(g-1)}{18}\sep$ & 0 & $\frac{5\sqrt2(g+2)}{18}\sep$ & $\frac{4(g-1)}{9}\sep$ & 0\\ 
$\ 2P_{1/2},\tfrac{1}{2},-\tfrac{1}{2}\ $ & 0 & $\frac{{\sqrt{5}}(g-1)}{9}\sep$ & $\frac{(g-4)}{9}\sep$ & 0 & $\frac{4(g-1)}{9}\sep$ & $\frac{\sqrt2(g-4)}{18}\sep$ & 0\\ 
$\ 2S_{1/2},\tfrac{1}{2},-\tfrac{1}{2}\ $ & 0 & 0 & 0 & $-\frac{g}{3}\sep$ & 0 & 0 & $-\frac{\sqrt2\,g}{6}\sep$ \\ \hline
\end{tabular}\\[15pt]
\centerline{\bf (Table \ref{t:D2.Mu}f)}
\end{table}
\begin{table}
\begin{tabular}{l||ccccccc|}
& \parbox{22mm}{\vskip5pt$\ 2P_{3/2},\tfrac{5}{2},-\tfrac{1}{2}\ $\vskip4pt} & $\ 2P_{3/2},\tfrac{3}{2},-\tfrac{1}{2}\ $ & $\ 2P_{1/2},\tfrac{3}{2},-\tfrac{1}{2}\ $ & $\ 2S_{1/2},\tfrac{3}{2},-\tfrac{1}{2}\ $ & $\ 2P_{3/2},\tfrac{1}{2},-\tfrac{1}{2}\ $ & $\ 2P_{1/2},\tfrac{1}{2},-\tfrac{1}{2}\ $ & $\ 2S_{1/2},\tfrac{1}{2},-\tfrac{1}{2}\ $\\ \hline\hline
$\ 2P_{3/2},\tfrac{5}{2},-\tfrac{1}{2}\ $ & $\frac{g+2}{10}\sez$ & $-\frac{g+2}{5}\sez$ & $\frac{g-1}{{\sqrt{5}}}\sez$ & 0 & 0 & 0 & 0\\ 
$\ 2P_{3/2},\tfrac{3}{2},-\tfrac{1}{2}\ $ & $-\frac{g+2}{5}\sez$ & $\frac{11(g+2)}{90}\sez$ & $\frac{2(g-1)}{9{\sqrt{5}}}\sez$ & 0 & $-\frac{{\sqrt{5}}(g+2)}{9}\sez$ & $\frac{{\sqrt{10}}(g-1)}{9}\sez$ & 0\\ 
$\ 2P_{1/2},\tfrac{3}{2},-\tfrac{1}{2}\ $ & $\frac{g-1}{{\sqrt{5}}}\sez$ & $\frac{2(g-1)}{9{\sqrt{5}}}\sez$ & $-\frac{g-4}{18}\sez$ & 0 & $-\frac{g-1}{9}\sez$ & $\frac{{\sqrt{2}}(g-4)}{9}\sez$ & 0\\ 
$\ 2S_{1/2},\tfrac{3}{2},-\tfrac{1}{2}\ $ & 0 & 0 & 0 & $\frac{g}{6}\sez$ & 0 & 0 & $-\frac{{\sqrt{2}}g}{3}\sez$ \\ 
$\ 2P_{3/2},\tfrac{1}{2},-\tfrac{1}{2}\ $ & 0 & $-\frac{{\sqrt{5}}(g+2)}{9}\sez$ & $-\frac{g-1}{9}\sez$ & 0 & $\frac{5(g+2)}{18}\sez$ & $\frac{2{\sqrt{2}}(g-1)}{9}\sez$ & 0\\ 
$\ 2P_{1/2},\tfrac{1}{2},-\tfrac{1}{2}\ $ & 0 & $\frac{{\sqrt{10}}(g-1)}{9}\sez$ & $\frac{{\sqrt{2}}(g-4)}{9}\sez$ & 0 & $\frac{2{\sqrt{2}}(g-1)}{9}\sez$ & $\frac{g-4}{18}\sez$ & 0\\ 
$\ 2S_{1/2},\tfrac{1}{2},-\tfrac{1}{2}\ $ & 0 & 0 & 0 & $-\frac{{\sqrt{2}}g}{3}\sez$ & 0 & 0 & $-\frac{g}{6}\sez$ \\ \hline
\end{tabular}\\[15pt]
\centerline{\bf (Table \ref{t:D2.Mu}g)}\vskip15pt
\begin{tabular}{l||cccc|}
& \parbox{22mm}{\vskip5pt$\ 2P_{3/2},\tfrac{5}{2},-\tfrac{3}{2}\ $\vskip4pt} & $\ 2P_{3/2},\tfrac{3}{2},-\tfrac{3}{2}\ $ & $\ 2P_{1/2},\tfrac{3}{2},-\tfrac{3}{2}\ $ & $\ 2S_{1/2},\tfrac{3}{2},-\tfrac{3}{2}\ $\\ \hline\hline
$\ 2P_{3/2},\tfrac{5}{2},-\tfrac{1}{2}\ $ & $-\frac{2(g+2)}{5}\sem$ & $\frac{\sqrt2(g+2)}{10 \sqrt{3}}\sem$ & $-\frac{g-1}{\sqrt{30}}\sem$ & 0 \\
$\ 2P_{3/2},\tfrac{3}{2},-\tfrac{1}{2}\ $ & $-\frac{g+2}{5}\sem$ & $-\frac{11\sqrt2(g+2)}{30 \sqrt{3}}\sem$ & $-\frac{\sqrt8(g-1)}{3 \sqrt{15}}\sem$ & 0 \\
$\ 2P_{1/2},\tfrac{3}{2},-\tfrac{1}{2}\ $ & $\frac{g-1}{\sqrt{5}}\sem$ & $-\frac{\sqrt8(g-1)}{3 \sqrt{15}}\sem$ & $\frac{\sqrt2(g-4)}{6 \sqrt{3}}\sem$ & 0 \\
$\ 2S_{1/2},\tfrac{3}{2},-\tfrac{1}{2}\ $ & 0 & 0 & 0 & $-\frac{g}{\sqrt{6}}\sem$ \\
$\ 2P_{3/2},\tfrac{1}{2},-\tfrac{1}{2}\ $ & 0 & $-\frac{\sqrt{10}(g+2)}{6\sqrt3}\sem$ & $-\frac{\sqrt2(g-1)}{6 \sqrt{3}}\sem$ & 0 \\
$\ 2P_{1/2},\tfrac{1}{2},-\tfrac{1}{2}\ $ & 0 & $\frac{\sqrt5(g-1)}{3\sqrt3}\sem$ & $\frac{g-4}{3 \sqrt{3}}\sem$ & 0 \\
$\ 2S_{1/2},\tfrac{1}{2},-\tfrac{1}{2}\ $ & 0 & 0 & 0 & $-\frac{g}{\sqrt{3}}\sem$ \\ \hline
\end{tabular}\\[15pt]
\centerline{\bf (Table \ref{t:D2.Mu}h)}
\end{table}
\begin{table}
\begin{tabular}{l||ccccccc|}
& \parbox{22mm}{\vskip5pt$\ 2P_{3/2},\tfrac{5}{2},-\tfrac{1}{2}\ $\vskip4pt} & $\ 2P_{3/2},\tfrac{3}{2},-\tfrac{1}{2}\ $ & $\ 2P_{1/2},\tfrac{3}{2},-\tfrac{1}{2}\ $ & $\ 2S_{1/2},\tfrac{3}{2},-\tfrac{1}{2}\ $ & $\ 2P_{3/2},\tfrac{1}{2},-\tfrac{1}{2}\ $ & $\ 2P_{1/2},\tfrac{1}{2},-\tfrac{1}{2}\ $ & $\ 2S_{1/2},\tfrac{1}{2},-\tfrac{1}{2}\ $\\ \hline\hline
$\ 2P_{3/2},\tfrac{5}{2},-\tfrac{3}{2}\ $ & $\frac{2(g+2)}{5}\sep$ & $\frac{g+2}{5}\sep$ & $-\frac{g-1}{\sqrt{5}}\sep$ & 0 & 0 & 0 & 0\\
$\ 2P_{3/2},\tfrac{3}{2},-\tfrac{3}{2}\ $ & $-\frac{\sqrt2(g+2)}{10 \sqrt{3}}\sep$ & $\frac{11\sqrt2(g+2)}{30 \sqrt{3}}\sep$ & $\frac{\sqrt8(g-1)}{3 \sqrt{15}}\sep$ & 0 & $\frac{\sqrt{10}(g+2)}{6\sqrt3}\sep$ & $-\frac{\sqrt5(g-1)}{3\sqrt3}\sep$ & 0\\
$\ 2P_{1/2},\tfrac{3}{2},-\tfrac{3}{2}\ $ & $\frac{g-1}{\sqrt{30}}\sep$ & $\frac{\sqrt8(g-1)}{3 \sqrt{15}}\sep$ & $-\frac{\sqrt2(g-4)}{6 \sqrt{3}}\sep$ & 0 & $\frac{\sqrt2(g-1)}{6 \sqrt{3}}\sep$ & $-\frac{g-4}{3 \sqrt{3}}\sep$ & 0\\
$\ 2S_{1/2},\tfrac{3}{2},-\tfrac{3}{2}\ $ & 0 & 0 & 0 & $\frac{g}{\sqrt{6}}\sep$ & 0 & 0 & $\frac{g}{\sqrt{3}}\sep$\\ \hline
\end{tabular}\\[15pt]
\centerline{\bf (Table \ref{t:D2.Mu}i)}\vskip15pt
\begin{tabular}{l||cccc|c|}
& $\ 2P_{3/2},\tfrac{5}{2},-\tfrac{3}{2}\ $ & $\ 2P_{3/2},\tfrac{3}{2},-\tfrac{3}{2}\ $ & $\ 2P_{1/2},\tfrac{3}{2},-\tfrac{3}{2}\ $ & $\ 2S_{1/2},\tfrac{3}{2},-\tfrac{3}{2}\ $ & \parbox{22mm}{\vskip5pt$\ 2P_{3/2},\tfrac{5}{2},-\tfrac{5}{2}\ $\vskip4pt}\\ \hline\hline
$\ 2P_{3/2},\tfrac{5}{2},-\tfrac{3}{2}\ $ & $\frac{3(g+2)}{10}\sez$ & $-\frac{\sqrt2(g+2)}{5\sqrt3}\sez$ & $\frac{\sqrt2(g-1)}{\sqrt{15}}\sez$ & 0 & $-\frac{g+2}{\sqrt{10}}\sem$\\
$\ 2P_{3/2},\tfrac{3}{2},-\tfrac{3}{2}\ $ & $-\frac{\sqrt2(g+2)}{5\sqrt3}\sez$ & $\frac{11(g+2)}{30}\sez$ & $\frac{2(g-1)}{3\sqrt5}\sez$ & 0 & $-\frac{g+2}{\sqrt{15}}\sem$\\  
$\ 2P_{1/2},\tfrac{3}{2},-\tfrac{3}{2}\ $ & $\frac{\sqrt2(g-1)}{\sqrt{15}}\sez$ & $\frac{2(g-1)}{3\sqrt5}\sez$ & $-\frac{g-4}{6}\sez$ & 0 & $\frac{g-1}{\sqrt3}\sem$\\  
$\ 2S_{1/2},\tfrac{3}{2},-\tfrac{3}{2}\ $ & 0 & 0 & 0 & $\frac g2\sez$ & 0\\ \hline
$\ 2P_{3/2},\tfrac{5}{2},-\tfrac{5}{2}\ $ & $\frac{g+2}{\sqrt{10}}\sep$ & $\frac{g+2}{\sqrt{15}}\sep$ & $-\frac{g-1}{\sqrt3}\sep$ & 0 & $\frac{g+2}2\sez$\\ \hline
\end{tabular}\\[15pt]
\centerline{\bf (Table \ref{t:D2.Mu}j)}
\end{table}}
\end{turnpage}
\clearpage
\newpage
\section{The adiabaticity condition}\label{s:Adiabaticity}

In this appendix we discuss the conditions, which have to be satisfied in order to apply the general results of I to the $n=2$ systems of hydrogen and deuterium.

The condition to have a group of metastable and a group of fast decaying states was already discussed in section \ref{s:AdiabaticLimit}. We found in (\ref{e2:E.bound}) that we have to require $|\vmc E| \lesssim 250\,\mathrm{V/cm}$
for $|\vmc B| = 0$. Turning on the $\vmc B$ field ($|\vmc B\leq 5\,\mathrm{mT}|$) one finds that the lifetimes of the $n=2$ states do not change substantially.

Now we come to the question how slow the variation in time of the $\vmc E$ and $\vmc B$ fields ought to be for the results of I to be applicable. In I we worked, for mathematical convenience, with a reduced time $\tau$ and a total time $T$. We studied the limit $T\to\infty$.

Consider, for example, the two-state system of section \extref{I.s:TwoStates}. The leading terms for $T\to\infty$ are given in (\extref{I.e2:longer.lived.amp}) and (\extref{I.e2:shorter.lived.amp}) and, with an estimate of the first correction terms, in (\extref{I.eA:estPsiLong}) and (\extref{I.eA:estPsiShort}). The first correction terms should be small and thus we should require
\begin{align}\label{B.1}
\frac{2 c_{12}}{T\,\Delta\Gamma_{\mathrm{min}}} &\ll 1\ ,\\ \label{B.2}
\frac{4 c_{12}c_{21}}{T\,\Delta\Gamma_{\mathrm{min}}} &\ll 1\ ,\\ \label{B.3}
\frac{2 c_{21}}{T\,\Delta\Gamma_{\mathrm{min}}} &\ll 1\ ,
\end{align}
where $c_{12}$ and $c_{21}$ are defined in (\extref{I.e2:Gmin.bound}) and (\extref{I.eA:Gmin.bound}), respectively.

Now we revert to time $t$ instead of the reduced time $\tau$. That is, we set $T=\tau_0$ in the following, see (\extref{I.e2:red.time}). It is clear that the typical observation time will correspond to the lifetime of the $2S$ states, that is, we can set
\begin{align}
T\approx \tau_S\ .
\end{align}
From the definitions (\extref{I.e2:Def.a}) and (\extref{I.e2:Gmin.bound}) we then find
\begin{align}\label{B.4}
c_{12} &\cong T\max_{0\leq t\leq T} |a_{12}(t)|\ ,\\
|a_{12}(t)| &= |\lrbra{1,t}\I\frac{\partial}{\partial t}\rket{2,t}|\ .
\end{align}
Now we insert for the state $\rket{1,t}$ the $2S_{1/2}$, for the state $\rket{2,t}$ the $2P_{1/2}$ states. We can then estimate the allowed speed of variation of the electric field using perturbation theory with perturbing term $\vec D\cdot\delta\vmc E$ in the mass matrix. We get in this way for constant magnetic field
\begin{align}
\begin{split}
\rket{2P_{1/2},\vmc E(t+\delta t),\vmc B} &\cong \rket{2P_{1/2},\vmc E(t),\vmc B}
+ \frac{1}{E(2P_{1/2})-E(2S_{1/2})}\rket{2S_{1/2},\vmc E(t),\vmc B}\\
&\quad\times \lrbra{2S_{1/2},\vmc E(t),\vmc B}\klr{-\vec D}\cdot\klr{\vmc E(t+\delta t) - \vmc E(t)}\rket{2P_{1/2},\vmc E(t),\vmc B}\ ,
\end{split}\\
\frac{\partial}{\partial t}\rket{2P_{1/2},\vmc E(t),\vmc B} &\cong
\frac1L\rket{2S_{1/2},\vmc E(t),\vmc B}
\lrbra{2S_{1/2},\vmc E(t),\vmc B}\vec D\cdot\frac{\partial\vmc E(t)}{\partial t}\rket{2P_{1/2},\vmc E(t),\vmc B}\ ,\\ \label{B.6}
\begin{split}
|a_{12}(t)| &\cong \left|\frac1L\lrbra{2S_{1/2},\vmc E(t),\vmc B}\vec D\cdot\frac{\partial\vmc E(t)}{\partial t}\rket{2P_{1/2},\vmc E(t),\vmc B}\right|\\
&\cong \frac{e\,r_B(1)}{L}\left|\frac{\partial\vmc E(t)}{\partial t}\right|\ .
\end{split}
\end{align}
For hydrogen and deuterium we have (see (3.16) and (3.19) of \cite{BoBrNa95})
\begin{align}\label{B.7}
\frac{e\,r_B(1)}L~\widehat{=}~\frac1{\sqrt3\,\mathcal E_0}\ ,\qquad \mathcal E_0 =~477.3\,\mathrm{V/cm}\ .
\end{align}
Thus we get
\begin{align}
|a_{12}(t)| \cong \frac{1}{\sqrt3\,\mathcal E_0}\left|\frac{\partial\vmc E(t)}{\partial t}\right|\ .
\end{align}
For the decay rates we have
\begin{align}
\Delta\Gamma_{\mathrm{min}} \cong \Gamma_P-\Gamma_S \cong \Gamma_P\ .
\end{align}
Inserting all this in (\ref{B.4}) and (\ref{B.1}) leads to the requirement
\begin{align}\label{B.10}
\begin{split}
\frac{2c_{12}}{T\,\Delta\Gamma_{\mathrm{min}}} &\cong \frac{1}{\Gamma_P}\max_{t\in[0,T]}\frac{1}{\mathcal E_0}\left|\frac{\partial \vmc E(t)}{\partial t}\right|\\
&= \tau_P\max_{t\in[0,T]}\frac{1}{\mcal E_0}\left|\frac{\partial\vmc E(t)}{\partial t}\right|\\ 
&\ll 1\ .
\end{split}
\end{align}
The same estimate is obtained from (\ref{B.3}). From (\ref{B.2}) we get
\begin{align}\label{B.11}
\begin{split}
\frac{4c_{12}c_{21}}{T\,\Delta\Gamma_{\mathrm{min}}} &= T\,\Delta\Gamma_{\mathrm{min}}
\klr{\frac{2c_{12}}{T\,\Delta\Gamma_{\mathrm{min}}}}
\klr{\frac{2c_{21}}{T\,\Delta\Gamma_{\mathrm{min}}}}\\
&\cong 
\klr{\tau_P\max_{t\in[0,T]}\frac{1}{\mcal E_0}\left|\frac{\partial\vmc E(t)}{\partial t}\right|}
\klr{T\max_{t\in[0,T]}\frac{1}{\mcal E_0}\left|\frac{\partial\vmc E(t)}{\partial t}\right|}\\
&\ll 1\ .
\end{split}
\end{align}
Both conditions, (\ref{B.10}) and (\ref{B.11}), are satisfied if
\begin{align}\label{B.12}
\max_{t\in[0,T]}\frac{1}{\mcal E_0}\left|\frac{\partial\vmc E(t)}{\partial t}\right| \ll \frac1{\tau_P}
\end{align}
and
\begin{align}\label{B.13}
\max_{t\in[0,T]}\frac{1}{\mcal E_0}\left|\frac{\partial\vmc E(t)}{\partial t}\right| < \frac1{T} \cong \frac1{\tau_S}\ .
\end{align}
Since we always have $\tau_S\gg \tau_P$ we see that (\ref{B.13}) already implies (\ref{B.12}).

Next we want to estimate the allowed rate of change for the magnetic field $\vmc B(t)$. Similarly to (\ref{B.6}) we can estimate here
\begin{align}\label{B.14}
|a_{12}(t)| &\cong \left|\frac1L\lrbra{2S_{1/2},\vmc E(t),\vmc B(t)}\vec\mu\cdot\frac{\partial\vmc B(t)}{\partial t}\rket{2P_{1/2},\vmc E(t),\vmc B(t)}\right|\ .
\end{align}
For the free $2S$ and $2P$ states - disregarding P violation - the matrix element in (\ref{B.14}) vanishes. But we are considering states in an electric field where there is Stark mixing. For electric fields satisfying (\ref{e2:E.bound}), $|\vmc E| \lesssim 250\,\mathrm{V/cm}$, this mixing is at most $\mcal{O}(1)$. We shall thus estimate from (\ref{B.14})
\begin{align}\label{B.15}
|a_{12}(t)| \lesssim \frac{\mu_B}{L} \left|\frac{\partial\vmc B(t)}{\partial t}\right|\ .
\end{align}
In analogy to (\ref{B.7}) we define $\mcal B_0$ by
\begin{align}\label{B.16}
\frac{\mu_B}L~\widehat=~\frac{1}{\sqrt3\,\mcal B_0}\ ,
\end{align}
which gives
\begin{align}\label{B.16a}
\mcal B_0 = 43.65\,\mathrm{mT}\ .
\end{align}
Repeating the same arguments as made above for the electric field we find the conditions
\begin{align}\label{B.17}
\max_{t\in[0,T]}\frac{1}{\mcal B_0}\left|\frac{\partial\vmc B(t)}{\partial t}\right| \ll \frac1{\tau_P}
\end{align}
and
\begin{align}\label{B.18}
\max_{t\in[0,T]}\frac{1}{\mcal B_0}\left|\frac{\partial\vmc B(t)}{\partial t}\right| < \frac1{T} \cong \frac1{\tau_S}\ .
\end{align}
Here again (\ref{B.18}) implies (\ref{B.17}) since we have $\tau_S\gg \tau_P$.

We summarise the adiabaticity conditions for the electric and magnetic fields found as follows.
The evolution of the --- of course mixed --- $2S$ states will decouple from that of the $2P$ states if the electric field satisfies (\ref{e2:E.bound}) which guarantees for the lifetimes $\tau_S\gg \tau_P$ and if the rate of change of the electric and magnetic fields normalised to $\mcal E_0$ and $\mcal B_0$, respectively, is smaller than the inverse lifetime $\tau^{-1}_S$ of the $2S$ states.

\newpage
\section{Geometric phases and flux densities}\label{s:GeometricFluxes}

In this appendix we will give the details of the calculations that lead to the results shown in sections \ref{s:AdiabaticLimit} and \ref{s:Results}. In section \ref{sB:Surface} we consider a closed path in parameter space and transform the integral of the geometric phase into a surface-integral using differential form algebra. In section \ref{sB:Fluxes} we introduce geometric flux densities. In these sections we discuss general properties of geometric phases and, therefore, suppress the PV parameters in the labels of the states etc. In section \ref{sB:Perturbation} we use perturbation theory to identify the PC and PV contributions to the geometric phase and we derive the geometric flux densities that are used in section \ref{s:Results} to visualise the geometric phase contributions.

\subsection{The geometric phase as surface-integral in parameter space}\label{sB:Surface}

Let $\vec R = \big(R_1,\ldots,R_r\big)$ be the parameter vectors and $\mcal C$ a closed curve in an $r$-dimensional parameter space $\mcal R$. Let $\umat M(\vec R)$ be a --- in general non-hermitian --- matrix function over $\mcal R$ with non-degenerate eigenvalues $E(\alpha,\vec R)$ ($\alpha = 1,\ldots,N$) for all $\vec R\in\mcal C$.

The abelian geometric phase (\ref{e3:BP.PS}) in parameter space reads (omitting the labels $\delta_{1,2}$)
\begin{align}\label{eB:BP.PS}
\gamma_{\alpha\alpha}(\mcal C) = \oint_{\mcal C}\mrmd\vec R\cdot\lrbra{\alpha,\vec R}\I\vec\nabla_{\vec R}\rket{\alpha,\vec R}\ ,
\end{align}
and can be written as an integral over a differential 1-form
\begin{align}\label{eB:BP.ED}
\gamma_{\alpha\alpha}(\mcal C) = \I\oint_{\mcal C}\lrbra{\alpha,\vec R}d\rket{\alpha,\vec R}\ ,
\end{align}
where the exterior derivative $d$ is defined as
\begin{align}\label{eB:ED}
d = \sum_{i=1}^{r}dR_i\,\frac{\partial}{\partial R_i}\ .
\end{align}
The exterior product (wedge product) of two 1-forms gives a 2-form. The wedge product $\wedge$ is antisymmetric, that is
\begin{align}\label{eB:wedge}
d R_i\wedge d R_j = -d R_j\wedge d R_i\ .
\end{align}
One can easily prove the following relations
\begin{align}\label{eB:Asym}
d R_i\wedge d R_i &= 0\ ,\\ \label{eB:Poincare}
dd &= 0\ ,\qquad\text{(Poincar\'{e} lemma)}\ .
\end{align}
For an introduction to differential forms see \cite{Fla63}. As was already pointed out by Berry in his original work \cite{Ber84}, for a parameter space of dimension $r>3$ one has to use the generalised Stokes' theorem to transform the integral (\ref{eB:BP.ED}) into a surface-integral. The generalised Stokes' theorem gives
\begin{align}\label{eB:Stokes}
\gamma_{\alpha\alpha}(\mcal C) = \I\oint_{\mcal C = \partial F}\lrbra{\alpha,\vec R}d\rket{\alpha,\vec R}
= \I\int_{\mcal F}d \lrbra{\alpha,\vec R}d\rket{\alpha,\vec R}
\end{align}
where $\mcal F$ is an arbitrary surface in parameter space bounded by the curve $\mcal C$ and lying fully in the regularity domain of $d\lrbra{\alpha,\vec R}d\rket{\alpha,\vec R}$. From the Poincar\'{e} lemma (\ref{eB:Poincare}) and the asymmetry of the wedge product (\ref{eB:wedge}) we obtain for the integrand in (\ref{eB:Stokes})
\begin{align}\label{eB:Integrand}
\begin{split}
d \lrbra{\alpha,\vec R}d\rket{\alpha,\vec R} &= \Big(d \lrbra{\alpha,\vec R}\Big)d\rket{\alpha,\vec R}\\
&= \sum_{\beta}\Big(d \lrbra{\alpha,\vec R}\Big)\rket{\beta,\vec R}\wedge
\lrbra{\beta,\vec R}d\rket{\alpha,\vec R}\\
&= -\sum_{\beta\neq\alpha}\Big(\lrbra{\alpha,\vec R}d\rket{\beta,\vec R}\Big)\wedge
\lrbra{\beta,\vec R}d\rket{\alpha,\vec R}\ .
\end{split}
\end{align}
Here we have used the completeness of the eigenstates of $\umat M(\vec R)$,
\begin{align}\label{eB:completeness}
\um = \sum_{\beta} \rket{\beta,\vec R}\lrbra{\beta,\vec R}\ ,
\end{align}
and the following relations, which are easy to prove,
\begin{align}\label{eB:rule1}
\lrbra{\alpha,\vec R}d\rket{\beta,\vec R} &= - \Big(d\lrbra{\alpha,\vec R}\Big)\rket{\beta,\vec R}\ ,\\ \label{eB:rule2}
0 &= \Big(\lrbra{\alpha,\vec R}d\rket{\alpha,\vec R}\Big)\wedge\lrbra{\alpha,\vec R}d\rket{\alpha,\vec R}\ .
\end{align}

Next we study the matrix elements $\lrbra{\beta,\vec R}d\rket{\alpha,\vec R}$. For each parameter vector $\vec R$ we have
\begin{align}\label{eB:EP}
\umat M(\vec R)\rket{\alpha,\vec R} &= E(\alpha,\vec R)\rket{\alpha,\vec R}\ ,\\ \label{eB:EP.left}
\lrbra{\alpha,\vec R}\umat M(\vec R) &= \lrbra{\alpha,\vec R}E(\alpha,\vec R)\ ,\\ \nonumber
(\alpha &= 1,\ldots,N)\ .
\end{align}
Applying the exterior derivative to (\ref{eB:EP}), multiplying from the left with $\lrbra{\beta,\vec R}$ and using the non-degeneracy of eigenvalues (which we supposed) we obtain for $\beta\neq\alpha$
\begin{align}\label{eB:d.Matrix}
\lrbra{\beta,\vec R}d\rket{\alpha,\vec R} = \frac{(d\umat M(\vec R))_{\beta\alpha}}{E(\alpha,\vec R) - E(\beta,\vec R)}
\end{align}
with the 1-forms
\begin{align}\label{eB:Def.dM}
(d\umat M(\vec R))_{\beta\alpha} \equiv \lrbra{\beta,\vec R}\Big(d\umat M(\vec R)\Big)\rket{\alpha,\vec R}\ .
\end{align}
Using (\ref{eB:BP.ED}), (\ref{eB:Stokes}), (\ref{eB:Integrand}) and (\ref{eB:d.Matrix}) we finally get
\begin{align}\label{eB:BP.Surface}
\gamma_{\alpha\alpha}(\mcal C) = \I\sum_{\substack{\beta=1\\ \beta\neq\alpha}}^N\int_{\mcal F}
\frac{(d\umat M(\vec R))_{\alpha\beta}\wedge (d\umat M(\vec R))_{\beta\alpha}}{\big(E(\alpha,\vec R) - E(\beta,\vec R)\big)^2}\ ,\quad
(\mcal C = \partial F;\ \alpha=1,\ldots,N)\ .
\end{align}
This is the geometric phase in the case of a non-hermitian mass matrix, written as a surface-integral in parameter space. The result (\ref{eB:BP.Surface}) is completely analogous to that in \cite{Ber84} for hermitian Hamiltonians. However, in the case of a non-hermitian mass matrix, the geometric phase will in general be complex and therefore contribute to the exponential decay of the corresponding eigenstate.

\subsection{Geometric Flux Densities}\label{sB:Fluxes}

The mass matrix $\umat M(\vec R)$ for an atom in an external electric ($\vmc E$) and magnetic ($\vmc B$) field has the form (see (\ref{e2:MM.full})-(\ref{e2:MM.free}))
\begin{align}\label{eB:MM}
\umat M(\vec R) = \umat{\tilde M}_0 - \uvec D\cdot\vmc E - \uvec\mu\cdot\vmc B\ ,
\end{align}
where the parameter vector $\vec R$ is
\begin{align}\label{eB:R}
\vec R = \klr{R_1,\ldots,R_6} = \klr{\mcal E_1,\mcal E_2,\mcal E_3,\mcal B_1,\mcal B_2,\mcal B_3}\ .
\end{align}
The exterior derivative of the mass matrix (\ref{eB:MM}) is
\begin{align}
d\umat M(\vec R) = -\uvec D\cdot d\vmc E - \uvec\mu\cdot d\vmc B\ .
\end{align}
Thus, the matrix elements (\ref{eB:Def.dM}) of $d\umat M(\vec R)$ are
\begin{align}\label{eB:dM.EB}
\begin{split}
(d\umat M(\vec R))_{\alpha\beta} &= \lrbra{\alpha,\vec R}(d\umat M)\rket{\beta,\vec R}\\
&= -\lrbra{\alpha,\vec R}\uvec D\rket{\beta,\vec R}\cdot d\vmc E 
- \lrbra{\alpha,\vec R}\uvec\mu\rket{\beta,\vec R}\cdot d\vmc B\\
&= -\uvec D_{\alpha\beta}(\vec R)\cdot d\vmc E - \uvec\mu_{\alpha\beta}(\vec R)\cdot d\vmc B\ .
\end{split}
\end{align}
The 2-form occurring in (\ref{eB:BP.Surface}) is
\begin{align}\label{eB:dM.dM}
\begin{split}
(d\umat M(\vec R))_{\alpha\beta}\wedge (d\umat M(\vec R))_{\beta\alpha} = \sum_{i,j=1}^{3}\phantom\wedge
&\klr{\underline D_{i,\alpha\beta}(\vec R)\,d\mcal E_i + \underline\mu_{i,\alpha\beta}(\vec R)\,d\mcal B_i}\\
\wedge&\klr{\underline D_{j,\beta\alpha}(\vec R)\,d\mcal E_j + \underline\mu_{j,\beta\alpha}(\vec R)\,d\mcal B_j}\ .
\end{split}
\end{align}
Expanding the wedge product in (\ref{eB:dM.dM}) and using (\ref{eB:wedge}) leads to three different contributions, proportional to $d\mcal E_i\wedge d\mcal E_j$, $d\mcal B_i\wedge d\mcal B_j$ and $d\mcal E_i\wedge d\mcal B_j$, respectively. The geometric phase (\ref{eB:BP.Surface}) can then be written in the form
\begin{align}\label{eB:BP.I}
\gamma_{\alpha\alpha}(\mcal C) = \int_{\mcal F}\mcal I^{(\vmc E)}_{\alpha\alpha}(\vec R)
+\int_{\mcal F}\mcal I^{(\vmc B)}_{\alpha\alpha}(\vec R)+\int_{\mcal F}\mcal I^{(\vmc E,\vmc B)}_{\alpha\alpha}(\vec R)
\end{align}
with the 2-forms
\begin{align}\label{eB:IE}
\mcal I^{(\vmc E)}_{\alpha\alpha}(\vec R) &= \frac\I2\sum_{\beta\neq\alpha}\sum_{i,j=1}^3
\frac{\underline D_{i,\alpha\beta}(\vec R)\underline D_{j,\beta\alpha}(\vec R) - (i\leftrightarrow j)}
{\big(E(\alpha,\vec R) - E(\beta,\vec R)\big)^2}d\mcal E_i\wedge d\mcal E_j\ ,\\ \label{eB:IB}
\mcal I^{(\vmc B)}_{\alpha\alpha}(\vec R) &= \frac\I2\sum_{\beta\neq\alpha}\sum_{i,j=1}^3
\frac{\underline\mu_{i,\alpha\beta}(\vec R)\underline\mu_{j,\beta\alpha}(\vec R) - (i\leftrightarrow j)}
{\big(E(\alpha,\vec R) - E(\beta,\vec R)\big)^2}d\mcal B_i\wedge d\mcal B_j\ ,\\ \label{eB:IEB}
\mcal I^{(\vmc E,\vmc B)}_{\alpha\alpha}(\vec R) &= \I\sum_{\beta\neq\alpha}\sum_{i,j=1}^3
\frac{\underline D_{i,\alpha\beta}(\vec R)\underline\mu_{j,\beta\alpha}(\vec R) - \underline\mu_{j,\alpha\beta}(\vec R)\underline D_{i,\beta\alpha}(\vec R)}
{\big(E(\alpha,\vec R) - E(\beta,\vec R)\big)^2}d\mcal E_i\wedge d\mcal B_j\ .
\end{align}
For general external field configurations, the geometric phase (\ref{eB:BP.I}) will get non-vanishing contributions from all three surface-integrals.

Consider now a closed curve $\mcal C$ in parameter space with constant magnetic field. If it is not prevented by singularities of the integrand we can choose also $\mcal F$ to correspond to this constant magnetic field. Then only the integral over $\mcal I^{(\vmc E)}_{\alpha\alpha}(\vec R)$ will be different from zero. The 2-form $\mcal I^{(\vmc E)}_{\alpha\alpha}(\vec R)$ (\ref{eB:IE}) is a product of two antisymmetric tensors.
Introducing
\begin{align}\label{eB:df.E}
df_\ell^{(\vmc E)} &= \frac12\sum_{i,j=1}^3\varepsilon_{ij\ell}\,d\mcal E_i\wedge d\mcal E_j\ ,\\ \label{eB:J.E}
\mcal J^{(\vmc E)}_{\ell,\alpha\alpha}(\vec R) &= 
\frac\I2\sum_{i,j=1}^3\varepsilon_{ij\ell}\klr{
\sum_{\beta\neq\alpha}\frac{\underline D_{i,\alpha\beta}(\vec R)\underline D_{j,\beta\alpha}(\vec R) - (i\leftrightarrow j)}
{\big(E(\alpha,\vec R) - E(\beta,\vec R)\big)^2}
}\ ,\\ \nonumber
(\ell &= 1,2,3)
\end{align}
we can write (\ref{eB:IE}) as a scalar product,
\begin{align}\label{eB:I.Jdf}
\mcal I^{(\vmc E)}_{\alpha\alpha}(\vec R) &= 
\vmc J^{(\vmc E)}_{\alpha\alpha}(\vec R)\cdot d\vec f^{(\vmc E)} = \sum_{\ell=1}^3
\mcal J_{\ell,\alpha\alpha}^{(\vmc E)}(\vec R)df_\ell^{(\vmc E)}\ ,
\end{align}
and obtain from (\ref{eB:BP.I})
\begin{align}\label{eB:BP.E}
\gamma_{\alpha\alpha}(\mcal C) = 
\int_{\mcal F}\vmc J^{(\vmc E)}_{\alpha\alpha}(\vec R)\cdot d\vec f^{(\vmc E)}\ .
\end{align}
Obviously, $d\vec f^{(\vmc E)}$ (\ref{eB:df.E}) is the oriented surface element in the 3-dimensional space of electric field strengths and $\vmc J_{\alpha\alpha}^{(\vmc E)}(\vec R)$ can be interpreted as a geometric flux density field in the space of electric field strengths. The geometric phase for a given closed path $\mcal C$ in the space of electric field strengths is given by the flux of the vector field $\vmc J^{(\vmc E)}_{\alpha\alpha}(\vec R)$ through $\mcal C$. The graphical visualisation of the flux density field (\ref{eB:J.E}) turns out to be a very useful tool for choosing closed curves $\mcal C$ in parameter space with desired properties, for instance minimising the decay rate for the metastable states along with maximising the geometric phase.

For a closed curve $\mcal C$ in parameter space with constant electric field the geometric phase $\gamma_{\alpha\alpha}(\mcal C)$ can be implemented in a similar way. We only have to replace in (\ref{eB:df.E}) to (\ref{eB:BP.E}) $\vmc E$ by $\vmc B$, the 2-form $\mcal I^{(\vmc E)}_{\alpha\alpha}(\vec R)$ by the 2-form $\mcal I^{(\vmc B)}_{\alpha\alpha}(\vec R)$, and $\uvec D$ by $\uvec\mu$. For general external field configurations where both $\vmc E$ and $\vmc B$ vary we still can define the flux density fields $\vmc J_{\alpha\alpha}^{(\vmc E)}(\vec R)$ and $\vmc J_{\alpha\alpha}^{(\vmc B)}(\vec R)$ and write the geometric phase (\ref{eB:BP.I}) as
\begin{align}\label{e:BP.surface}
\gamma_{\alpha\alpha}(\mcal C) = 
  \int_{\mcal F}\vmc J^{(\vmc E)}_{\alpha\alpha}(\vec R)\cdot d\vec f^{(\vmc E)}
+ \int_{\mcal F}\vmc J^{(\vmc B)}_{\alpha\alpha}(\vec R)\cdot d\vec f^{(\vmc B)}
+ \int_{\mcal F}\mcal I^{(\vmc E,\vmc B)}_{\alpha\alpha}(\vec R)
\end{align}
with $\mcal I^{(\vmc E,\vmc B)}_{\alpha\alpha}(\vec R)$ from (\ref{eB:IEB}). However, visualisation in a single three dimensional parameter space is no longer possible.

\subsection{P-conserving and P-violating flux densities}\label{sB:Perturbation}

In this section we discuss the PC and PV contributions to the flux densities and geometric phases. Therefore, the dependence of quantities on the PV parameters is kept explicitly in this section. The mass matrix for the atom in external fields and including P-violation is given in (\ref{e2:MM.full}). For ease of notation we define
\begin{align}\label{eB:Def.delta}
\delta &= \klr{\delta_1^2 + \delta_2^2}^{1/2}\ ,\\ \label{e2:Def.MPV}
\umat M_\PV &= \sum_{i=1}^2\frac{\delta_i}{\delta}\umat M_\PV^{(i)}\ ,
\end{align}
where $\delta_{1,2}$ are given in (\ref{e2:deltaPV}) to (\ref{e2:deltaPV12D}). We can now write (\ref{e2:MM.full}) and (\ref{eB:MM}) in the form
\begin{align}\label{eB:M0.short}
\umat{\tilde M}_0(\delta) &= \umat M_0 + \delta\umat M_{\PV}\ ,\\ \label{eB:MM.short}
\umat M(\vec R,\delta) &= \umat M_0 + \delta\umat M_\PV - \uvec D\cdot\vmc E - \uvec\mu\cdot\vmc B
\end{align}
where $\umat M_0$ is the mass matrix for zero external fields and no PV contribution. We use perturbation theory to obtain a power series expansion in the small parameter $\delta$ for the eigenstates of $\umat M(\vec R,\delta)$. The details of the formalism which applies to non-hermitian mass matrices can be found in \cite{BoBrNa95}, appendix C and \cite{BrGaNa99}, appendix C.

The perturbation series expansion of the eigenvalues of $\umat M(\vec R,\delta)$ reads
\begin{align}\label{eB:E.PE}
E(\alpha,\vec R,\delta) = E(\alpha^{(0)},\vec R) + \mcal O(\delta^2)\ .
\end{align}
There are no $\delta$-linear terms in these eigenvalues due to time reversal symmetry invariance, see section 1.3 in \cite{BrGaNa99}. The perturbation series expansion for the left and right eigenvectors of $\umat M(\vec R,\delta)$ reads
\begin{align}\label{eB:rket.PE}
\rket{\alpha,\vec R,\delta} &= \rket{\alpha^{(0)},\vec R} + \delta\rket{\alpha^{(1)},\vec R} + \mcal O(\delta^2)\ ,\\ \label{eB:rket.PE1}
\rket{\alpha^{(1)},\vec R} &= \sum_{\beta\neq\alpha}\rket{\beta^{(0)},\vec R}\frac{\umat M_{\PV,\beta\alpha}(\vec R)}{E(\alpha^{(0)},\vec R) - E(\beta^{(0)},\vec R)}\ ,\\ \label{eB:lrbra.PE}
\lrbra{\alpha,\vec R,\delta} &= \lrbra{\alpha^{(0)},\vec R} + \delta\lrbra{\alpha^{(1)},\vec R} + \mcal O(\delta^2)\ ,\\ \label{eB:lrbra.PE1}
\lrbra{\alpha^{(1)},\vec R} &= \sum_{\beta\neq\alpha}\frac{\umat M_{\PV,\alpha\beta}(\vec R)}{E(\alpha^{(0)},\vec R) - E(\beta^{(0)},\vec R)}\lrbra{\beta^{(0)},\vec R}\ .
\end{align}
In zeroth order we are left with the eigenstates $\rket{\alpha^{(0)},\vec R}$ of the mass matrix $\umat M(\vec R,0)$ of (\ref{eB:MM.short}). In first order contributions proportional to the matrix elements of $\umat M_\PV$ occur which are defined as
\begin{align}\label{eB:MPV.Matelem}
\umat M_{\PV,\alpha\beta}(\vec R) = \lrbra{\alpha^{(0)},\vec R}\umat M_\PV\rket{\beta^{(0)},\vec R}\ .
\end{align}
The geometric flux density (\ref{eB:J.E}) which has PC and PV contributions, is given by
\begin{align}\label{eB:J.E.2}
\mcal J^{(\vmc E)}_{\ell,\alpha\alpha}(\vec R,\delta) &= 
\frac\I2\sum_{i,j=1}^3\varepsilon_{ij\ell}\klr{
\sum_{\beta\neq\alpha}\frac{\underline D_{i,\alpha\beta}(\vec R,\delta)\underline D_{j,\beta\alpha}(\vec R,\delta) - (i\leftrightarrow j)}
{\big(E(\alpha^{(0)},\vec R) - E(\beta^{(0)},\vec R)\big)^2}
}\ ,\\ \nonumber
(\ell &= 1,2,3)\ .
\end{align}
The matrix elements of the dipole operator $\uvec D$ in (\ref{eB:J.E.2}) are
\begin{align}
\uvec D_{\alpha\beta}(\vec R,\delta) = \lrbra{\alpha,\vec R,\delta}\uvec D\rket{\beta,\vec R,\delta}
\end{align}
and by using (\ref{eB:rket.PE})ff. we obtain, to first order in $\delta$,
\begin{align}\label{eB:Dip.PCPV}
\uvec D_{\alpha\beta}(\vec R,\delta) &= \uvec D^\PC_{\alpha\beta}(\vec R) 
+ \delta\uvec D^\PV_{\alpha\beta}(\vec R)\ ,\\ \label{eB:Dip.PC}
\uvec D^\PC_{\alpha\beta}(\vec R) &= \lrbra{\alpha^{(0)},\vec R}\uvec D\rket{\beta^{(0)},\vec R}\ ,\\ \label{eB:Dip.PV}
\uvec D^{\PV}_{\alpha\beta}(\vec R) &= \sum_{\gamma\neq\alpha}
\frac{\umat M_{\PV,\alpha\gamma}(\vec R)\uvec D^\PC_{\gamma\beta}(\vec R)}{E(\alpha^{(0)},\vec R) - E(\gamma^{(0)},\vec R)}
+\sum_{\gamma\neq\beta}
\frac{\uvec D^\PC_{\alpha\gamma}(\vec R)\umat M_{\PV,\gamma\beta}(\vec R)}{E(\beta^{(0)},\vec R) - E(\gamma^{(0)},\vec R)}
\ .
\end{align}

Using (\ref{eB:Dip.PCPV}) to (\ref{eB:Dip.PV}), we are able to separate the PC and PV contributions to the geometric flux density (\ref{eB:J.E.2}) and find
\begin{align}\label{eB:J.E.PCPV}
\mcal J^{(\vmc E)}_{\ell,\alpha\alpha}(\vec R,\delta) &= 
\mcal J^{(\vmc E,\PC)}_{\ell,\alpha\alpha}(\vec R) 
+ \delta\mcal J^{(\vmc E,\PV)}_{\ell,\alpha\alpha}(\vec R)\ ,\\ \label{eB:J.E.PC}
\mcal J^{(\vmc E,\PC)}_{\ell,\alpha\alpha}(\vec R) &= 
\frac\I2\sum_{i,j=1}^3\varepsilon_{ij\ell}\klr{
\sum_{\beta\neq\alpha}\frac{\underline D^\PC_{i,\alpha\beta}(\vec R)\underline D^\PC_{j,\beta\alpha}(\vec R) - (i\leftrightarrow j)}{\big(E(\alpha^{(0)},\vec R) - E(\beta^{(0)},\vec R)\big)^2}}\ ,\\ \label{eB:J.E.PV}
\begin{split}
\mcal J^{(\vmc E,\PV)}_{\ell,\alpha\alpha}(\vec R) &= \frac\I2\sum_{i,j=1}^3\varepsilon_{ij\ell}
\Bigg(\sum_{\beta\neq\alpha}
\frac{\underline D^\PC_{i,\alpha\beta}(\vec R)\underline D^\PV_{j,\beta\alpha}(\vec R) 
+ \underline D^\PV_{i,\alpha\beta}(\vec R)\underline D^\PC_{j,\beta\alpha}(\vec R) - (i\leftrightarrow j)
}{\big(E(\alpha^{(0)},\vec R) - E(\beta^{(0)},\vec R)\big)^2}\Bigg)\ .
\end{split}
\end{align}
The corresponding results for the flux density field $\vmc J^{(\vmc B)}_{\alpha\alpha}(\vec R,\delta)$ follow from straightforward substitution of $\vmc E$ by $\vmc B$ and $\uvec D$ by $\uvec\mu$ in (\ref{eB:J.E.2}) to (\ref{eB:J.E.PV}).

We will now investigate the PC and PV character of the geometric flux densities (\ref{eB:J.E.PC}) and (\ref{eB:J.E.PV}). We define the quasiprojectors for the mass matrix $\umat M(\vec R,0) = \umat M(\vmc E,\vmc B,0)$ (\ref{eB:MM.short}) as
\begin{align}\label{e:QPro}
\Pro^{(0)}_{\alpha}(\vmc E,\vmc B) = \rket{\alpha^{(0)},\vmc E,\vmc B}\lrbra{\alpha^{(0)},\vmc E,\vmc B}\ .
\end{align}
Here and in the following we write out $\vec R$ as $(\vmc E,\vmc B)$, see (\ref{e3:R}). With the completeness relation
\begin{align}
 \um = \sum_{\alpha}\Pro^{(0)}_{\alpha}(\vmc E,\vmc B)
\end{align}
we obtain from (\ref{eB:J.E.PC}) and (\ref{eB:Dip.PC})
\begin{align}\label{e:J.E.PC.Pro}
\begin{split}
\mcal J^{(\vmc E,\PC)}_{\ell,\alpha\alpha}(\vmc E,\vmc B) = 
\frac\I2\sum_{i,j=1}^3\varepsilon_{ij\ell}\Tr\Bigg[&\underline D_i\klr{
\sum_{\beta\neq\alpha}\frac{\Pro_\beta^{(0)}(\vmc E,\vmc B)}{\big(E(\alpha^{(0)},\vmc E,\vmc B) - E(\beta^{(0)},\vmc E,\vmc B)\big)^2}}\\
\times~&\underline D_j\Pro_\alpha^{(0)}(\vmc E,\vmc B)-(i\leftrightarrow j)\Bigg]\ .
\end{split}
\end{align}
Consider now the parity transformation operator $P$ defined by
\begin{align}
 P:\quad \vec x\ \longrightarrow\ -\vec x
\end{align}
and its matrix representation $\umc P$ in the $(n=2)$ subspace of atomic states,
\begin{align}
\umc P\rket{2L_J,F,F_3} = (-1)^L\rket{2L_J,F,F_3}\ .
\end{align}
With $\umc P$ we have
\begin{align}
 \umc P\,\uvec x\,\umc P^\dag &= -\uvec x\ ,\\ \label{e:P.One}
\umc P^\dag\umc P &= \um\ ,\\ \label{e:P.D}
\umc P\,\uvec D\,\umc P^\dag &= -\uvec D\ ,\\ \label{e:P.mu}
\umc P\,\uvec\mu\umc P^\dag &= \uvec\mu\ .
\end{align}
Thus, the mass matrix (\ref{eB:MM.short}) satisfies for $\delta=0$
\begin{align}\label{e:P.on.M}
\umc P\,\umat M(\vmc E,\vmc B,0)\,\umc P^\dag = \umc P\,\klr{\umat M_0 - \uvec D\cdot\vmc E - \uvec\mu\cdot\vmc B}\,\umc P^\dag = \umat M(-\vmc E,\vmc B,0)\ .
\end{align}
We will now use an argumentation analogous to the one in \cite{BoBrNa95}, (3.43)ff.
Assuming non-degenerate eigenvalues of the mass matrix for the considered field configuration, the quasiprojectors (\ref{e:QPro}) are the residues of the poles of the resolvent
\begin{align}\label{e:resolvent}
\frac{1}{\umat M(\vmc E,\vmc B,0) - \xi\um} = \sum_\alpha \frac{\Pro_\alpha^{(0)}(\vmc E,\vmc B)}{E(\alpha^{(0)},\vmc E,\vmc B) - \xi}\ ,
\end{align}
where $\xi$ is an arbitrary complex parameter. Applying the parity transformation to the left hand side of (\ref{e:resolvent}) we obtain with (\ref{e:P.on.M})
\begin{align}
\umc P\,\frac{1}{\umat M(\vmc E,\vmc B,0)-\xi\um}\,\umc P^\dag = \frac{1}{\umat M(-\vmc E,\vmc B,0)-\xi\um}
= \sum_\alpha \frac{\Pro_\alpha^{(0)}(-\vmc E,\vmc B)}{E(\alpha^{(0)},-\vmc E,\vmc B) - \xi}\ .
\end{align}
Of course this must be equal to the parity transformation of the right hand side of (\ref{e:resolvent}), and we get
\begin{align}
\umc P\,\sum_\alpha \frac{\Pro_\alpha^{(0)}(\vmc E,\vmc B)}{E(\alpha^{(0)},\vmc E,\vmc B) - \xi}\,\umc P^\dag
= \sum_\alpha \frac{\Pro_\alpha^{(0)}(-\vmc E,\vmc B)}{E(\alpha^{(0)},-\vmc E,\vmc B) - \xi}\ .
\end{align}
Clearly, with our choice of $\vmc B$ (see (\ref{eA:B3})) and our numbering scheme, table \ref{t:state.labels}, the states $\rket{\alpha^{(0)},\vmc E=0,\vmc B}$ are just the states $\rket{2L_J,F,F_3,\vmc E=0,\vmc B}$ which are eigenstates of $\umc P$ with eigenvalue $(-1)^L$. Turning on the electric field we find from continuity arguments that with the numbering scheme of table \ref{t:state.labels} we have
\begin{align}\label{e:P.E}
E(\alpha^{(0)},\vmc E,\vmc B) &= E(\alpha^{(0)},-\vmc E,\vmc B)\ ,\\ \label{e:P.Pro}
\umc P\,\Pro_\alpha^{(0)}(\vmc E,\vmc B)\,\umc P^\dag &= \Pro_\alpha^{(0)}(-\vmc E,\vmc B)\ .
\end{align}
By inserting the identity (\ref{e:P.One}) between each dipole operator and quasiprojector in (\ref{e:J.E.PC.Pro}) and by using (\ref{e:P.D}), (\ref{e:P.E}) and (\ref{e:P.Pro}) we find
\begin{align}\label{e:P.J.E.PC}
\vmc J^{(\vmc E,\PC)}_{\alpha\alpha}(\vmc E,\vmc B) = \vmc J^{(\vmc E,\PC)}_{\alpha\alpha}(-\vmc E,\vmc B)\ .
\end{align}
Thus, the PC geometric flux fields are indeed invariant under parity transformation of the external fields,
\begin{align}
P:\quad(\vmc E,\vmc B)\ \longrightarrow\ (-\vmc E,\vmc B)\ .
\end{align}
For the transformation properties of the PV geometric flux density fields we use
\begin{align}\label{e:P.MPV}
\umc P\,\umat M_\PV\,\umc P^\dag = -\umat M_\PV
\end{align}
and find with the same methods as for the PC flux density fields the result
\begin{align}\label{e:P.J.E.PV}
\vmc J^{(\vmc E,\PV)}_{\alpha\alpha}(\vmc E,\vmc B) = -\vmc J^{(\vmc E,\PV)}_{\alpha\alpha}(-\vmc E,\vmc B)\ .
\end{align}
Similar relations can be derived easily for the flux densities in magnetic field space,
\begin{align}\label{e:P.J.B.PC}
\vmc J^{(\vmc B,\PC)}_{\alpha\alpha}(\vmc E,\vmc B) &= \vmc J^{(\vmc B,\PC)}_{\alpha\alpha}(-\vmc E,\vmc B)\ ,\\ \label{e:P.J.B.PV}
\vmc J^{(\vmc B,\PV)}_{\alpha\alpha}(\vmc E,\vmc B) &= -\vmc J^{(\vmc B,\PV)}_{\alpha\alpha}(-\vmc E,\vmc B)\ .
\end{align}

Finally, we consider the matrix elements of the time derivative
\begin{align}
\umat D_{\alpha\beta}(\vec R(t),\delta) = \lrbra{\alpha,\vec R(t),\delta}\frac{\partial}{\partial t}\rket{\beta,\vec R(t),\delta}\ ,
\end{align}
see (\ref{e3:D}). Inserting the expansions (\ref{eB:rket.PE}) and (\ref{eB:lrbra.PE}) we get
\begin{align}
\umat D_{\alpha\beta}(\vec R(t),\delta) = \umat D_{\PC,\alpha\beta}(\vec R(t)) + \delta\umat D_{\PV,\alpha\beta}(\vec R(t)) + \mcal O(\delta^2)\ ,
\end{align}
where
\begin{align}\label{eB:D.PC}
\umat D_{\PC,\alpha\beta}(\vec R(t)) &= \lrbra{\alpha^{(0)},\vec R(t)}\frac{\partial}{\partial t}\rket{\beta^{(0)},\vec R(t)}\ ,\\ \label{eB:D.PV}
\umat D_{\PV,\alpha\beta}(\vec R(t)) &= \lrbra{\alpha^{(1)},\vec R(t)}\frac{\partial}{\partial t}\rket{\beta^{(0)},\vec R(t)}
+ \lrbra{\alpha^{(0)},\vec R(t)}\frac{\partial}{\partial t}\rket{\beta^{(1)},\vec R(t)}\ .
\end{align}
Now we consider $\alpha=\beta$. We have from (\ref{eB:rket.PE1}) for all $t$
\begin{align}
\lrbracket{\alpha^{(0)},\vec R(t)}{\alpha^{(1)},\vec R(t)} = 0
\end{align}
which implies
\begin{align}
\lrbra{\alpha^{(0)},\vec R(t)}\frac{\partial}{\partial t}\rket{\alpha^{(1)},\vec R(t)}
+ \lrbra{\alpha^{(0)},\vec R(t)}\frac{\overset{\leftharpoonup}\partial}{\partial t}\rket{\alpha^{(1)},\vec R(t)} = 0\ .
\end{align}
Inserting this in (\ref{eB:D.PV}) for $\alpha=\beta$ and using (\ref{eB:rket.PE1}) and (\ref{eB:lrbra.PE1}) we get
\begin{align}\label{eB:D.PV.step}
\begin{split}
\umat D_{\PV,\alpha\alpha}(\vec R(t)) &= \sum_{\gamma\neq\alpha}\frac{1}{E(\alpha^{(0)},\vec R(t)) - E(\gamma^{(0)},\vec R(t))}\\
&\quad\times \Bigg\{
\umat M_{\PV,\alpha\gamma}(\vec R(t))\lrbra{\gamma^{(0)},\vec R(t)}\frac{\partial}{\partial t}\rket{\alpha^{(0)},\vec R(t)}\\
&\quad\qquad -\lrbra{\alpha^{(0)},\vec R(t)}\frac{\overset{\leftharpoonup}\partial}{\partial t}\rket{\gamma^{(0)},\vec R(t)}\umat M_{\PV,\gamma\alpha}(\vec R(t))
\Bigg\}\ .
\end{split}
\end{align}
From
\begin{align}
\lrbracket{\alpha^{(0)},\vec R(t)}{\gamma^{(0)},\vec R(t)} = \delta_{\alpha\gamma}
\end{align}
we get
\begin{align}
\lrbra{\alpha^{(0)},\vec R(t)}\frac{\overset{\leftharpoonup}\partial}{\partial t}\rket{\gamma^{(0)},\vec R(t)}
+\lrbra{\alpha^{(0)},\vec R(t)}\frac{\partial}{\partial t}\rket{\gamma^{(0)},\vec R(t)} = 0\ .
\end{align}
Inserting this in (\ref{eB:D.PV.step}) and using (\ref{eB:D.PC}) leads to
\begin{align}\label{eB:D.PV.step2}
\begin{split}
\umat D_{\PV,\alpha\alpha}(\vec R(t)) &= \sum_{\gamma\neq\alpha}\frac{1}{E(\alpha^{(0)},\vec R(t)) - E(\gamma^{(0)},\vec R(t))}\\
&\quad\times \Bigg\{
\umat M_{\PV,\alpha\gamma}(\vec R(t))\umat D_{\PC,\gamma\alpha}(\vec R(t)) + 
\umat D_{\PC,\alpha\gamma}(\vec R(t))\umat M_{\PV,\gamma\alpha}(\vec R(t))
\Bigg\}\ .
\end{split}
\end{align}
Splitting up $\delta\umat D_{\PV,\alpha\alpha}(\vec R(t))$ into the contributions proportional to $\delta_1$ and $\delta_2$ we write
\begin{align}
\delta\umat D_{\PV,\alpha\alpha}(\vec R(t)) = \delta_1\umat D_{\PV,\alpha\alpha}^{(1)}(\vec R(t)) + \delta_2\umat D_{\PV,\alpha\alpha}^{(2)}(\vec R(t))\ .
\end{align}
Inserting here (\ref{eB:D.PV.step2}) leads to the results (\ref{e3:D.PV}) and (\ref{e3:M.PV}).


\bibliography{myapvbib}

\begin{thebibliography}{25}
\expandafter\ifx\csname natexlab\endcsname\relax\def\natexlab#1{#1}\fi
\expandafter\ifx\csname bibnamefont\endcsname\relax
  \def\bibnamefont#1{#1}\fi
\expandafter\ifx\csname bibfnamefont\endcsname\relax
  \def\bibfnamefont#1{#1}\fi
\expandafter\ifx\csname citenamefont\endcsname\relax
  \def\citenamefont#1{#1}\fi
\expandafter\ifx\csname url\endcsname\relax
  \def\url#1{\texttt{#1}}\fi
\expandafter\ifx\csname urlprefix\endcsname\relax\def\urlprefix{URL }\fi
\providecommand{\bibinfo}[2]{#2}
\providecommand{\eprint}[2][]{\url{#2}}

\bibitem[{\citenamefont{Bergmann et~al.}(2007)\citenamefont{Bergmann, Gasenzer,
  and Nachtmann}}]{BeGaNa07_I}
\bibinfo{author}{\bibfnamefont{T.}~\bibnamefont{Bergmann}},
  \bibinfo{author}{\bibfnamefont{T.}~\bibnamefont{Gasenzer}}, \bibnamefont{and}
  \bibinfo{author}{\bibfnamefont{O.}~\bibnamefont{Nachtmann}},
  \emph{\bibinfo{title}{Metastable states, the adiabatic theorem and parity violating geometric phases I}}, 
  \bibinfo{journal}{HD--THEP--07--08, physics/0703275}  (\bibinfo{year}{2007}).

\bibitem[{\citenamefont{DeKieviet et~al.}(1995)\citenamefont{DeKieviet,
  Dubbers, Schmidt, Scholz, and Spinola}}]{ABSE95}
\bibinfo{author}{\bibfnamefont{M.}~\bibnamefont{DeKieviet}},
  \bibinfo{author}{\bibfnamefont{D.}~\bibnamefont{Dubbers}},
  \bibinfo{author}{\bibfnamefont{C.}~\bibnamefont{Schmidt}},
  \bibinfo{author}{\bibfnamefont{D.}~\bibnamefont{Scholz}}, \bibnamefont{and}
  \bibinfo{author}{\bibfnamefont{U.}~\bibnamefont{Spinola}},
  \bibinfo{journal}{Phys. Rev. Lett.} \textbf{\bibinfo{volume}{75}},
  \bibinfo{pages}{1919} (\bibinfo{year}{1995}).

\bibitem[{\citenamefont{Weinberg}(1967)}]{Wei67}
\bibinfo{author}{\bibfnamefont{S.}~\bibnamefont{Weinberg}},
  \bibinfo{journal}{Phys. Rev. Lett.} \textbf{\bibinfo{volume}{19}},
  \bibinfo{pages}{1264} (\bibinfo{year}{1967}).

\bibitem[{\citenamefont{Salam}(1968)}]{Sal68}
\bibinfo{author}{\bibfnamefont{A.}~\bibnamefont{Salam}}, in
  \emph{\bibinfo{booktitle}{Proceedings of the Eighth Nobel Symposium on
  Elementary Particle Theory}}, edited by
  \bibinfo{editor}{\bibfnamefont{N.}~\bibnamefont{Svartholm}}
  (\bibinfo{publisher}{Almquist and Wiskell}, \bibinfo{address}{Stockholm},
  \bibinfo{year}{1968}).

\bibitem[{\citenamefont{Glashow et~al.}(1970)\citenamefont{Glashow, Iliopoulos,
  and Maiani}}]{Gla70}
\bibinfo{author}{\bibfnamefont{S.}~\bibnamefont{Glashow}},
  \bibinfo{author}{\bibfnamefont{L.}~\bibnamefont{Iliopoulos}},
  \bibnamefont{and} \bibinfo{author}{\bibfnamefont{L.}~\bibnamefont{Maiani}},
  \bibinfo{journal}{Phys. Rev. {\bf D}} \textbf{\bibinfo{volume}{2}},
  \bibinfo{pages}{1285} (\bibinfo{year}{1970}).

\bibitem[{\citenamefont{Nachtmann}(1990)}]{Nac90}
\bibinfo{author}{\bibfnamefont{O.}~\bibnamefont{Nachtmann}},
  \emph{\bibinfo{title}{Elementary Particle Physics, Concepts and Phenomena}}
  (\bibinfo{publisher}{Springer}, \bibinfo{address}{Berlin},
  \bibinfo{year}{1990}).

\bibitem[{\citenamefont{Botz et~al.}(1995)\citenamefont{Botz, Bruss, and
  Nachtmann}}]{BoBrNa95}
\bibinfo{author}{\bibfnamefont{G.~W.} \bibnamefont{Botz}},
  \bibinfo{author}{\bibfnamefont{D.}~\bibnamefont{Bruss}}, \bibnamefont{and}
  \bibinfo{author}{\bibfnamefont{O.}~\bibnamefont{Nachtmann}},
  \bibinfo{journal}{Ann. Phys. (NY)} \textbf{\bibinfo{volume}{240}},
  \bibinfo{pages}{107} (\bibinfo{year}{1995}), \eprint{hep-ph/9406222}.

\bibitem[{\citenamefont{Guena et~al.}(2005)\citenamefont{Guena, Lintz, and
  Bouchiat}}]{Bou05}
\bibinfo{author}{\bibfnamefont{J.}~\bibnamefont{Guena}},
  \bibinfo{author}{\bibfnamefont{M.}~\bibnamefont{Lintz}}, \bibnamefont{and}
  \bibinfo{author}{\bibfnamefont{M.}~\bibnamefont{Bouchiat}},
  \bibinfo{journal}{Mod. Phys. Lett. {\bf A}} \textbf{\bibinfo{volume}{20}},
  \bibinfo{pages}{375} (\bibinfo{year}{2005}).

\bibitem[{\citenamefont{Yao et~al.}(2006)}]{PDG06}
\bibinfo{author}{\bibfnamefont{W.~M.} \bibnamefont{Yao}} \bibnamefont{et~al.}
  (\bibinfo{collaboration}{Particle Data Group}), \bibinfo{journal}{J. Phys.}
  \textbf{\bibinfo{volume}{G33}}, \bibinfo{pages}{1} (\bibinfo{year}{2006}).

\bibitem[{\citenamefont{Airapetian et~al.}(2007)}]{Air07}
\bibinfo{author}{\bibfnamefont{A.}~\bibnamefont{Airapetian}}
  \bibnamefont{et~al.} (\bibinfo{collaboration}{HERMES coll.}),
  \bibinfo{journal}{Phys. Rev.} \textbf{\bibinfo{volume}{D75}},
  \bibinfo{pages}{012007} (\bibinfo{year}{2007}).

\bibitem[{\citenamefont{Bass}(2005)}]{Bas05}
\bibinfo{author}{\bibfnamefont{S.~D.} \bibnamefont{Bass}},
  \bibinfo{journal}{Rev. Mod. Phys.} \textbf{\bibinfo{volume}{77}},
  \bibinfo{pages}{1257} (\bibinfo{year}{2005}), \eprint{hep-ph/0411005}.

\bibitem[{\citenamefont{Aschenauer and Flegel}(2006)}]{AsFl06}
\bibinfo{author}{\bibfnamefont{E.}~\bibnamefont{Aschenauer}} \bibnamefont{and}
  \bibinfo{author}{\bibfnamefont{I.}~\bibnamefont{Flegel}},
  \bibinfo{journal}{CERN Cour.} \textbf{\bibinfo{volume}{46N3}},
  \bibinfo{pages}{26} (\bibinfo{year}{2006}).

\bibitem[{\citenamefont{Bradamante et~al.}(2006)\citenamefont{Bradamante,
  Magnon, and Mallot}}]{Bra06}
\bibinfo{author}{\bibfnamefont{F.}~\bibnamefont{Bradamante}},
  \bibinfo{author}{\bibfnamefont{A.}~\bibnamefont{Magnon}}, \bibnamefont{and}
  \bibinfo{author}{\bibfnamefont{G.}~\bibnamefont{Mallot}},
  \bibinfo{journal}{CERN Cour.} \textbf{\bibinfo{volume}{46N6}},
  \bibinfo{pages}{15} (\bibinfo{year}{2006}).

\bibitem[{\citenamefont{Labzowsky et~al.}(2005)\citenamefont{Labzowsky, Shonin,
  and Solovyev}}]{LaShSo05}
\bibinfo{author}{\bibfnamefont{L.~N.} \bibnamefont{Labzowsky}},
  \bibinfo{author}{\bibfnamefont{A.~V.} \bibnamefont{Shonin}},
  \bibnamefont{and} \bibinfo{author}{\bibfnamefont{D.~A.}
  \bibnamefont{Solovyev}}, \bibinfo{journal}{J. Phys. B: At. Mol. Opt. Phys.}
  \textbf{\bibinfo{volume}{38}}, \bibinfo{pages}{265} (\bibinfo{year}{2005}).

\bibitem[{\citenamefont{Sapirstein et~al.}(2004)\citenamefont{Sapirstein,
  Pachucki, and Cheng}}]{Sap04}
\bibinfo{author}{\bibfnamefont{J.}~\bibnamefont{Sapirstein}},
  \bibinfo{author}{\bibfnamefont{K.}~\bibnamefont{Pachucki}}, \bibnamefont{and}
  \bibinfo{author}{\bibfnamefont{K.~T.} \bibnamefont{Cheng}},
  \bibinfo{journal}{Phys. Rev. A} \textbf{\bibinfo{volume}{69}},
  \bibinfo{eid}{022113} (\bibinfo{year}{2004}).

\bibitem[{\citenamefont{Bruss et~al.}(1999)\citenamefont{Bruss, Gasenzer, and
  Nachtmann}}]{BrGaNa99}
\bibinfo{author}{\bibfnamefont{D.}~\bibnamefont{Bruss}},
  \bibinfo{author}{\bibfnamefont{T.}~\bibnamefont{Gasenzer}}, \bibnamefont{and}
  \bibinfo{author}{\bibfnamefont{O.}~\bibnamefont{Nachtmann}},
  \bibinfo{journal}{Eur. Phys. J. direct} \textbf{\bibinfo{volume}{D2}},
  \bibinfo{pages}{1} (\bibinfo{year}{1999}), \eprint{hep-ph/9802317}.

\bibitem[{Plo()}]{Plots}
\urlprefix\url{http://www.thphys.uni-heidelberg.de/~bergmann/fluxdensities.pdf%
}.

\bibitem[{\citenamefont{Gasenzer}(1998)}]{DissTG}
\bibinfo{author}{\bibfnamefont{T.}~\bibnamefont{Gasenzer}}, Ph.D. thesis,
  \bibinfo{school}{Ruprecht-Karls-Universit\"at}, \bibinfo{address}{Heidelberg}
  (\bibinfo{year}{1998}), \bibinfo{note}{(unpublished)}.

\bibitem[{\citenamefont{Jackson}(2006)}]{Jac06}
\bibinfo{author}{\bibfnamefont{H.~E.} \bibnamefont{Jackson}}
  (\bibinfo{collaboration}{HERMES coll.}), \bibinfo{journal}{AIP Conf. Proc.}
  \textbf{\bibinfo{volume}{842}}, \bibinfo{pages}{363} (\bibinfo{year}{2006}),
  \eprint{hep-ex/0601006}.

\bibitem[{\citenamefont{Jentschura et~al.}(2005)\citenamefont{Jentschura,
  Kotochigova, LeBigot, Mohr, and Taylor}}]{NISTData}
\bibinfo{author}{\bibfnamefont{U.}~\bibnamefont{Jentschura}},
  \bibinfo{author}{\bibfnamefont{S.}~\bibnamefont{Kotochigova}},
  \bibinfo{author}{\bibfnamefont{E.}~\bibnamefont{LeBigot}},
  \bibinfo{author}{\bibfnamefont{P.}~\bibnamefont{Mohr}}, \bibnamefont{and}
  \bibinfo{author}{\bibfnamefont{B.}~\bibnamefont{Taylor}},
  \emph{\bibinfo{title}{The energy levels of hydrogen and deuterium (version
  2.1)}}, \bibinfo{howpublished}{National Institute of Standards and
  Technology, Gaithersburg, MD} (\bibinfo{year}{2005}),
  \urlprefix\url{http://physics.nist.gov/HDEL}.

\bibitem[{\citenamefont{Karshenboim}(2005)}]{Kar05}
\bibinfo{author}{\bibfnamefont{S.~G.} \bibnamefont{Karshenboim}},
  \bibinfo{journal}{Phys. Rep.} \textbf{\bibinfo{volume}{422}},
  \bibinfo{pages}{1} (\bibinfo{year}{2005}), \eprint{hep-ph/0509010}.

\bibitem[{\citenamefont{Erler and Ramsey-Musolf}(2005)}]{ErRa05}
\bibinfo{author}{\bibfnamefont{J.}~\bibnamefont{Erler}} \bibnamefont{and}
  \bibinfo{author}{\bibfnamefont{M.~J.} \bibnamefont{Ramsey-Musolf}},
  \bibinfo{journal}{Phys. Rev.} \textbf{\bibinfo{volume}{D 72}},
  \bibinfo{pages}{073003} (\bibinfo{year}{2005}).

\bibitem[{\citenamefont{Mohr and Taylor}(2005)}]{Moh05}
\bibinfo{author}{\bibfnamefont{P.~J.} \bibnamefont{Mohr}} \bibnamefont{and}
  \bibinfo{author}{\bibfnamefont{B.~N.} \bibnamefont{Taylor}},
  \bibinfo{journal}{Rev. Mod. Phys.} \textbf{\bibinfo{volume}{77}},
  \bibinfo{eid}{1} (\bibinfo{year}{2005}).

\bibitem[{\citenamefont{Flanders}(1963)}]{Fla63}
\bibinfo{author}{\bibfnamefont{H.}~\bibnamefont{Flanders}},
  \emph{\bibinfo{title}{Differential Forms}}, vol.~\bibinfo{volume}{11} of
  \emph{\bibinfo{series}{Mathematics in Science and Engineering}}
  (\bibinfo{publisher}{Academic Press}, \bibinfo{address}{New York},
  \bibinfo{year}{1963}).

\bibitem[{\citenamefont{Berry}(1984)}]{Ber84}
\bibinfo{author}{\bibfnamefont{M.~V.} \bibnamefont{Berry}},
  \bibinfo{journal}{Proc. R. Soc. Lond.} \textbf{\bibinfo{volume}{A392}},
  \bibinfo{pages}{45} (\bibinfo{year}{1984}).

\end{thebibliography}

\end{document}